\documentclass[sigconf]{acmart}

\usepackage{amsfonts}  
\usepackage{amsmath}
\usepackage{booktabs}  
\usepackage{pifont}

\usepackage{xcolor}
\usepackage{multirow}
\usepackage{bm}
\usepackage{array}
\usepackage{color}
\usepackage{colortbl}
\usepackage{textcomp}
\usepackage{stfloats}
\usepackage{hyperref}
\usepackage{verbatim}
\usepackage{graphicx}
\usepackage{booktabs}
\usepackage{subfig}
\usepackage{paralist}

\allowdisplaybreaks 
\usepackage{arydshln} 
\usepackage{float}
\usepackage{tcolorbox}
\usepackage{adjustbox}
\usepackage{CJKutf8}
\usepackage{tcolorbox}
\usepackage{pgfplots}

\AtBeginDocument{%
 }

\definecolor{lightblue}{rgb}{0.93, 0.95, 1.0} 

\tcbuselibrary{breakable}
\definecolor{red}{RGB}{255,44,0}
\definecolor{ired}{RGB}{229,72,72}
\definecolor{igreen}{RGB}{80,219,144}

\newcommand{\paratitle}[1]{\vspace{1.2ex}\noindent \textbf{#1}}
\newcommand{\customfont}{\linespread{0.85}\selectfont}

\copyrightyear{2025}
\acmYear{2025}
\setcopyright{acmlicensed}
\acmConference[WWW '25] {Proceedings of the ACM Web Conference 2025}{April 28--May 2, 2025}{Sydney, NSW, Australia.}
\acmBooktitle{Proceedings of the ACM Web Conference 2025 (WWW '25), April 28--May 2, 2025, Sydney, NSW, Australia}
\acmDOI{10.1145/3696410.3714739}
\acmISBN{979-8-4007-1274-6/25/04}
\begin{document}

\title[Multimodal Empathetic Response Generation]{Towards Multimodal Empathetic Response Generation:\\ A Rich Text-Speech-Vision Avatar-based Benchmark}


\author{Han Zhang}
\affiliation{%
 \institution{School of Electronic Engineering, Xidian University}
 \city{Xi'an}
 \state{Shaanxi}
 \country{China}}
\email{22021110280@stu.xidian.edu.cn}

\author{Zixiang Meng}
\affiliation{%
  \institution{School of Cyber Science and Engineering, Wuhan University}
  \city{Wuhan}
  \country{China}}
\email{zixiangmeng@whu.edu.cn}

\author{Meng Luo}
\affiliation{%
 \institution{National University of Singapore}
 \city{Singapore}
 \country{Singapore}}
\email{mluo@u.nus.edu}

\author{Hong Han}
\affiliation{%
 \institution{School of Electronic Engineering, Xidian University}
 \city{Xi'an}
 \state{Shaanxi}
 \country{China}}
\email{hanh@mail.xidian.edu.cn}

\author{Lizi Liao}
\affiliation{%
 \institution{Singapore Management University}
 \city{Singapore}
 \country{Singapore}}
\email{lzliao@smu.edu.sg}

\author{Erik Cambria}
\affiliation{%
 \institution{Nanyang Technological University}
 \city{Singapore}
 \country{Singapore}}
\email{cambria@ntu.edu.sg}

\author{Hao Fei}
\authornote{Hao Fei is the corresponding author.}
\affiliation{%
  \institution{National University of Singapore}
  \city{Singapore}
  \country{Singapore}}
\email{haofei37@nus.edu.sg}
\renewcommand{\shortauthors}{Han Zhang et al.}

\begin{abstract}
Empathetic Response Generation (ERG) is one of the key tasks of the affective computing area, which aims to produce emotionally nuanced and compassionate responses to user's queries.
However, existing ERG research is predominantly confined to the singleton text modality, limiting its effectiveness since human emotions are inherently conveyed through multiple modalities.
To combat this, we introduce an avatar-based Multimodal ERG (MERG) task, entailing rich text, speech, and facial vision information.
We first present a large-scale high-quality benchmark dataset, \textbf{AvaMERG}, which extends traditional text ERG by incorporating authentic human speech audio and dynamic talking-face avatar videos, encompassing a diverse range of avatar profiles and broadly covering various topics of real-world scenarios.
Further, we deliberately tailor a system, named \textbf{Empatheia}, for MERG. 
Built upon a Multimodal Large Language Model (MLLM) with multimodal encoder, speech and avatar generators, Empatheia performs end-to-end MERG, with Chain-of-Empathetic reasoning mechanism integrated for enhanced empathy understanding and reasoning.
Finally, we devise a list of empathetic-enhanced tuning strategies, strengthening the capabilities of emotional accuracy and content, avatar-profile consistency across modalities.
Experimental results on AvaMERG data demonstrate that Empatheia consistently shows superior performance than baseline methods on both textual ERG and MERG.
All data and code are open at \url{https://AvaMERG.github.io/}.
\end{abstract}



\maketitle

\vspace{-3mm}
\section{Introduction}

\begin{figure}[t]
 \centering
 \includegraphics[width=\columnwidth]{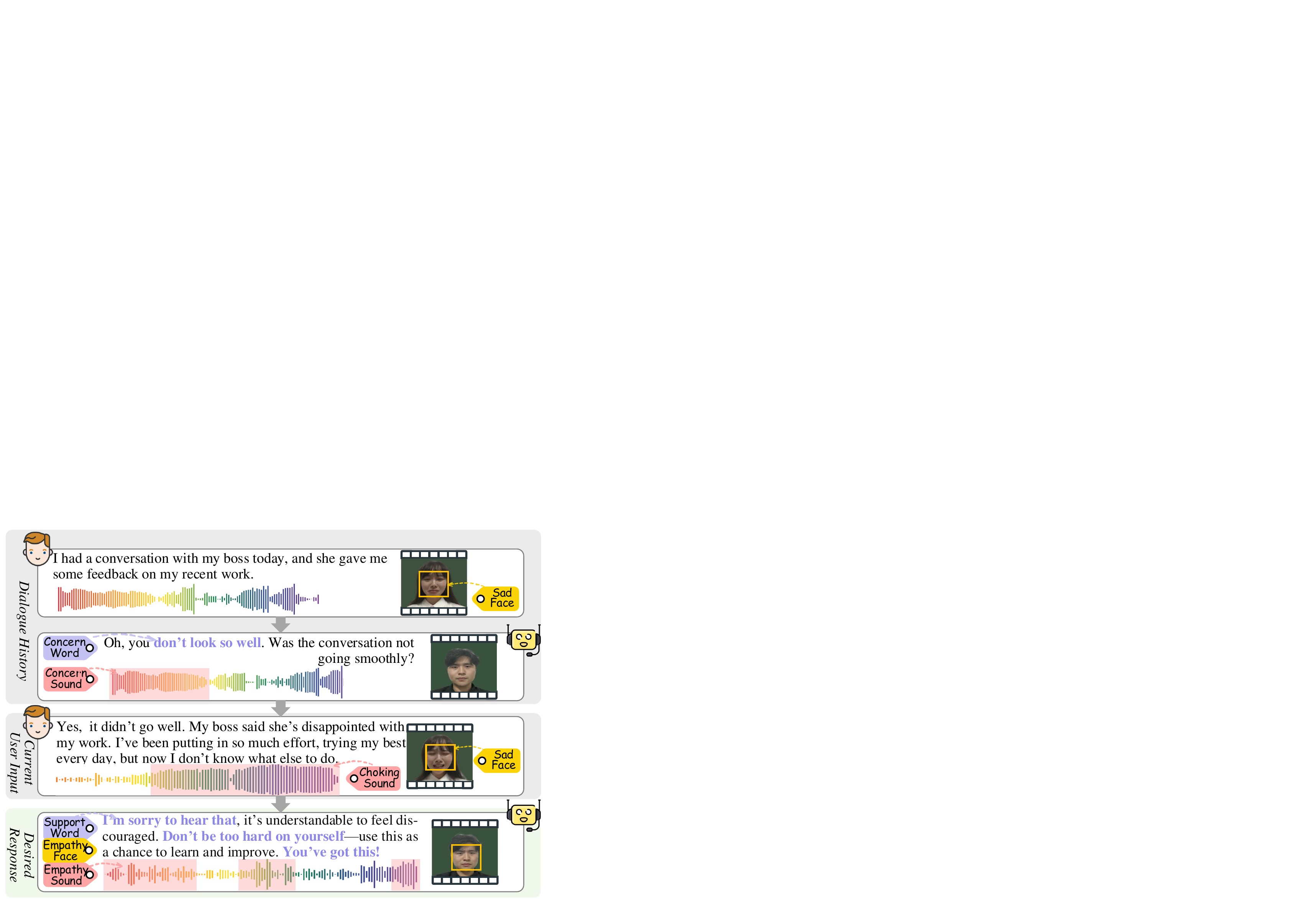}
 \caption{A snippet of avatar-based Multimodal Empathetic Response Generation (MERG) with rich multimodal signals: text (dialogue), audio (acoustic speech) and vision (dynamic talking-head avatar). 
 }
 \label{fig:intro}
 \vspace{-4mm}
\end{figure}

In recent years, the advent of Large Language Models (LLMs)~\cite{zhao2023survey,li2024survey,fei2024vitron,fei2024enhancing,wu2024towards,fei2024multimodal} has endowed machines with unprecedented levels of intelligence, bringing us closer to the realization of Artificial General Intelligence (AGI). 
However, the true essence of AGI extends beyond merely achieving human-level intelligent abilities; it must also encompass emotional understanding and empathetic capabilities comparable to those of humans. 
For instance, during human-machine interactions, it is crucial for machines to comprehend human emotions and intentions~\cite{luo2024panosent,luo2024nus,zheng2023ecqed,fei2020latent,li2022diaasq}. 
This necessity has driven the development of Empathetic Response Generation (ERG)~\cite{rashkin2018towards}, a task aimed at enabling machines to produce emotionally nuanced and compassionate responses to user queries, thereby facilitating emotion-aware conversations. 
Over the past decade, ERG has garnered significant research attention~\cite{majumder2020mime,sabour2022cem,yang2024exploiting}. 
Due to its ability to support emotional interactions with humans, ERG has been applied in various practical scenarios, such as psychological therapy and elderly companionship dialogue systems.

However, current ERG research might encounter significant challenges due to its confinement to a singleton textual modality as task definition. 
It is worthwhile to reflect on how humans naturally express emotions; in many cases, the subtleties of emotions are more effectively and comprehensively conveyed through non-textual modalities.
Specifically, in dynamic visual contexts, subtle facial expressions and body movements can communicate richer emotions and intentions.
Simultaneously, in the auditory domain, variations in speech intonation and pitch can also convey emotional states that text alone cannot express.
Figure \ref{fig:intro} demonstrates a multimodal empathetic dialogue process. 
Existing text-based ERG tasks are restricted to providing users with mere textual responses, which lack enough warmth and emotional resonance inherent in human interactions, thereby falling short of achieving adequate empathetic effects.
Furthermore, from the user's perspective, there is a desire to express emotions directly through speech or talking-facial video rather than being confined to text-based queries. 
In practical applications, numerous ERG scenarios require the ability to accept multimodal signal inputs and generate empathetic responses in multimodalities, such as in psychological therapy, companion robots, and electronic personal assistants. 
Unfortunately, there has yet to be any research on avatar-based Multimodal Empathetic Response Generation (MERG) within the community.

To bridge this gap, in this paper we present an \textbf{Avatar-based Multimodal Empathetic Response Generation} benchmark dataset (namely, \textbf{AvaMERG}). 
Building upon existing text-based ERG benchmark~\cite{rashkin2018towards}, we further augment the dataset to include multimodal signals and annotations. 
Specifically, for each utterance in the dialogue, we provide 1) authentic human-reading speech and 2) dynamic talking-face avatar videos (2D facial modeling) that both correspond to the intended emotion.
AvaMERG features a wide variety of avatar profiles and covers broad common topics of real-world scenarios, including multiple age groups, genders, vocal tones, intonations, and appearances, thereby effectively simulating a diverse range of multimodal empathetic dialogue scenarios in realistic environments.
We maintain the high quality of annotations through meticulous manual verification, guaranteeing the emotional accuracy and consistency of both the avatars' speech and video. 
Finally, we compile 33,048 annotated dialogues with 152,021 multimodal utterances, establishing a foundation for MERG research.

A direct approach to generating multimodal empathetic responses can be first producing the textual part of the response using existing text-based ERG models (e.g., high-performing LLMs), and then through a pipeline paradigm to invoke external well-trained speech generator and talking-head generator (e.g., diffusion-based models) to generate the corresponding multimodal content. 
However, there can be several non-trivial issues and inherent challenges.
\textbf{First}, ensuring the emotional accuracy across the text, audio, and video is the most fundamental capability. 
\textbf{Second}, it is essential to maintain synchronization and consistency among the three modalities in terms of content, emotion, and style. 
Pipeline models often suffer from inadequate interaction between different modules, making it difficult to guarantee consistency. 
For example, the generated speech may convey the emotion of a happy girl, while the corresponding avatar depicts a crying boy. 
\textbf{Third}, the discrete approach (where LLMs invoke external audio and video generators) can largely lead to the quality decrease of the generated content due to error propagation.

To achieve high-quality MERG, we thus propose a novel Multimodal LLM, termed \textbf{Empatheia}. 
Architecturally, we employ a multimodal encoder to feed all input signals into the central LLM for comprehension and reasoning. 
We then utilize StyleTTS2~\cite{li2024styletts} as the speech generation module and DreamTalk~\cite{ma2023dreamtalk} as the Talking Face Generation module. 
By using continuous embeddings as the medium for message passing, we connect the LLM to the frontend encoders and backend cross-modal generation modules, resulting in a full end-to-end system.
Next, we optimize Empatheia by implementing a series of tuning strategies. 
We first devise a \emph{Chain-of-Empathetic Inference} to assist the LLM to reason step-by-step, from understanding the emotion to identifying the underlying rationale and intent, and ultimately determining how to respond to the user's input. 
Then, we introduce \emph{Content Consistency Learning}, which encourages the LLM to guide the two backend modules to produce speech and talking-face avatar videos that align with the empathetic textual content. 
Further, we propose a \emph{Style-aware Alignment and Consistency Learning} mechanism to accurately identify the style signals transmitted by the central LLM, and ensure consistency in the style of both speech and video avatars, including emotion and profile. 
Finally, we perform overall MERG tuning to achieve overall high-quality multimodal empathetic responses.

We conduct experiments on the AvaMERG dataset, where the results demonstrate that our Empatheia system generates both textual and multimodal empathetic responses of higher quality compared to baseline models. 
In-depth analyses further reveal the underlying rationales for our model's advancements. 
Overall, this work pioneers the research of MERG, contributing a benchmark dataset and a strong-performing end-to-end MERG model, laying a solid foundation for future exploration in multimodal empathetic response generation.

\vspace{-4mm}
\section{Related Work}

ERG~\cite{raamkumar2022empathetic,qian2023harnessing} is one of the crucial tasks within the field of affective computing, which aims at enabling dialogue models to produce responses imbued with empathy during human-machine conversations. 
Due to its significant practical applications, ERG has attracted substantial and sustained prior research attention~\cite{lin2019moel,zhou2022case,fei2024empathyear}. 
Existing studies have developed various methods to enhance the performance of ERG systems~\cite{gao2021improving,yang2024enhancing,sabour2022cem,chen2024empathetic}. 

Yet current ERG approaches can be limited to a single text modality, which significantly restricts their effectiveness. 
In real-world dialogue scenarios, multiple modalities are often involved. 
As previously emphasized, multimodal information is crucial for generating more empathetic responses. 
Therefore, this paper tries to pioneer the research of Multimodal Empathetic Response Generation (MERG) by presenting a novel benchmark. 
It is also noteworthy that several recent related works have also touched upon multimodal ERG~\cite{yan-etal-2024-talk,zhang-etal-2024-stickerconv}. 

However, we emphasize that these studies do not fully address or cover all the modalities most relevant to empathy. 
Intuitively, both audio (capturing variations in a person’s tone) and visual (capturing facial expressions) modalities can be important, and need to be simultaneously addressed. 
Moreover, it is insufficient to rely solely on emoticon-type visual features. 
Effective ERG that closely aligns with real-world application scenarios should present authentic facial visual signals.

Unlike existing text-based ERG models and methods, achieving multimodal emotional understanding and generating multimodal signals requires the utilization of multimodal-related technologies. 
First, our approach is related to research on Multimodal Large Language Models (MLLMs), with our system being based on a backbone MLLM. 
Various MLLMs, such as LLaVA~\cite{liu2024visual}, MiniGPT-4~\cite{zhu2023minigpt}, have been investigated and widely validated for their strong semantic understanding capabilities. 
However, most MLLMs are limited to multimodal information comprehension yet do not support the flexible generation of diverse modal content beyond text~\cite{li2023blip,su2023pandagpt,bai2023qwen}, such as audio and visual outputs.
Although there are a few MLLMs that support the generation of various modal signals, such as NExT-GPT~\cite{wu24next} and Unified-IO 2~\cite{lu2024unified}, these models, unfortunately, are only capable of understanding and generating signals in general scenarios. 
They lack sufficient capabilities in emotion detection and emotional content generation. 
In other words, these MLLMs are unable to generate emotionally expressive speech or talking-face avatars.
Therefore, we consider developing a novel MLLM for MERG, which is able to accurately generate emotionally charged speech and talking-face avatar videos. 
Additionally, we design a series of emotion-enhancement training strategies to ensure that our MLLM possesses highly-performing MERG capabilities.

\section{AvaMERG Benchmark}

\begin{figure}[!t]
\centering
\hspace{-4mm}
\subfloat[Distribution of the utterance.]{
		\includegraphics[width=0.62\columnwidth]{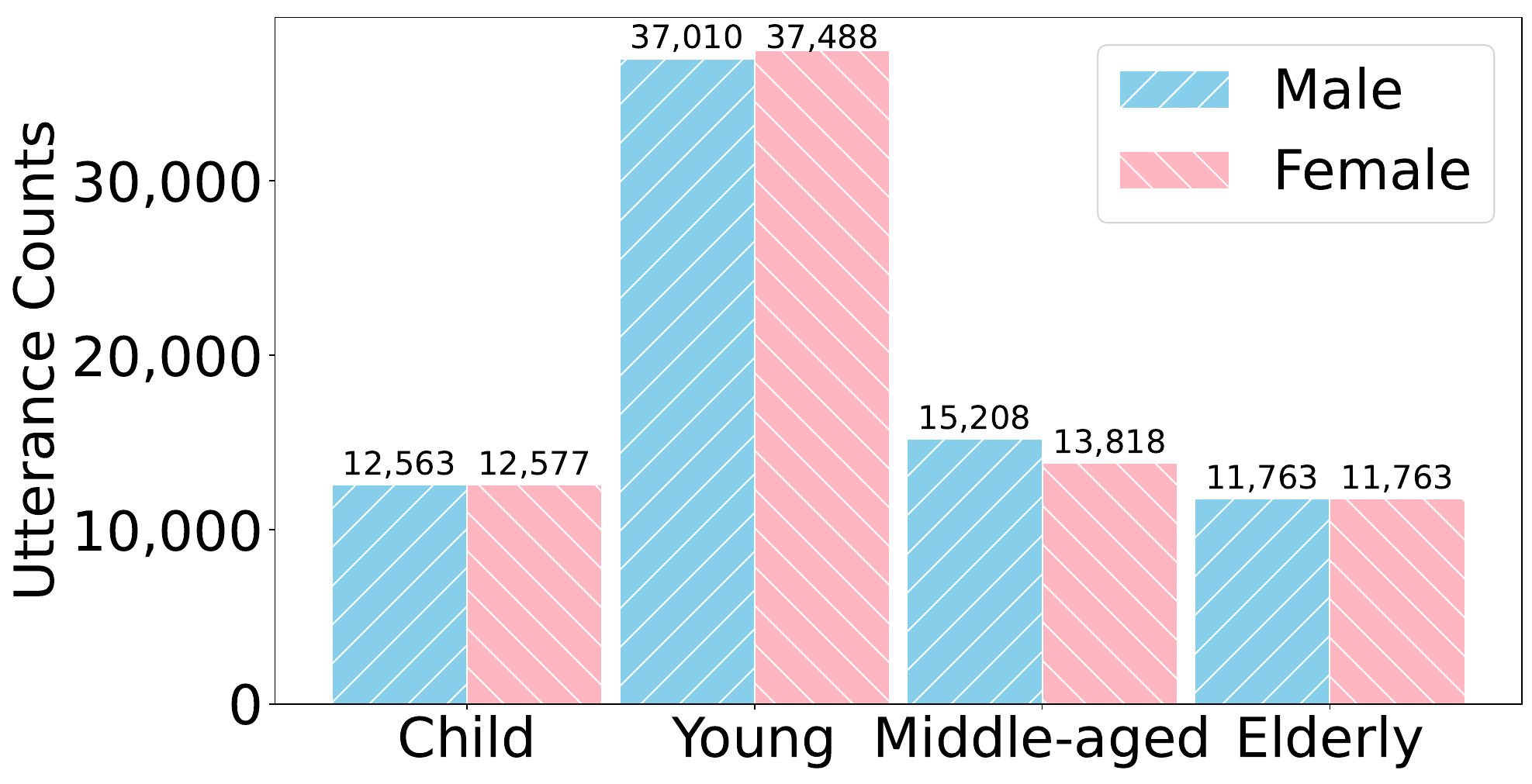}}
\subfloat[Eemotion distribution.]{
		\includegraphics[width=0.38\columnwidth]{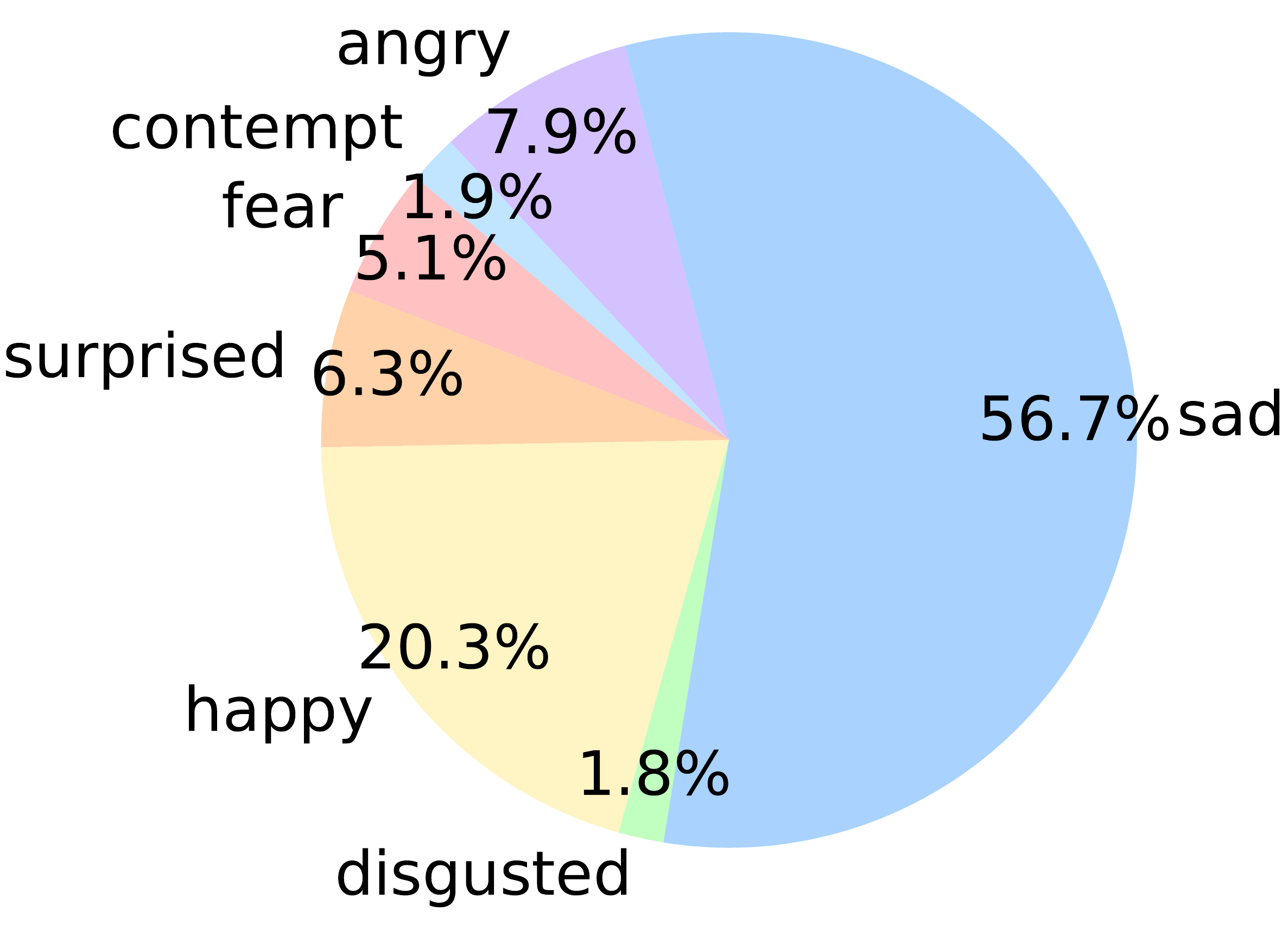}}
\vspace{-2mm}
\caption{Visualized statistics of AvaMERG dataset.}
\label{fig_statistic}
\end{figure}

\begin{table}[!t]
\fontsize{8}{9}\selectfont
\setlength{\tabcolsep}{3.5mm}
\centering
\vspace{-2mm}
\caption{Statistics of AvaMERG dataset. 
}
\vspace{-3mm}

\begin{tabular}{llr}
\toprule
\multicolumn{2}{c}{\textbf{Item}} & \bf{Stats} \\
\midrule
\multirow{6}{*}{\bf Dialogue}
 & \#Train Set & 24,696\\
 & \#Valid Set & 4,373\\
 & \#Test Set & 3,979\\
 & \#Total & 33,048\\
 \cdashline{2-3}
 & Avg. Words Per Utterance & 14.68 \\
 & Avg. Utterance Per Dialigue & 4.6 \\ 
\hdashline
\multirow{4}{*}{\bf Modality}
 & Utterance Text & 152,021\\
 & Speech Audio & 152,021\\
 & Talking-head Video & 152,021\\
 \cdashline{2-3}
 & Avg. Length (Sec) Per Aud/Vid & 5.67 \\
\hdashline
\multirow{5}{*}{\bf Avatar}
 & Child (Male/Female) & 3/3 \\
 & Young (Male/Female) & 25/17 \\
 & Middle-aged (Male/Female) & 4/4 \\
 & Elderly (Male/Female) & 5/4 \\
 \cdashline{2-3}
 & Tone (Emphatic/Mild/Gentle) & 14/38/13 \\
 \cdashline{2-3}
 & Race & 5\\ 
\hdashline
\multirow{1}{*}{\bf Emotion}
& Text/Multimodal & 32/7\\ 
\hdashline
\multicolumn{2}{l}{\multirow{1}{*}{\bf Topic\&Scenario}}
& 10\\ 
\bottomrule
\end{tabular}
\vspace{-3mm}
\label{tab:data statistic}
\end{table}

\subsection{Task Definition of MERG}

Given a multimodal dialogue $\hat{D}$=$(Q_i | D_{<i})$, where $Q_i$ denotes the current $i$-th round multimodal user query input, and $D_{<i}$ represents the dialogue history, MERG task is to produce a contextually appropriate and empathetic multimodal response $R_i$ for $Q_i$, with each utterance (i.e., $Q_i$ and $R_i$) consisting of three content-synchronized modalities: text $t_i$, speech audio $s_i$, and talking-face video $v_i$, i.e., $Q_i$/$R_i$=$(t^{q/r}_i, s^{q/r}_i, v^{q/r}_i)$.
This results in $D_i$=$\{(Q_1, R_1),\dots,$ $(Q_{i}, R_{i})\}$, a total of $i$ round of a multimodal dialogue, includes the user query $Q_{i}$ and model response $R_{i}$.
The task requires maintaining coherence and emotional congruence across these modalities to ensure that the generated response $R_{i}$ well aligns with the emotional cues in user input and also context.

\vspace{-2mm}
\subsection{Dataset Construction}

We construct our \emph{Ava-MERG} dataset by augmenting the existing pure-text ERG dataset, \emph{Empathetic Dialogue} (ED)~\cite{rashkin2018towards}, where the textual empathetic response $t_i \in R_i$ with the query's corresponding emotion categories.
First, we consider enriching the data with the identity information for both participants in the dialogue, including ages, genders, and also tone, such that MERG models can learn the correct avatar profile for both audio and video.

As the OpenAI GPT-4\footnote{\url{https://openai.com/index/gpt-4/}, June, 2024} has been validated for its remarkable performance in context understanding and thus extensively employed for data generation~\cite{wu24next,luo2024panosent}, here we also adopt GPT-4 for our annotation.
We define four age periods (\emph{child}, \emph{young}, \emph{middle-aged}, \emph{elderly}), 
binary genders (\emph{male}, \emph{female}),
and three vocal tones (\emph{emphatic}, \emph{mild}, \emph{gentle}).
We ask GPT-4 to determine the above labels for each utterance in ED.
Since the data in the raw ED is ill-balanced, e.g., most of the dialogues occurred between young or middle-aged participants, we further employ GPT-4 to produce more dialogue of ERG with above meta-information.
Also, GPT-4 will detect the dialogue topics.
Human annotators with 3-person cross-checking are recruited here to carefully check if the dialogue content, the meta-profile, and the topics are correct and of high quality.
This led to the textual part of our AvaMERG data.

Next, we create the multimodal part of the information.
First, we recruit a big number of English-speaking volunteers of the above different ages, genders, and vocal characteristics, and also different races (i.e., Asian, Caucasian, African, Latino, Indian).
Then, we assign and group different pairs of two participants according to the profile determined in the AvaMERG dialogue.
Next, we let these annotators carefully read the utterance text, with the correct emotional performance, including the tone, pitch, timbre and micro-facial expressions, where we then record their vocal speeches and talking-head videos.
After the recordings, we recruit another group of well-trained annotators to evaluate each dialogue for content accuracy and emotional accuracy with same 3-person cross-checking.
We ask each annotator to check:
1) whether the speech and video content match the content in textual utterance;
2) whether the speech and video style (including age, gender, tone, emotion) are consistent.
Only the instance will be accepted where all three annotators vote for approval.
This results in the final AvaMERG dataset.

\vspace{-2mm}
\subsection{Dataset Highlight}

The data statistics are detailed in \autoref{tab:data statistic} and \autoref{fig_statistic}.
Here we summarize the data characteristics that are key to MERG.
Due to the space limitation, we show the complete data description and statistics in Appendix \S \ref{More Details of Datasets}.

\paratitle{Large Scale and High Quality.} AvaMERG comprises a total of 33,048 dialogues with 152,021 utterances, which is large-scale enough to uncover the immense potential of the task.
Also the construction undergoes a rigorous manual checking involving both textual and multimodal content verification, ensuring its high quality.

\paratitle{Multimodal Dialogue.} Dialogues in AvaMERG cover three modalities: text, speech, and avatar video, which overcome the limitation of single-modality in existing textual ERG benchmarks.

\paratitle{Avatar Profile Diversity.} The avatars encompass 4 distinct age groups, with each represented by male and female in 3 different vocal tones.
Also avatars come from different races.
This rich diversity of avatar profiles ensures the robustness of the MERG.

\paratitle{Emotion Diversity.} AvaMERG includes 7 commonly occurred emotions: \emph{sad}, \emph{disgusted}, \emph{surprised}, \emph{contempt}, \emph{happy}, \emph{fear}, and \emph{angry}.

\paratitle{Broad Topic Coverage.} 
AvaMERG covers 10 primary common topics of real empathetic dialogue, along with hundreds of specific subtopics, fully covering the wide range of potential real-world applications for ERG.

\vspace{-2mm}
\section{Empatheia: MERG System}

\begin{figure}[t]
 \centering
 \includegraphics[width=0.99\columnwidth]{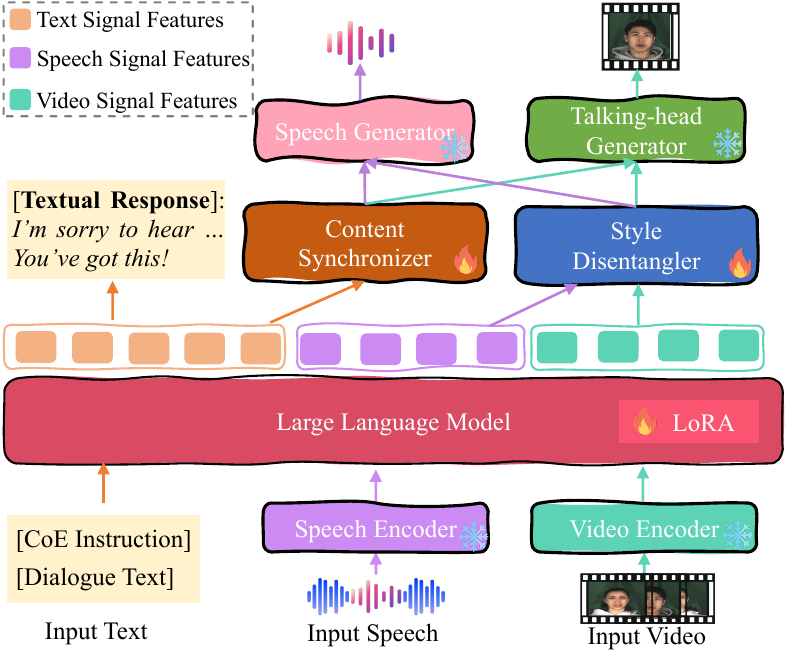}
 \vspace{-3mm}
 \caption{Architecture of our Empatheia MLLM for MERG.}
 \label{fig:overview}
 \vspace{-4mm}
\end{figure}

Figure \ref{fig:overview} illustrates the overall architecture of our Empatheia system. 
Overall, Empatheia consists of three main blocks: multimodal encoding layer, 
LLM-based core reasoning layer,
and multimodal generation layer.

\vspace{-2mm}
\subsection{Multimodal Encoder}
To perceive the multimodal dialogue inputs, we employ the HuBERT~\cite{hsu2021hubert} and CLIP ViT-L/14@336px~\cite{radford2021learning} as the speech encoder and avatar video encoder.
Essentially, the latent representations of synchronous text, speech, and talking face video should convey consistent semantics, meaning that ideally, their embeddings are aligned. 
We thus align the speech and avatar encoders' representation into the LLM's language semantic space via projections.

\vspace{-2mm}
\subsection{LLM-based Core Reasoner}

\vspace{-1mm}
\paratitle{LLM Backbone.} The LLM serves as the ``brain'' of our system, responsible for understanding multimodal signals, reasoning about appropriate empathetic responses, and sending signals for multimodal generation. 
Given that Vicuna~\cite{vicuna2023} is widely adopted as a baseline for MLLMs~\cite{dong2023dreamllm,lin2023video} and demonstrates superior performance, we select it as our backbone LLM. 
After encoding the input multimodal dialogue $\hat{D}$, LLM is expected to output the representations of 1) text tokens $r^t_i$, 
2) speech signal tokens $r^s_i$, 
and 3) video signal tokens $r^v_i$.
Here $r^s_i$ and $r^v_i$ entail rich emotion and style features, which all will be used for controlling the follow-up modules.

\paratitle{Chain-of-Empathy Reasoning.}
Empathy is an advanced human capability that is challenging to interpret, and individuals often engage in several steps of contemplation before responding as listeners.
Inspired by Chain-of-Thought~\cite{xu2024faithful,fei2023reasoning}, we design a Chain-of-Empathy (CoE) reasoning mechanism.
Specifically, we guide the LLM to think through the following progressive steps to gradually derive the final empathetic responses more accurately and more interpretably.
\vspace{-1mm}
\begin{tcolorbox}[breakable, fontupper=\customfont]
\vspace{-2mm}
{\small
\noindent$\bullet$ \textbf{\normalsize CoE Instruction}:\\
You are an empathetic conversational agent. 
Your goal is to understand the user's emotions and intentions, and respond or comfort them with appropriate language that helps them feel understood and cared for. 
Avoid rushing into your response; instead, carefully consider each step before replying by following these steps, one by one:
\setdefaultleftmargin{0.5em}{1em}{}{}{}{}
\vspace{-1mm}
\begin{compactenum} 
  \setlength{\itemsep}{-3pt}
  \item[$\blacktriangleright$] \paratitle{Step-1.} \emph{Event scenario}.
  Reflect on the event scenarios that arise from the ongoing dialogue. 
  \item[$\blacktriangleright$] \paratitle{Step-2.} \emph{User's emotion}.
  Analyze both the implicit and explicit emotions conveyed by the user.
  \item[$\blacktriangleright$] \paratitle{Step-3.} \emph{Emotion cause}.
  Infer the underlying reasons for the user’s emotions.
  \item[$\blacktriangleright$] \paratitle{Step-4.} \emph{Goal to response}.
  Determine the goal of your response in this particular instance, such as alleviating anxiety, offering reassurance, or expressing understanding.
  \item[$\blacktriangleright$] \paratitle{Step-5.} \emph{Generating empathetic response}.
  Formulate a response that addresses the user's emotions and situation, ensuring it reflects the reasoning from the previous steps. The output should be purely focused on providing a thoughtful and empathetic reply.
\end{compactenum}
}
\vspace{-2mm}
\end{tcolorbox}
\vspace{-1mm}
These steps simulate the thought process that humans typically engage in.
In the following \S \ref{Chain-of-Empathetic Learning} we expand the training of the CoE reasoning on our system.

\vspace{-2mm}
\subsection{Multimodal Generation}

\vspace{-2mm}
\paratitle{Multimodal Generator Backbones.}
Following the signal features ($r^t_i$, $r^s_i$, $r^v_i$) from LLM, the backbone speech generator and talking-head generator will produce the non-textual contents, respectively.
To ensure high-quality multimodal generation, we employ the current state-of-the-art StyleTTS2~\cite{li2024styletts} and DreamTalk~\cite{ma2023dreamtalk}, respectively.
Note that these generators are well-trained before integrating into our system.
However, directly generating speeches and dynamic avatars would largely lead to the issues of inconsistency of both content and style.
That is, two aspects of consistency are required:
1) \textbf{Consistency of content,} both the speech should be synchronized with the talking-head video, both of which should be further aligned with the textual response;
2) \textbf{Stylistic Coherence,} the style within text/speech/vision, including both the emotion and profile (age, gender, tone, appearance), should be kept consistent.
For natural and accurate MERG, maintaining synchronized content and style across modalities is crucial. 

For these purposes, we further design two modules before the two generators: content synchronizer and style disentangler.

\begin{figure}[t]
 \centering
 \includegraphics[width=1.0\columnwidth]{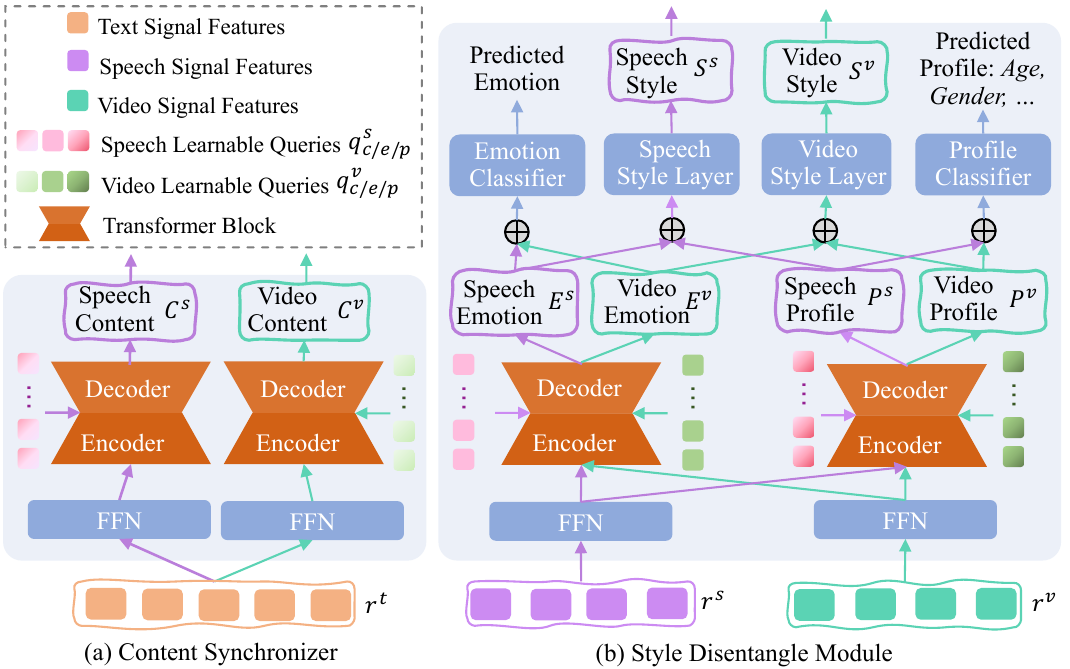}
 \caption{Illustration of the Content Synchronizer and Style Disentangle modules.}
 \label{fig:Fig2.pdf}
\vspace{-4mm}
\end{figure}

\paratitle{Content Synchronizer.}
The content synchronizer (CS) aims to ensure that the speech and vision generators receive the correct response content information.
As shown in \autoref{fig:Fig2.pdf}(a), the module is essentially a Transformer-based~\cite{vaswani2017attention} variational auto-encoder (VAE)~\cite{kingma2013auto}.
mainly consists of two transformer blocks, which
CS encodes the $r^t$ into latent representation $z_c$, from which the decoder reconstructs the content of speech $C^{s}$ and vision $C^{v}$.

\setlength\abovedisplayskip{3pt}
\setlength\belowdisplayskip{3pt}
\begin{align}
  z_c^{s/v} &= \text{Enc}^{\text{CS}}(\text{FFN}(r^t), q_c^{s/v}) \,, \\
  C^{s/v} &= \text{Dec}^{\text{CS}}(\text{FFN}(z_c^{s/v}), q_c^{s/v}) \,,
\end{align}
where $q_c^s$ and $q_c^v$ represent learnable content query features for two modalities, which are fed into the decoder along with the output from the encoder.
$C^{s}$ guides the speech generator to produce speech that correctly delivers the response text, while $C^{v}$ guides the talking-head generator to generate accurate mouth movements reflecting the response text.


\paratitle{Style Disentangler.}
Style features (including emotions and profiles) can be subtly different in speech module and vision module.
The style disentangler (SD) module thus aims to disentangle the style features from the LLM-output $r^s_i$ and $r^v_i$, for two modules, respectively.
As shown in \autoref{fig:Fig2.pdf}(b), similar to CS module, SD also uses VAE blocks to
disentangle the emotion and profile representations for speech and video:
\setlength\abovedisplayskip{3pt}
\setlength\belowdisplayskip{3pt}
\begin{align}
  z_e^{s/v} &= \text{Enc}^{\text{SD}}(\text{FFN}(r^s), q_e^{s/v}) \,, \\
  E^{s/v} &= \text{Dec}^{\text{SD}}(\text{FFN}(z_e^{s/v}), q_e^{s/v}) \,, \\
  z_p^{s/v} &= \text{Enc}^{\text{SD}}(\text{FFN}(r^s), q_p^{s/v}) \,, \\
  P^{s/v} &= \text{Dec}^{\text{SD}}(\text{FFN}(z_p^{s/v}), q_p^{s/v}) \,,
\end{align}
where $E^{s/v}$ are the disentangled emotion features.
$P^{s/v}$ are the corresponding profile features. 
$q_e^{s/v}$ and $q_p^{s/v}$ denote the learnable query features.
Then, we fuse the $E^{s/v}$ and $P^{s/v}$ by a speech/video style layer, and obtain the final speech/video style feature:
\setlength\abovedisplayskip{3pt}
\setlength\belowdisplayskip{3pt}
\begin{equation}\label{style-label}
  S^{s/v}=E^{s/v}\oplus P^{s/v} \,,
\end{equation}
which will be passed to two generators separately.
To further regulate the successful extraction of emotional and profile-aware features, we also fuse the emotion feature $E^s$ and $E^v$ into $E$, and the profile feature $P^s$ and $P^v$ into $P$.
Then we use an emotion classifier and a set of profile classifiers to predict the labels of emotion, avatar's age, gender, and tone.

\begin{figure*}[t]
 \centering
 \includegraphics[width=0.9\textwidth]{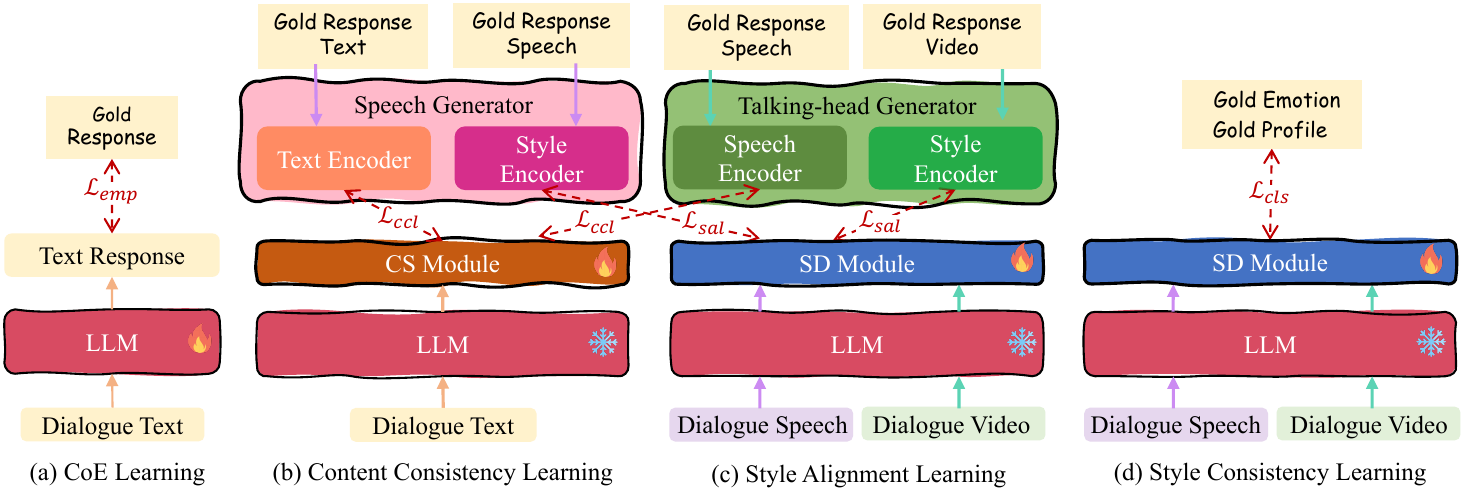}
 \vspace{-2mm}
 \caption{Illustrations of the proposed training strategies. 
 }
 \label{fig:training_stage}
\vspace{-3mm}
\end{figure*}

\vspace{-2mm}
\section{Empathetic-enhanced Training Strategy} 
With the above Empatheia model architecture, we now empower it with effective MERG capability via a series of training strategies.


\vspace{-2mm}
\subsection{Chain-of-Empathy Training}
\label{Chain-of-Empathetic Learning}
For the first stage, to teach Empatheia to learn how to perform CoE, we perform supervised fine-tuning.
For this training, we annotate a set of CoE labels based on a subset of the Ava-MERG training data.
Then, as shown in \autoref{fig:training_stage}(a), this training only updates the core LLM part for text generation, with Lora~\cite{hu2021lora} technique.
\begin{equation}
  \mathcal{L}_{emp} = - \sum_{i=1}^N \log P(x_i|x_1, \cdots, x_{i-1}) \,,
\end{equation}
where $x_i$ denotes the output token of the LLM at $i$-th time step.
Upon completion of training, the LLM is capable of not only generating empathetic responses but also providing a comprehensive CoE reasoning process.


\vspace{-2mm}
\subsection{Content Consistency Learning}
The aim of the second training stage is to encourage the content signals output by CS module to guide the multimodal generator in producing content-consistent speech and video.
This requires aligning the content representations of both sides.
Therefore, as shown in \autoref{fig:training_stage}(b), we minimize the Euclidean distance between $C^s$ and the text embedding $\hat{C^s}$ encoded by the text encoder in the speech generator, as well as the distance between $C^v$ and the audio embedding $\hat{C^v}$ encoded by the audio encoder in the talking-head generator:
\setlength\abovedisplayskip{3pt}
\setlength\belowdisplayskip{3pt}
\begin{equation}
  \mathcal{L}_{ccl} = \|C^s-\hat{C^s}\|_2^2 + \|C^v-\hat{C^v}\|_2^2 \,.
\end{equation}

Since the input text for the speech generator and the input audio for the video generator are well paired, the CS module naturally produces consistent multimodal content signal features after training.
In this stage, we keep the LLM frozen to prevent it from forgetting the empathetic response capability.

\vspace{-2mm}
\subsection{Style Alignment and Consistency Learning}

\vspace{-1mm}
\paratitle{Style Alignment Learning.}
For the third stage, on the one hand, we aim to align the style features, ensuring that the multimodal generators accurately interpret the style signals provided by the SD module. 
As illustrated in \autoref{fig:training_stage}(c), we minimize the Euclidean distance between $S^s$ (\autoref{style-label}) and the audio style features $\hat{S^s}$ encoded by the style encoder in the speech generator, as well as between $S^v$ and the video style features $\hat{S^v}$:
\begin{equation}
  \mathcal{L}_{sal} = \|S^s-\hat{S^s}\|_2^2 + \|S^v-\hat{S^v}\|_2^2 \,.
\end{equation}

\paratitle{Style Consistency Learning.}
On the other hand, the target style features are not only exclusively composed of the predefined emotion and profile features, but also include additional modality-specific representations. 
For example, video style features may depict facial variations under specific emotional states. 
To further ensure style consistency across modalities, we constrain the SD to disentangle pure emotion and profile representations. 
We here introduce two classification losses for emotion and profile prediction:
\begin{equation}
  \mathcal{L}_{cls} = \frac{1}{N}\sum_i^N(\sum_c^{M_e} y_{i,c}\log(p_{i,c}) + \sum_p^{M_p}\sum_c^{p} y_{i,c}\log(p_{i,c})) \,,
\end{equation}
where $M_e$ represents the number of emotion categories, and $M_p$ is the set of categories for gender, age, and tone. 
In this stage, we also fix the LLM to prevent loss of previously acquired capabilities.
In summary, the total loss for the third stage is:
\begin{equation}
  \mathcal{L}_{sac} = \mathcal{L}_{sal} + \mathcal{L}_{cls} \,.
\end{equation}

\vspace{-2mm}
\subsection{Overall MERG Tuning}

The previous training steps effectively decompose the MERG task into sub-processes of separate capabilities. 
To enhance the overall performance of MERG, comprehensive end-to-end fine-tuning is necessary. 
In this stage, we integrate all previous training processes, and jointly fine-tune the LLM, CS, and SD modules.
The overall loss can be denoted as:
\begin{equation}
    \mathcal{L}_{oal} = \mathcal{L}_{emp} + \alpha\mathcal{L}_{ccl} +\beta\mathcal{L}_{sac} \,.
\end{equation}

By jointly optimizing the components, we aim to improve the consistency and accuracy of the generated speech and video outputs, while maintaining the empathetic dialogue capabilities learned in earlier stages. 
Furthermore, this unified fine-tuning stage allows the model to leverage cross-modal interactions more effectively, resulting in a more robust and coherent multimodal generation system tailored to the MERG task.

\vspace{-2mm}
\section{Experiment}

\subsection{Settings}

\vspace{-2mm}
\paratitle{Baseline.}
In our preliminary experiment, to identify the most suitable backbone LLM, we compare Flan-T5 XXL~\cite{chung2024scaling}, ChatGLM3-6B~\cite{team2024chatglm}, and Vicuna-7B~\cite{vicuna2023}.
Besides MERG, we also compare the text ERG performance with existing models, including KEMP~\cite{li2022knowledge}, CEM~\cite{sabour2022cem} and CASE~\cite{zhou2022case}, where we evaluate our Empatheia using only text queries for generating textual responses only. 
Since no prior work addresses the MERG task, for the speech and video generation, we develop a pipeline-based baseline, where the LLM only outputs the invocation commands for the two backend multimodal generators, without feature embedding passing and end-to-end joint training.
It first generates response text from the LLM, then passes the text into StyleTTS2~\cite{li2024styletts} to synthesize speech, and then processes the speech using DreamTalk~\cite{ma2023dreamtalk} to generate the corresponding talking-head video.

\vspace{-1mm}
\paratitle{Evaluation Metrics.}
For the text ERG task, we employ three evaluation metrics: Emotion Accuracy (Acc), and Distinct metrics (Dist-1 and Dist-2)~\cite{li2015diversity}. 
For speech generation, we use the 5-scale Mean Opinion Score (MOS)~\cite{viswanathan2005measuring} and Similarity MOS (SMOS)~\cite{lorenzo2018voice}. 
For talking head generation, we adopt the Cumulative Probability of Blur Detection (CPBD)~\cite{narvekar2011no}, Structural Similarity Index Measure (SSIM)~\cite{wang2002universal} and SyncNet confidence score (Sync$_{cf}$)~\cite{chung2017out}.

We also consider human evaluations.
For textual ERG, we employ 4 human evaluation metrics: Empathy (Emp.), Coherence (Coh.), Informativity (Inf.), and Fluency (Flu.). 
For MERG, we newly define 6 metrics: {Speech Content Accuracy (SCA)}, {Video Content Accuracy (VCA)}, {Speech Style Accuracy (SSA)}, {Video Style Accuracy (VSA)}, {Multimodal Content Consistency (MCC)}, and {Multimodal Style Consistency (MSC)}.

\vspace{-1mm}
\paratitle{Implementation Details.}
We fine-tune our model using LoRA~\cite{hu2021lora} and DeepSpeed~\cite{rasley2020deepspeed} techniques on a single 80GB A100 GPU. 
Each Transformer block comprises four encoder-decoder modules in CS and SD modules. 
To minimize training time and costs, we utilize BF16 precision and gradient accumulation. 
Also, we pre-extract content and style features for each speech and audio sample in the training set. Due to the space limitation, we leave more experimental settings in Appendix \S \ref{Extended Experiment Settings}.

\subsection{Automatic Evaluation Results}

First, we compare the performance of different methods on textual ERG in Table \ref{tab:main-1}, where we find that the Empatheia model performs the best. 
When we remove the speech and talking-face video information, a decline in performance is observed (though it still outperforms the baseline), indicating that multimodal information aids in better empathetic understanding. 
Also, removing the CoE strategy has the greatest impact on the response text, reflecting the importance of CoE.
Next, we examine the performance of MERG in multimodal content generation, where we present the results of speech generation and avatar generation in Table \ref{tab:main-2} and Table \ref{tab:main-3}, respectively. 
It is evident that our Empatheia model consistently outperforms the pipeline system across all metrics for both speech and avatar video generation.
We also analyze the model's ablation results. 
Firstly, when using different LLMs as backbones, we observe that Vicuna achieves better performance compared to ChatGLM3 and Flan-T5, so our subsequent evaluations are based on Vicuna. 
Then, when we remove the CS and SD modules individually, we observe a degradation in results, demonstrating the importance of both modules. 
Finally, we evaluate the impact of different learning strategies, where each causes varying degrees of performance decline, thus validating their effectiveness.

\begin{table}[!t]
\fontsize{8}{8}\selectfont
\setlength{\tabcolsep}{4mm}
\centering
\caption{Comparisons of textual ERG on AvaMERG data.
$\uparrow$: the higher the better;
$\downarrow$: the lower the better.
}
\vspace{-3mm}
\begin{tabular}{lccc}
\toprule
 \bf Model & \bf Acc $\uparrow$ & \bf Dis-1 $\uparrow$ & \bf Dis-2 $\uparrow$\\
\midrule
KEMP~\cite{li2022knowledge} & 35.87 & 0.41 & 1.78 \\
CEM~\cite{sabour2022cem} & 37.32 & 0.50 & 2.07 \\
CASE~\cite{zhou2022case} & 40.96 & 0.54 & 2.14 \\
\hline
\rowcolor{lightblue} Empatheia & \bf 48.51 & \bf 2.69 & \bf 14.76 \\
w/o CoE & 46.62 & 2.49 & 12.77 \\
w/o SPC\&VID & 45.89 & 2.43 & 12.56 \\
\bottomrule
\end{tabular}
\vspace{-5mm}
\label{tab:main-1}
\end{table}

\begin{table}[!t]
\fontsize{8}{8}\selectfont
\setlength{\tabcolsep}{0.6mm}
\centering
\caption{Performance of MERG on AvaMERG for speech and talking-head avatar generation.}
\vspace{-3mm}
\begin{tabular}{lcccccc}
\toprule
\multirow{2}{*}{\bf Model} & \multicolumn{2}{c}{\bf Speech} & \multicolumn{3}{c}{\bf Talking-head Avatar} \\
\cmidrule(lr){2-3} \cmidrule(lr){4-6}
 & \bf MOS $\uparrow$ & \bf SMOS $\uparrow$ & \bf CPBD $\uparrow$ & \bf SSIM $\uparrow$ & \bf $Sync_{cf}$ $\uparrow$ \\
\midrule
Ground-Truth & 4.35 & 4.81 & 0.20 & 1 & 3.93 \\
Pipeline & 3.88 & 3.97 & 0.08 & 0.43 & 1.95 \\
\hline
Empatheia (ChatGLM3) & 3.99 & 4.08 & 0.14 & 0.45 & 2.41 \\
Empatheia (Flan-T5) & 4.07 & 4.09 & 0.14 & 0.46 & 2.26 \\
\rowcolor{lightblue} Empatheia (Vicuna) & \bf 4.16 & \bf 4.33  & \bf 0.15 & \bf 0.49 & \bf 2.76 \\
\hdashline
w/o CS & 3.90 & 4.07 & 0.08 & 0.44 & 2.21 \\
w/o SD & 3.83 & 4.10 & 0.11 & 0.41 & 2.16 \\
\hdashline
w/o $\mathcal{L}_{emp} +\mathcal{L}_{ccl} +\mathcal{L}_{sac}$ & 3.90 & 4.11 & 0.10 & 0.33 & 2.14 \\
w/o $\mathcal{L}_{ccl}$ & 4.04 & 4.25 & 0.13 & 0.45 & 2.36 \\
w/o $\mathcal{L}_{sac}$ & 4.10 & 4.29 & 0.11 & 0.41 & 2.45 \\
\bottomrule
\end{tabular}
\vspace{-3mm}
\label{tab:main-2}
\end{table}

\vspace{-2mm}
\subsection{Human Evaluation Results}

Since emotions represent a form of high-level human information, the above automatic evaluation metrics might be insufficient for assessing empathy-related capacities. 
Thus, we further present the results of human evaluations on textual ERG and MERG in Table \ref{tab:human-reult-1} and Table \ref{tab:human-reult-2}. 
It is evident that Empatheia system significantly outperforms the baselines. 
Also, the model ablation results exhibit trends similar to those observed in the automatic evaluations. 
As seen, multimodal information contributes to enhanced empathetic understanding and generation. 
The effectiveness of the CoE mechanism is further confirmed. 
Moreover, the proposed CS and SD modules, along with various sophisticated training strategies, influence the overall system performance consistently, again revealing their efficacy and importance.

\begin{table}[!t]
\fontsize{8}{8}\selectfont
\setlength{\tabcolsep}{3.4mm}
\centering
\caption{Human evaluation on textual ERG.
}
\vspace{-3mm}
\begin{tabular}{lcccc}
\toprule
 \bf Model & \bf Emp. $\uparrow$ & \bf Coh. $\uparrow$ & \bf Inf. $\uparrow$ & \bf Flu. $\uparrow$\\
\midrule
KEMP~\cite{li2022knowledge} & 2.97 & 3.11 & 2.80 & 4.13\\
CEM~\cite{sabour2022cem} & 3.18 & 3.17 & 3.15 & 4.39\\
CASE~\cite{zhou2022case} & 3.03 & 3.21 & 3.14 & 4.31\\ 
\hline
\rowcolor{lightblue} Empatheia & \bf 4.33 & \bf 4.02 & \bf 3.95 & \bf 4.67 \\
w/o SPC\&VID & 4.12 & 3.98 & 3.67 & 4.49 \\
w/o CoE & 4.03 & 3.77 & 3.49 & 4.35 \\
\bottomrule
\end{tabular}
\vspace{-1mm}
\label{tab:human-reult-1}
\end{table}

\begin{table}[!t]
\fontsize{8}{8}\selectfont
\setlength{\tabcolsep}{0.7mm}
\centering
\caption{Human evaluation on MERG.
}
\vspace{-3mm}
\begin{tabular}{@{} lcccccc @{}}
\toprule
\bf Model & \bf SCA $\uparrow$ & \bf VCA $\uparrow$ & \bf SEA $\uparrow$ & \bf VEA $\uparrow$ & \bf MCC $\uparrow$ & \bf MSC$\uparrow$ \\
\midrule
Pipeline & 3.23 & 3.28 & 3.75 & 3.62 & 3.10 & 3.19\\
\hline
\rowcolor{lightblue} Empatheia & \bf 3.92 & \bf 3.85 & \bf 4.39 & \bf 4.46 & \bf 3.98 & \bf 3.91\\
\hdashline
w/o CS & 3.46 & 3.34 & 3.78 & 3.63 & 3.29 & 3.30 \\
w/o SD & 3.55 & 3.53 & 3.84 & 3.77 & 3.45 & 3.55\\
\hdashline
w/o $\mathcal{L}_{emp}+\mathcal{L}_{ccl}+\mathcal{L}_{sac}$ & 3.33 & 3.47 & 3.92 & 3.78 & 3.51 & 3.70\\
w/o $\mathcal{L}_{ccl}$ & 3.67 & 3.50 & 4.14 & 4.25 & 3.74 & 3.79\\
w/o $\mathcal{L}_{sac}$ & 3.88 & 3.82 & 3.99 & 4.04 & 3.81 & 3.74\\
\bottomrule
\end{tabular}
\vspace{-1mm}
\label{tab:human-reult-2}
\end{table}

\begin{figure}[!t]
\centering
\includegraphics[width=0.98\columnwidth]{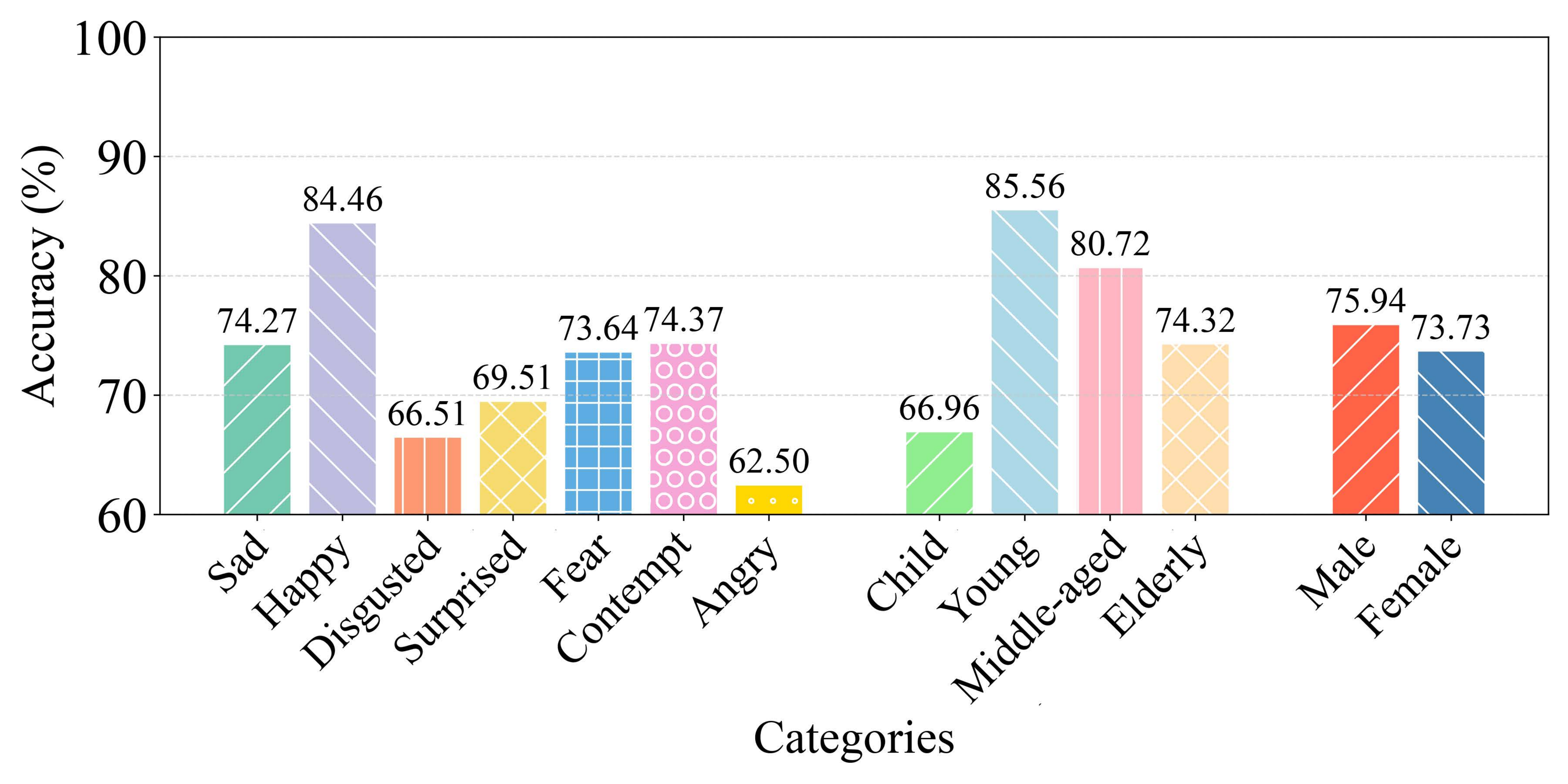}
\vspace{-4mm}
\caption{Results on various emotions, ages, and genders.}
\label{fine_grained_acc}
\vspace{-3mm}
\end{figure}

\subsection{Analyses and Discussions}

We now conduct more in-depth analyses of several key aspects of Empatheia, offering further insights for better understanding.

\paratitle{Q1. How does Empatheia perform across different emotions, genders, and age groups?}
Emotion prediction accuracy serves as an indirect measure of the model's capacity for empathetic understanding.
We first study the emotion accuracy of Empatheia under varying emotions, genders, and age groups.
As shown in \autoref{fine_grained_acc}, Empatheia is most sensitive to \emph{sad} emotions.

In terms of gender, we observe that the model performs slightly better for males compared to females, which might be attributed to the higher number of male avatars compared to female avatars in the training set.
Regarding age groups, Empatheia's accuracy in recognizing children's emotions is relatively low, potentially because children's facial expressions are more dynamic, or their emotional expression patterns differ significantly from adults.

\begin{figure}[!t]
\centering
\includegraphics[width=0.9\columnwidth]{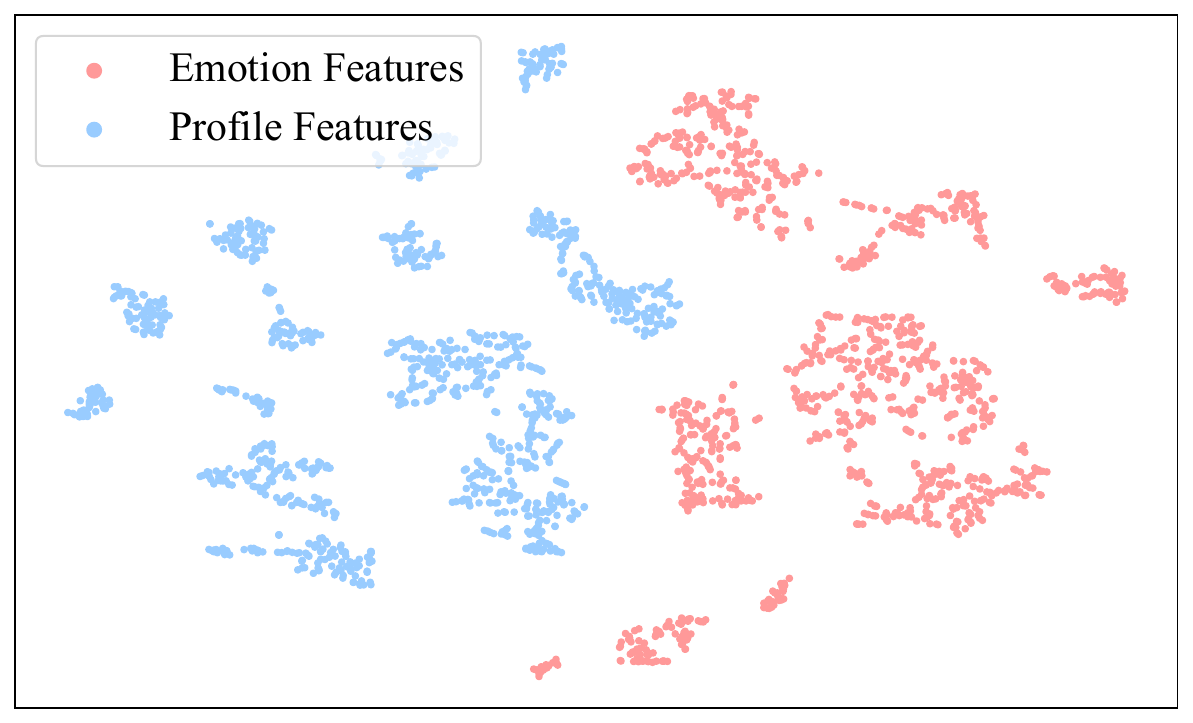}
\vspace{-3mm}
\caption{T-SNE visualization of emotion and profile features.}
\label{T-SNE-Disentangle}
\vspace{-3mm}
\end{figure}

\paratitle{Q2. Has SD module successfully disentangled emotion and profile features?}
While previous ablation experiments have validated the efficacy of the SD module, it remains uncertain whether it has fully achieved the intended goal of separating emotion and profile features.
To explore this, we present the t-SNE~\cite{van2008visualizing} visualization on the fused multimodal emotion representations in \autoref{T-SNE-Disentangle}, where we select 500 samples with varying emotions from AvaMERG.

As shown, the results indicate that SD module significantly increases the separation between different emotion categories while clustering the representations of the same emotion. 
Similarly, the patterns on profile features confirm that SD has successfully disentangled the non-emotion avatar features.

\subsection{Qualitative Case Study}

\begin{figure}[!t]
\centering
\includegraphics[width=0.99\columnwidth]{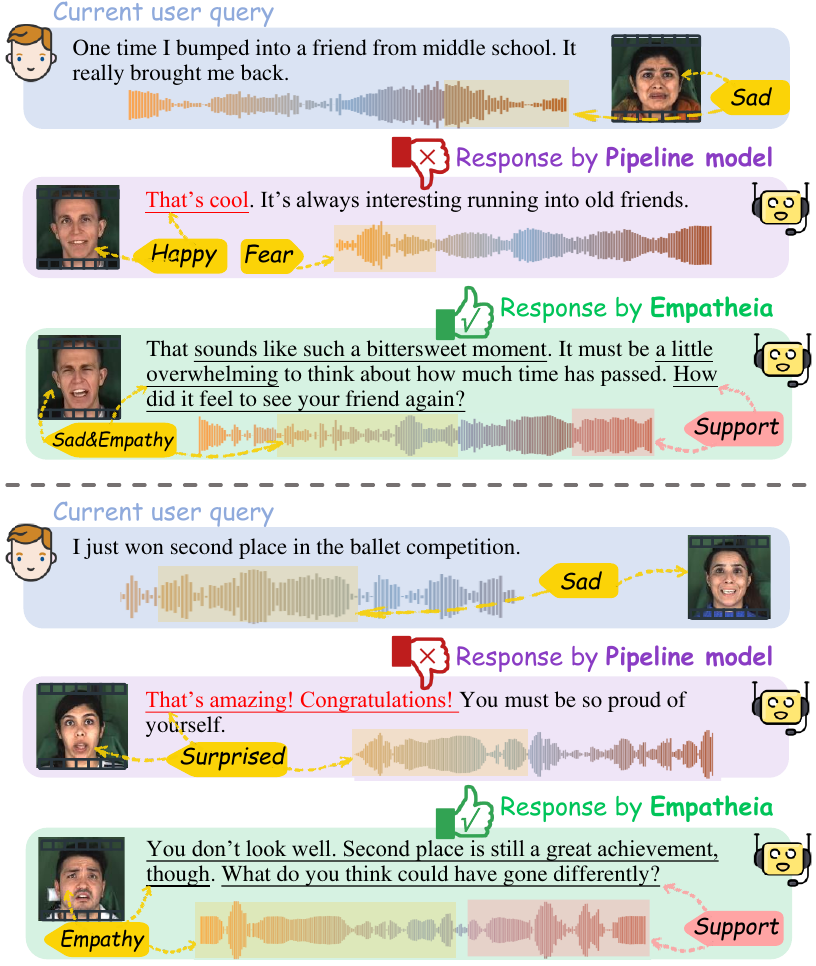}
\vspace{-3mm}
\caption{Qualitative results of two testing instances.}
\label{fig:case-study-main}
\vspace{-3mm}
\end{figure}

Finally, we present two case studies to further demonstrate the specific multimodal empathetic generation capabilities of Empatheia, as illustrated in \autoref{fig:case-study-main}, where we compare the outputs of the Pipeline baseline (without CoE). 
In the first instance, the user's text does not exhibit an explicit emotional inclination. However, the accompanying sad speech and facial expressions suggest that the user may feel sentimental about ``\emph{meeting a friend from middle school}''.
The Pipeline model, lacking the integration of the CoE strategy, generates an unempathetic response. 
Also, due to the absence of a style synchronization mechanism, there are inconsistencies in the emotions conveyed between the video and audio components.
In contrast, our Empatheia system not only produces high-quality empathetic response content but also ensures that the speech and talking avatar exhibit correct and consistent emotional expressions. 
Similarly, in the second example, the Pipeline system erroneously interprets the user's emotion, mistakenly assuming that the user is happy about securing second place, whereas Empatheia accurately identifies the user's true emotional state through comprehensive multimodal understanding. 
Furthermore, the Pipeline incorrectly assigns the avatar's identity, presenting a male voice paired with a female avatar.
On the contrary, our Empatheia shows outstanding capability in correctly handling the avatar profile consistency challenge.
In Appendix \S \ref{More Case Study} we showcase more instances for more sufficient case studies.

\section{Conclusion}

In this paper, we pioneer a novel task of avatar-based MERG.
We first introduce AvaMERG, a large-scale high-quality benchmark dataset for MERG, which extends traditional text-based ERG by integrating authentic human speech audio and dynamic talking-face avatar videos. 
AvaMERG encompasses a diverse range of avatar profiles and covers various real-world scenarios, providing a robust foundation for multimodal empathetic dialogue research.
Further, we present Empatheia, a benchmark system tailored for MERG. 
Based on a backbone LLM as the core reasoner, Empatheia leverages a multimodal encoder, speech generator, and talking-face avatar generator, forming an end-to-end system. 
We further enhance Empatheia with a Chain-of-Empathetic reasoning mechanism, and implement a series of empathetic-enhanced tuning strategies, including content consistency learning and style-aware alignment and consistency learning, to ensure emotional accuracy and content/profile consistency across modalities.
Experimental results demonstrate that Empatheia consistently outperforms baseline methods in both textual ERG and MERG tasks, highlighting the efficacy of our approach.

\clearpage


\bibliographystyle{ACM-Reference-Format}
\bibliography{reference}


\begin{thebibliography}{55}


\ifx \showCODEN    \undefined \def \showCODEN     #1{\unskip}     \fi
\ifx \showDOI      \undefined \def \showDOI       #1{#1}\fi
\ifx \showISBNx    \undefined \def \showISBNx     #1{\unskip}     \fi
\ifx \showISBNxiii \undefined \def \showISBNxiii  #1{\unskip}     \fi
\ifx \showISSN     \undefined \def \showISSN      #1{\unskip}     \fi
\ifx \showLCCN     \undefined \def \showLCCN      #1{\unskip}     \fi
\ifx \shownote     \undefined \def \shownote      #1{#1}          \fi
\ifx \showarticletitle \undefined \def \showarticletitle #1{#1}   \fi
\ifx \showURL      \undefined \def \showURL       {\relax}        \fi
\providecommand\bibfield[2]{#2}
\providecommand\bibinfo[2]{#2}
\providecommand\natexlab[1]{#1}
\providecommand\showeprint[2][]{arXiv:#2}

\bibitem[Bai et~al\mbox{.}(2023)]%
        {bai2023qwen}
\bibfield{author}{\bibinfo{person}{Jinze Bai}, \bibinfo{person}{Shuai Bai}, \bibinfo{person}{Yunfei Chu}, \bibinfo{person}{Zeyu Cui}, \bibinfo{person}{Kai Dang}, \bibinfo{person}{Xiaodong Deng}, \bibinfo{person}{Yang Fan}, \bibinfo{person}{Wenbin Ge}, \bibinfo{person}{Yu Han}, \bibinfo{person}{Fei Huang}, {et~al\mbox{.}}} \bibinfo{year}{2023}\natexlab{}.
\newblock \showarticletitle{Qwen technical report}.
\newblock \bibinfo{journal}{\emph{arXiv preprint arXiv:2309.16609}} (\bibinfo{year}{2023}).
\newblock


\bibitem[Chen et~al\mbox{.}(2024)]%
        {chen2024empathetic}
\bibfield{author}{\bibinfo{person}{Changyu Chen}, \bibinfo{person}{Yanran Li}, \bibinfo{person}{Chen Wei}, \bibinfo{person}{Jianwei Cui}, \bibinfo{person}{Bin Wang}, {and} \bibinfo{person}{Rui Yan}.} \bibinfo{year}{2024}\natexlab{}.
\newblock \showarticletitle{Empathetic Response Generation with Relation-aware Commonsense Knowledge}. In \bibinfo{booktitle}{\emph{Proceedings of the 17th ACM International Conference on Web Search and Data Mining}}. \bibinfo{pages}{87--95}.
\newblock


\bibitem[Chiang et~al\mbox{.}(2023)]%
        {vicuna2023}
\bibfield{author}{\bibinfo{person}{Wei-Lin Chiang}, \bibinfo{person}{Zhuohan Li}, \bibinfo{person}{Zi Lin}, \bibinfo{person}{Ying Sheng}, \bibinfo{person}{Zhanghao Wu}, \bibinfo{person}{Hao Zhang}, \bibinfo{person}{Lianmin Zheng}, \bibinfo{person}{Siyuan Zhuang}, \bibinfo{person}{Yonghao Zhuang}, \bibinfo{person}{Joseph~E. Gonzalez}, \bibinfo{person}{Ion Stoica}, {and} \bibinfo{person}{Eric~P. Xing}.} \bibinfo{year}{2023}\natexlab{}.
\newblock \bibinfo{title}{Vicuna: An Open-Source Chatbot Impressing GPT-4 with 90\%* ChatGPT Quality}.
\newblock
\newblock
\urldef\tempurl%
\url{https://lmsys.org/blog/2023-03-30-vicuna/}
\showURL{%
\tempurl}


\bibitem[Chung et~al\mbox{.}(2024)]%
        {chung2024scaling}
\bibfield{author}{\bibinfo{person}{Hyung~Won Chung}, \bibinfo{person}{Le Hou}, \bibinfo{person}{Shayne Longpre}, \bibinfo{person}{Barret Zoph}, \bibinfo{person}{Yi Tay}, \bibinfo{person}{William Fedus}, \bibinfo{person}{Yunxuan Li}, \bibinfo{person}{Xuezhi Wang}, \bibinfo{person}{Mostafa Dehghani}, \bibinfo{person}{Siddhartha Brahma}, {et~al\mbox{.}}} \bibinfo{year}{2024}\natexlab{}.
\newblock \showarticletitle{Scaling instruction-finetuned language models}.
\newblock \bibinfo{journal}{\emph{Journal of Machine Learning Research}} \bibinfo{volume}{25}, \bibinfo{number}{70} (\bibinfo{year}{2024}), \bibinfo{pages}{1--53}.
\newblock


\bibitem[Chung and Zisserman(2017)]%
        {chung2017out}
\bibfield{author}{\bibinfo{person}{Joon~Son Chung} {and} \bibinfo{person}{Andrew Zisserman}.} \bibinfo{year}{2017}\natexlab{}.
\newblock \showarticletitle{Out of time: automated lip sync in the wild}. In \bibinfo{booktitle}{\emph{Computer Vision--ACCV 2016 Workshops: ACCV 2016 International Workshops, Taipei, Taiwan, November 20-24, 2016, Revised Selected Papers, Part II 13}}. \bibinfo{pages}{251--263}.
\newblock


\bibitem[Dong et~al\mbox{.}(2023)]%
        {dong2023dreamllm}
\bibfield{author}{\bibinfo{person}{Runpei Dong}, \bibinfo{person}{Chunrui Han}, \bibinfo{person}{Yuang Peng}, \bibinfo{person}{Zekun Qi}, \bibinfo{person}{Zheng Ge}, \bibinfo{person}{Jinrong Yang}, \bibinfo{person}{Liang Zhao}, \bibinfo{person}{Jianjian Sun}, \bibinfo{person}{Hongyu Zhou}, \bibinfo{person}{Haoran Wei}, {et~al\mbox{.}}} \bibinfo{year}{2023}\natexlab{}.
\newblock \showarticletitle{Dreamllm: Synergistic multimodal comprehension and creation}.
\newblock \bibinfo{journal}{\emph{arXiv preprint arXiv:2309.11499}} (\bibinfo{year}{2023}).
\newblock


\bibitem[Fei et~al\mbox{.}(2023)]%
        {fei2023reasoning}
\bibfield{author}{\bibinfo{person}{Hao Fei}, \bibinfo{person}{Bobo Li}, \bibinfo{person}{Qian Liu}, \bibinfo{person}{Lidong Bing}, \bibinfo{person}{Fei Li}, {and} \bibinfo{person}{Tat-Seng Chua}.} \bibinfo{year}{2023}\natexlab{}.
\newblock \showarticletitle{Reasoning Implicit Sentiment with Chain-of-Thought Prompting}. In \bibinfo{booktitle}{\emph{Proceedings of the 61st Annual Meeting of the Association for Computational Linguistics (Volume 2: Short Papers)}}. \bibinfo{pages}{1171--1182}.
\newblock


\bibitem[Fei et~al\mbox{.}(2024a)]%
        {fei2024vitron}
\bibfield{author}{\bibinfo{person}{Hao Fei}, \bibinfo{person}{Shengqiong Wu}, \bibinfo{person}{Hanwang Zhang}, \bibinfo{person}{Tat-Seng Chua}, {and} \bibinfo{person}{Shuicheng Yan}.} \bibinfo{year}{2024}\natexlab{a}.
\newblock \showarticletitle{VITRON: A Unified Pixel-level Vision LLM for Understanding, Generating, Segmenting, Editing}.
\newblock \bibinfo{journal}{\emph{Proceedings of the Advances in neural information processing systems}}.
\newblock


\bibitem[Fei et~al\mbox{.}(2024b)]%
        {fei2024enhancing}
\bibfield{author}{\bibinfo{person}{Hao Fei}, \bibinfo{person}{Shengqiong Wu}, \bibinfo{person}{Meishan Zhang}, \bibinfo{person}{Min Zhang}, \bibinfo{person}{Tat-Seng Chua}, {and} \bibinfo{person}{Shuicheng Yan}.} \bibinfo{year}{2024}\natexlab{b}.
\newblock \showarticletitle{Enhancing video-language representations with structural spatio-temporal alignment}.
\newblock \bibinfo{journal}{\emph{IEEE Transactions on Pattern Analysis and Machine Intelligence}} (\bibinfo{year}{2024}).
\newblock


\bibitem[Fei et~al\mbox{.}(2024c)]%
        {fei2024multimodal}
\bibfield{author}{\bibinfo{person}{Hao Fei}, \bibinfo{person}{Yuan Yao}, \bibinfo{person}{Zhuosheng Zhang}, \bibinfo{person}{Fuxiao Liu}, \bibinfo{person}{Ao Zhang}, {and} \bibinfo{person}{Tat-Seng Chua}.} \bibinfo{year}{2024}\natexlab{c}.
\newblock \showarticletitle{From Multimodal LLM to Human-level AI: Modality, Instruction, Reasoning, Efficiency and Beyond}. In \bibinfo{booktitle}{\emph{Proceedings of the 2024 Joint International Conference on Computational Linguistics, Language Resources and Evaluation (LREC-COLING 2024): Tutorial Summaries}}. \bibinfo{pages}{1--8}.
\newblock


\bibitem[Fei et~al\mbox{.}(2024d)]%
        {fei2024empathyear}
\bibfield{author}{\bibinfo{person}{Hao Fei}, \bibinfo{person}{Han Zhang}, \bibinfo{person}{Bin Wang}, \bibinfo{person}{Lizi Liao}, \bibinfo{person}{Qian Liu}, {and} \bibinfo{person}{Erik Cambria}.} \bibinfo{year}{2024}\natexlab{d}.
\newblock \showarticletitle{EmpathyEar: An Open-source Avatar Multimodal Empathetic Chatbot}.
\newblock \bibinfo{journal}{\emph{arXiv preprint arXiv:2406.15177}} (\bibinfo{year}{2024}).
\newblock


\bibitem[Fei et~al\mbox{.}(2020)]%
        {fei2020latent}
\bibfield{author}{\bibinfo{person}{Hao Fei}, \bibinfo{person}{Yue Zhang}, \bibinfo{person}{Yafeng Ren}, {and} \bibinfo{person}{Donghong Ji}.} \bibinfo{year}{2020}\natexlab{}.
\newblock \showarticletitle{Latent emotion memory for multi-label emotion classification}. In \bibinfo{booktitle}{\emph{Proceedings of the AAAI conference on artificial intelligence}}, Vol.~\bibinfo{volume}{34}. \bibinfo{pages}{7692--7699}.
\newblock


\bibitem[Gao et~al\mbox{.}(2021)]%
        {gao2021improving}
\bibfield{author}{\bibinfo{person}{Jun Gao}, \bibinfo{person}{Yuhan Liu}, \bibinfo{person}{Haolin Deng}, \bibinfo{person}{Wei Wang}, \bibinfo{person}{Yu Cao}, \bibinfo{person}{Jiachen Du}, {and} \bibinfo{person}{Ruifeng Xu}.} \bibinfo{year}{2021}\natexlab{}.
\newblock \showarticletitle{Improving empathetic response generation by recognizing emotion cause in conversations}. In \bibinfo{booktitle}{\emph{Findings of the association for computational linguistics: EMNLP 2021}}. \bibinfo{pages}{807--819}.
\newblock


\bibitem[Hsu et~al\mbox{.}(2021)]%
        {hsu2021hubert}
\bibfield{author}{\bibinfo{person}{Wei-Ning Hsu}, \bibinfo{person}{Benjamin Bolte}, \bibinfo{person}{Yao-Hung~Hubert Tsai}, \bibinfo{person}{Kushal Lakhotia}, \bibinfo{person}{Ruslan Salakhutdinov}, {and} \bibinfo{person}{Abdelrahman Mohamed}.} \bibinfo{year}{2021}\natexlab{}.
\newblock \showarticletitle{Hubert: Self-supervised speech representation learning by masked prediction of hidden units}.
\newblock \bibinfo{journal}{\emph{IEEE/ACM transactions on audio, speech, and language processing}}  \bibinfo{volume}{29} (\bibinfo{year}{2021}), \bibinfo{pages}{3451--3460}.
\newblock


\bibitem[Hu et~al\mbox{.}(2021)]%
        {hu2021lora}
\bibfield{author}{\bibinfo{person}{Edward~J Hu}, \bibinfo{person}{Yelong Shen}, \bibinfo{person}{Phillip Wallis}, \bibinfo{person}{Zeyuan Allen-Zhu}, \bibinfo{person}{Yuanzhi Li}, \bibinfo{person}{Shean Wang}, \bibinfo{person}{Lu Wang}, {and} \bibinfo{person}{Weizhu Chen}.} \bibinfo{year}{2021}\natexlab{}.
\newblock \showarticletitle{Lora: Low-rank adaptation of large language models}.
\newblock \bibinfo{journal}{\emph{arXiv preprint arXiv:2106.09685}} (\bibinfo{year}{2021}).
\newblock


\bibitem[Kingma(2013)]%
        {kingma2013auto}
\bibfield{author}{\bibinfo{person}{Diederik~P Kingma}.} \bibinfo{year}{2013}\natexlab{}.
\newblock \showarticletitle{Auto-encoding variational bayes}.
\newblock \bibinfo{journal}{\emph{arXiv preprint arXiv:1312.6114}} (\bibinfo{year}{2013}).
\newblock


\bibitem[Li et~al\mbox{.}(2022a)]%
        {li2022diaasq}
\bibfield{author}{\bibinfo{person}{Bobo Li}, \bibinfo{person}{Hao Fei}, \bibinfo{person}{Fei Li}, \bibinfo{person}{Yuhan Wu}, \bibinfo{person}{Jinsong Zhang}, \bibinfo{person}{Shengqiong Wu}, \bibinfo{person}{Jingye Li}, \bibinfo{person}{Yijiang Liu}, \bibinfo{person}{Lizi Liao}, \bibinfo{person}{Tat-Seng Chua}, {et~al\mbox{.}}} \bibinfo{year}{2022}\natexlab{a}.
\newblock \showarticletitle{Diaasq: A benchmark of conversational aspect-based sentiment quadruple analysis}.
\newblock \bibinfo{journal}{\emph{arXiv preprint arXiv:2211.05705}} (\bibinfo{year}{2022}).
\newblock


\bibitem[Li et~al\mbox{.}(2015)]%
        {li2015diversity}
\bibfield{author}{\bibinfo{person}{Jiwei Li}, \bibinfo{person}{Michel Galley}, \bibinfo{person}{Chris Brockett}, \bibinfo{person}{Jianfeng Gao}, {and} \bibinfo{person}{Bill Dolan}.} \bibinfo{year}{2015}\natexlab{}.
\newblock \showarticletitle{A diversity-promoting objective function for neural conversation models}.
\newblock \bibinfo{journal}{\emph{arXiv preprint arXiv:1510.03055}} (\bibinfo{year}{2015}).
\newblock


\bibitem[Li et~al\mbox{.}(2023)]%
        {li2023blip}
\bibfield{author}{\bibinfo{person}{Junnan Li}, \bibinfo{person}{Dongxu Li}, \bibinfo{person}{Silvio Savarese}, {and} \bibinfo{person}{Steven Hoi}.} \bibinfo{year}{2023}\natexlab{}.
\newblock \showarticletitle{Blip-2: Bootstrapping language-image pre-training with frozen image encoders and large language models}. In \bibinfo{booktitle}{\emph{International conference on machine learning}}. \bibinfo{pages}{19730--19742}.
\newblock


\bibitem[Li and Lu(2024)]%
        {li2024survey}
\bibfield{author}{\bibinfo{person}{Jian Li} {and} \bibinfo{person}{Weiheng Lu}.} \bibinfo{year}{2024}\natexlab{}.
\newblock \showarticletitle{A Survey on Benchmarks of Multimodal Large Language Models}.
\newblock \bibinfo{journal}{\emph{arXiv preprint arXiv:2408.08632}} (\bibinfo{year}{2024}).
\newblock


\bibitem[Li et~al\mbox{.}(2022b)]%
        {li2022knowledge}
\bibfield{author}{\bibinfo{person}{Qintong Li}, \bibinfo{person}{Piji Li}, \bibinfo{person}{Zhaochun Ren}, \bibinfo{person}{Pengjie Ren}, {and} \bibinfo{person}{Zhumin Chen}.} \bibinfo{year}{2022}\natexlab{b}.
\newblock \showarticletitle{Knowledge bridging for empathetic dialogue generation}. In \bibinfo{booktitle}{\emph{Proceedings of the AAAI conference on artificial intelligence}}. \bibinfo{pages}{10993--11001}.
\newblock


\bibitem[Li et~al\mbox{.}(2024)]%
        {li2024styletts}
\bibfield{author}{\bibinfo{person}{Yinghao~Aaron Li}, \bibinfo{person}{Cong Han}, \bibinfo{person}{Vinay Raghavan}, \bibinfo{person}{Gavin Mischler}, {and} \bibinfo{person}{Nima Mesgarani}.} \bibinfo{year}{2024}\natexlab{}.
\newblock \showarticletitle{Styletts 2: Towards human-level text-to-speech through style diffusion and adversarial training with large speech language models}.
\newblock \bibinfo{journal}{\emph{Advances in Neural Information Processing Systems}}  \bibinfo{volume}{36} (\bibinfo{year}{2024}).
\newblock


\bibitem[Lin et~al\mbox{.}(2023)]%
        {lin2023video}
\bibfield{author}{\bibinfo{person}{Bin Lin}, \bibinfo{person}{Bin Zhu}, \bibinfo{person}{Yang Ye}, \bibinfo{person}{Munan Ning}, \bibinfo{person}{Peng Jin}, {and} \bibinfo{person}{Li Yuan}.} \bibinfo{year}{2023}\natexlab{}.
\newblock \showarticletitle{Video-llava: Learning united visual representation by alignment before projection}.
\newblock \bibinfo{journal}{\emph{arXiv preprint arXiv:2311.10122}} (\bibinfo{year}{2023}).
\newblock


\bibitem[Lin et~al\mbox{.}(2019)]%
        {lin2019moel}
\bibfield{author}{\bibinfo{person}{Zhaojiang Lin}, \bibinfo{person}{Andrea Madotto}, \bibinfo{person}{Jamin Shin}, \bibinfo{person}{Peng Xu}, {and} \bibinfo{person}{Pascale Fung}.} \bibinfo{year}{2019}\natexlab{}.
\newblock \showarticletitle{Moel: Mixture of empathetic listeners}.
\newblock \bibinfo{journal}{\emph{arXiv preprint arXiv:1908.07687}} (\bibinfo{year}{2019}).
\newblock


\bibitem[Liu et~al\mbox{.}(2024)]%
        {liu2024visual}
\bibfield{author}{\bibinfo{person}{Haotian Liu}, \bibinfo{person}{Chunyuan Li}, \bibinfo{person}{Qingyang Wu}, {and} \bibinfo{person}{Yong~Jae Lee}.} \bibinfo{year}{2024}\natexlab{}.
\newblock \showarticletitle{Visual instruction tuning}.
\newblock \bibinfo{journal}{\emph{Advances in neural information processing systems}}  \bibinfo{volume}{36} (\bibinfo{year}{2024}).
\newblock


\bibitem[Lorenzo-Trueba et~al\mbox{.}(2018)]%
        {lorenzo2018voice}
\bibfield{author}{\bibinfo{person}{Jaime Lorenzo-Trueba}, \bibinfo{person}{Junichi Yamagishi}, \bibinfo{person}{Tomoki Toda}, \bibinfo{person}{Daisuke Saito}, \bibinfo{person}{Fernando Villavicencio}, \bibinfo{person}{Tomi Kinnunen}, {and} \bibinfo{person}{Zhenhua Ling}.} \bibinfo{year}{2018}\natexlab{}.
\newblock \showarticletitle{The voice conversion challenge 2018: Promoting development of parallel and nonparallel methods}.
\newblock \bibinfo{journal}{\emph{arXiv preprint arXiv:1804.04262}} (\bibinfo{year}{2018}).
\newblock


\bibitem[Lu et~al\mbox{.}(2024)]%
        {lu2024unified}
\bibfield{author}{\bibinfo{person}{Jiasen Lu}, \bibinfo{person}{Christopher Clark}, \bibinfo{person}{Sangho Lee}, \bibinfo{person}{Zichen Zhang}, \bibinfo{person}{Savya Khosla}, \bibinfo{person}{Ryan Marten}, \bibinfo{person}{Derek Hoiem}, {and} \bibinfo{person}{Aniruddha Kembhavi}.} \bibinfo{year}{2024}\natexlab{}.
\newblock \showarticletitle{Unified-IO 2: Scaling Autoregressive Multimodal Models with Vision Language Audio and Action}. In \bibinfo{booktitle}{\emph{Proceedings of the IEEE/CVF Conference on Computer Vision and Pattern Recognition}}. \bibinfo{pages}{26439--26455}.
\newblock


\bibitem[Luo et~al\mbox{.}(2024a)]%
        {luo2024panosent}
\bibfield{author}{\bibinfo{person}{Meng Luo}, \bibinfo{person}{Hao Fei}, \bibinfo{person}{Bobo Li}, \bibinfo{person}{Shengqiong Wu}, \bibinfo{person}{Qian Liu}, \bibinfo{person}{Soujanya Poria}, \bibinfo{person}{Erik Cambria}, \bibinfo{person}{Mong-Li Lee}, {and} \bibinfo{person}{Wynne Hsu}.} \bibinfo{year}{2024}\natexlab{a}.
\newblock \showarticletitle{PanoSent: A Panoptic Sextuple Extraction Benchmark for Multimodal Conversational Aspect-based Sentiment Analysis}.
\newblock \bibinfo{journal}{\emph{arXiv preprint arXiv:2408.09481}} (\bibinfo{year}{2024}).
\newblock


\bibitem[Luo et~al\mbox{.}(2024b)]%
        {luo2024nus}
\bibfield{author}{\bibinfo{person}{Meng Luo}, \bibinfo{person}{Han Zhang}, \bibinfo{person}{Shengqiong Wu}, \bibinfo{person}{Bobo Li}, \bibinfo{person}{Hong Han}, {and} \bibinfo{person}{Hao Fei}.} \bibinfo{year}{2024}\natexlab{b}.
\newblock \showarticletitle{NUS-Emo at SemEval-2024 Task 3: Instruction-Tuning LLM for Multimodal Emotion-Cause Analysis in Conversations}. In \bibinfo{booktitle}{\emph{Proceedings of the 18th International Workshop on Semantic Evaluation (SemEval-2024)}}. \bibinfo{pages}{1599--1606}.
\newblock


\bibitem[Ma et~al\mbox{.}(2023)]%
        {ma2023dreamtalk}
\bibfield{author}{\bibinfo{person}{Yifeng Ma}, \bibinfo{person}{Shiwei Zhang}, \bibinfo{person}{Jiayu Wang}, \bibinfo{person}{Xiang Wang}, \bibinfo{person}{Yingya Zhang}, {and} \bibinfo{person}{Zhidong Deng}.} \bibinfo{year}{2023}\natexlab{}.
\newblock \showarticletitle{Dreamtalk: When expressive talking head generation meets diffusion probabilistic models}.
\newblock \bibinfo{journal}{\emph{arXiv preprint arXiv:2312.09767}} (\bibinfo{year}{2023}).
\newblock


\bibitem[Majumder et~al\mbox{.}(2020)]%
        {majumder2020mime}
\bibfield{author}{\bibinfo{person}{Navonil Majumder}, \bibinfo{person}{Pengfei Hong}, \bibinfo{person}{Shanshan Peng}, \bibinfo{person}{Jiankun Lu}, \bibinfo{person}{Deepanway Ghosal}, \bibinfo{person}{Alexander Gelbukh}, \bibinfo{person}{Rada Mihalcea}, {and} \bibinfo{person}{Soujanya Poria}.} \bibinfo{year}{2020}\natexlab{}.
\newblock \showarticletitle{MIME: MIMicking emotions for empathetic response generation}.
\newblock \bibinfo{journal}{\emph{arXiv preprint arXiv:2010.01454}} (\bibinfo{year}{2020}).
\newblock


\bibitem[Narvekar and Karam(2011)]%
        {narvekar2011no}
\bibfield{author}{\bibinfo{person}{Niranjan~D Narvekar} {and} \bibinfo{person}{Lina~J Karam}.} \bibinfo{year}{2011}\natexlab{}.
\newblock \showarticletitle{A no-reference image blur metric based on the cumulative probability of blur detection (CPBD)}.
\newblock \bibinfo{journal}{\emph{IEEE Transactions on Image Processing}} \bibinfo{volume}{20}, \bibinfo{number}{9} (\bibinfo{year}{2011}), \bibinfo{pages}{2678--2683}.
\newblock


\bibitem[Qian et~al\mbox{.}(2023)]%
        {qian2023harnessing}
\bibfield{author}{\bibinfo{person}{Yushan Qian}, \bibinfo{person}{Wei-Nan Zhang}, {and} \bibinfo{person}{Ting Liu}.} \bibinfo{year}{2023}\natexlab{}.
\newblock \showarticletitle{Harnessing the power of large language models for empathetic response generation: Empirical investigations and improvements}.
\newblock \bibinfo{journal}{\emph{arXiv preprint arXiv:2310.05140}} (\bibinfo{year}{2023}).
\newblock


\bibitem[Raamkumar and Yang(2022)]%
        {raamkumar2022empathetic}
\bibfield{author}{\bibinfo{person}{Aravind~Sesagiri Raamkumar} {and} \bibinfo{person}{Yinping Yang}.} \bibinfo{year}{2022}\natexlab{}.
\newblock \showarticletitle{Empathetic conversational systems: A review of current advances, gaps, and opportunities}.
\newblock \bibinfo{journal}{\emph{IEEE Transactions on Affective Computing}} (\bibinfo{year}{2022}), \bibinfo{pages}{2722--2739}.
\newblock


\bibitem[Radford et~al\mbox{.}(2021)]%
        {radford2021learning}
\bibfield{author}{\bibinfo{person}{Alec Radford}, \bibinfo{person}{Jong~Wook Kim}, \bibinfo{person}{Chris Hallacy}, \bibinfo{person}{Aditya Ramesh}, \bibinfo{person}{Gabriel Goh}, \bibinfo{person}{Sandhini Agarwal}, \bibinfo{person}{Girish Sastry}, \bibinfo{person}{Amanda Askell}, \bibinfo{person}{Pamela Mishkin}, \bibinfo{person}{Jack Clark}, {et~al\mbox{.}}} \bibinfo{year}{2021}\natexlab{}.
\newblock \showarticletitle{Learning transferable visual models from natural language supervision}. In \bibinfo{booktitle}{\emph{International conference on machine learning}}. \bibinfo{pages}{8748--8763}.
\newblock


\bibitem[Rashkin(2018)]%
        {rashkin2018towards}
\bibfield{author}{\bibinfo{person}{Hannah Rashkin}.} \bibinfo{year}{2018}\natexlab{}.
\newblock \showarticletitle{Towards empathetic open-domain conversation models: A new benchmark and dataset}.
\newblock \bibinfo{journal}{\emph{arXiv preprint arXiv:1811.00207}} (\bibinfo{year}{2018}).
\newblock


\bibitem[Rasley et~al\mbox{.}(2020)]%
        {rasley2020deepspeed}
\bibfield{author}{\bibinfo{person}{Jeff Rasley}, \bibinfo{person}{Samyam Rajbhandari}, \bibinfo{person}{Olatunji Ruwase}, {and} \bibinfo{person}{Yuxiong He}.} \bibinfo{year}{2020}\natexlab{}.
\newblock \showarticletitle{Deepspeed: System optimizations enable training deep learning models with over 100 billion parameters}. In \bibinfo{booktitle}{\emph{Proceedings of the 26th ACM SIGKDD International Conference on Knowledge Discovery \& Data Mining}}. \bibinfo{pages}{3505--3506}.
\newblock


\bibitem[Sabour et~al\mbox{.}(2022)]%
        {sabour2022cem}
\bibfield{author}{\bibinfo{person}{Sahand Sabour}, \bibinfo{person}{Chujie Zheng}, {and} \bibinfo{person}{Minlie Huang}.} \bibinfo{year}{2022}\natexlab{}.
\newblock \showarticletitle{Cem: Commonsense-aware empathetic response generation}. In \bibinfo{booktitle}{\emph{Proceedings of the AAAI Conference on Artificial Intelligence}}. \bibinfo{pages}{11229--11237}.
\newblock


\bibitem[Su et~al\mbox{.}(2023)]%
        {su2023pandagpt}
\bibfield{author}{\bibinfo{person}{Yixuan Su}, \bibinfo{person}{Tian Lan}, \bibinfo{person}{Huayang Li}, \bibinfo{person}{Jialu Xu}, \bibinfo{person}{Yan Wang}, {and} \bibinfo{person}{Deng Cai}.} \bibinfo{year}{2023}\natexlab{}.
\newblock \showarticletitle{Pandagpt: One model to instruction-follow them all}.
\newblock \bibinfo{journal}{\emph{arXiv preprint arXiv:2305.16355}} (\bibinfo{year}{2023}).
\newblock


\bibitem[Team et~al\mbox{.}(2024)]%
        {team2024chatglm}
\bibfield{author}{\bibinfo{person}{GLM Team}, \bibinfo{person}{Aohan Zeng}, \bibinfo{person}{Bin Xu}, \bibinfo{person}{Bowen Wang}, \bibinfo{person}{Chenhui Zhang}, \bibinfo{person}{Da Yin}, \bibinfo{person}{Diego Rojas}, \bibinfo{person}{Guanyu Feng}, \bibinfo{person}{Hanlin Zhao}, \bibinfo{person}{Hanyu Lai}, {et~al\mbox{.}}} \bibinfo{year}{2024}\natexlab{}.
\newblock \showarticletitle{Chatglm: A family of large language models from glm-130b to glm-4 all tools}.
\newblock \bibinfo{journal}{\emph{arXiv e-prints}} (\bibinfo{year}{2024}), \bibinfo{pages}{arXiv--2406}.
\newblock


\bibitem[Van~der Maaten and Hinton(2008)]%
        {van2008visualizing}
\bibfield{author}{\bibinfo{person}{Laurens Van~der Maaten} {and} \bibinfo{person}{Geoffrey Hinton}.} \bibinfo{year}{2008}\natexlab{}.
\newblock \showarticletitle{Visualizing data using t-SNE.}
\newblock \bibinfo{journal}{\emph{Journal of machine learning research}} \bibinfo{volume}{9}, \bibinfo{number}{11} (\bibinfo{year}{2008}).
\newblock


\bibitem[Vaswani(2017)]%
        {vaswani2017attention}
\bibfield{author}{\bibinfo{person}{A Vaswani}.} \bibinfo{year}{2017}\natexlab{}.
\newblock \showarticletitle{Attention is all you need}.
\newblock \bibinfo{journal}{\emph{Advances in Neural Information Processing Systems}} (\bibinfo{year}{2017}).
\newblock


\bibitem[Viswanathan and Viswanathan(2005)]%
        {viswanathan2005measuring}
\bibfield{author}{\bibinfo{person}{Mahesh Viswanathan} {and} \bibinfo{person}{Madhubalan Viswanathan}.} \bibinfo{year}{2005}\natexlab{}.
\newblock \showarticletitle{Measuring speech quality for text-to-speech systems: development and assessment of a modified mean opinion score (MOS) scale}.
\newblock \bibinfo{journal}{\emph{Computer speech \& language}} \bibinfo{volume}{19}, \bibinfo{number}{1} (\bibinfo{year}{2005}), \bibinfo{pages}{55--83}.
\newblock


\bibitem[Wang and Bovik(2002)]%
        {wang2002universal}
\bibfield{author}{\bibinfo{person}{Zhou Wang} {and} \bibinfo{person}{Alan~C Bovik}.} \bibinfo{year}{2002}\natexlab{}.
\newblock \showarticletitle{A universal image quality index}.
\newblock \bibinfo{journal}{\emph{IEEE signal processing letters}} \bibinfo{volume}{9}, \bibinfo{number}{3} (\bibinfo{year}{2002}), \bibinfo{pages}{81--84}.
\newblock


\bibitem[Wu et~al\mbox{.}(2024a)]%
        {wu2024towards}
\bibfield{author}{\bibinfo{person}{Shengqiong Wu}, \bibinfo{person}{Hao Fei}, \bibinfo{person}{Xiangtai Li}, \bibinfo{person}{Jiayi Ji}, \bibinfo{person}{Hanwang Zhang}, \bibinfo{person}{Tat-Seng Chua}, {and} \bibinfo{person}{Shuicheng Yan}.} \bibinfo{year}{2024}\natexlab{a}.
\newblock \showarticletitle{Towards Semantic Equivalence of Tokenization in Multimodal LLM}.
\newblock \bibinfo{journal}{\emph{arXiv preprint arXiv:2406.05127}} (\bibinfo{year}{2024}).
\newblock


\bibitem[Wu et~al\mbox{.}(2024b)]%
        {wu24next}
\bibfield{author}{\bibinfo{person}{Shengqiong Wu}, \bibinfo{person}{Hao Fei}, \bibinfo{person}{Leigang Qu}, \bibinfo{person}{Wei Ji}, {and} \bibinfo{person}{Tat-Seng Chua}.} \bibinfo{year}{2024}\natexlab{b}.
\newblock \showarticletitle{{NE}x{T}-{GPT}: Any-to-Any Multimodal {LLM}}. In \bibinfo{booktitle}{\emph{Proceedings of the International Conference on Machine Learning}}. \bibinfo{pages}{53366--53397}.
\newblock


\bibitem[Xu et~al\mbox{.}(2024)]%
        {xu2024faithful}
\bibfield{author}{\bibinfo{person}{Jundong Xu}, \bibinfo{person}{Hao Fei}, \bibinfo{person}{Liangming Pan}, \bibinfo{person}{Qian Liu}, \bibinfo{person}{Mong-Li Lee}, {and} \bibinfo{person}{Wynne Hsu}.} \bibinfo{year}{2024}\natexlab{}.
\newblock \showarticletitle{Faithful Logical Reasoning via Symbolic Chain-of-Thought}.
\newblock \bibinfo{journal}{\emph{arXiv preprint arXiv:2405.18357}} (\bibinfo{year}{2024}).
\newblock


\bibitem[Yan et~al\mbox{.}(2024)]%
        {yan-etal-2024-talk}
\bibfield{author}{\bibinfo{person}{Haoqiu Yan}, \bibinfo{person}{Yongxin Zhu}, \bibinfo{person}{Kai Zheng}, \bibinfo{person}{Bing Liu}, \bibinfo{person}{Haoyu Cao}, \bibinfo{person}{Deqiang Jiang}, {and} \bibinfo{person}{Linli Xu}.} \bibinfo{year}{2024}\natexlab{}.
\newblock \showarticletitle{Talk With Human-like Agents: Empathetic Dialogue Through Perceptible Acoustic Reception and Reaction}. In \bibinfo{booktitle}{\emph{Proceedings of the 62nd Annual Meeting of the Association for Computational Linguistics (Volume 1: Long Papers)}}. \bibinfo{pages}{15009--15022}.
\newblock


\bibitem[Yang et~al\mbox{.}(2024a)]%
        {yang2024exploiting}
\bibfield{author}{\bibinfo{person}{Zhou Yang}, \bibinfo{person}{Zhaochun Ren}, \bibinfo{person}{Yufeng Wang}, \bibinfo{person}{Xiaofei Zhu}, \bibinfo{person}{Zhihao Chen}, \bibinfo{person}{Tiecheng Cai}, \bibinfo{person}{Yunbing Wu}, \bibinfo{person}{Yisong Su}, \bibinfo{person}{Sibo Ju}, {and} \bibinfo{person}{Xiangwen Liao}.} \bibinfo{year}{2024}\natexlab{a}.
\newblock \showarticletitle{Exploiting emotion-semantic correlations for empathetic response generation}.
\newblock \bibinfo{journal}{\emph{arXiv preprint arXiv:2402.17437}} (\bibinfo{year}{2024}).
\newblock


\bibitem[Yang et~al\mbox{.}(2024b)]%
        {yang2024enhancing}
\bibfield{author}{\bibinfo{person}{Zhou Yang}, \bibinfo{person}{Zhaochun Ren}, \bibinfo{person}{Wang Yufeng}, \bibinfo{person}{Shizhong Peng}, \bibinfo{person}{Haizhou Sun}, \bibinfo{person}{Xiaofei Zhu}, {and} \bibinfo{person}{Xiangwen Liao}.} \bibinfo{year}{2024}\natexlab{b}.
\newblock \showarticletitle{Enhancing Empathetic Response Generation by Augmenting LLMs with Small-scale Empathetic Models}.
\newblock \bibinfo{journal}{\emph{arXiv preprint arXiv:2402.11801}} (\bibinfo{year}{2024}).
\newblock


\bibitem[Zhang et~al\mbox{.}(2024)]%
        {zhang-etal-2024-stickerconv}
\bibfield{author}{\bibinfo{person}{Yiqun Zhang}, \bibinfo{person}{Fanheng Kong}, \bibinfo{person}{Peidong Wang}, \bibinfo{person}{Shuang Sun}, \bibinfo{person}{SWangLing SWangLing}, \bibinfo{person}{Shi Feng}, \bibinfo{person}{Daling Wang}, \bibinfo{person}{Yifei Zhang}, {and} \bibinfo{person}{Kaisong Song}.} \bibinfo{year}{2024}\natexlab{}.
\newblock \showarticletitle{{STICKERCONV}: Generating Multimodal Empathetic Responses from Scratch}. In \bibinfo{booktitle}{\emph{Proceedings of the 62nd Annual Meeting of the Association for Computational Linguistics (Volume 1: Long Papers)}}. \bibinfo{pages}{7707--7733}.
\newblock


\bibitem[Zhao et~al\mbox{.}(2023)]%
        {zhao2023survey}
\bibfield{author}{\bibinfo{person}{Wayne~Xin Zhao}, \bibinfo{person}{Kun Zhou}, \bibinfo{person}{Junyi Li}, \bibinfo{person}{Tianyi Tang}, \bibinfo{person}{Xiaolei Wang}, \bibinfo{person}{Yupeng Hou}, \bibinfo{person}{Yingqian Min}, \bibinfo{person}{Beichen Zhang}, \bibinfo{person}{Junjie Zhang}, \bibinfo{person}{Zican Dong}, {et~al\mbox{.}}} \bibinfo{year}{2023}\natexlab{}.
\newblock \showarticletitle{A survey of large language models}.
\newblock \bibinfo{journal}{\emph{arXiv preprint arXiv:2303.18223}} (\bibinfo{year}{2023}).
\newblock


\bibitem[Zheng et~al\mbox{.}(2023)]%
        {zheng2023ecqed}
\bibfield{author}{\bibinfo{person}{Li Zheng}, \bibinfo{person}{Donghong Ji}, \bibinfo{person}{Fei Li}, \bibinfo{person}{Hao Fei}, \bibinfo{person}{Shengqiong Wu}, \bibinfo{person}{Jingye Li}, \bibinfo{person}{Bobo Li}, {and} \bibinfo{person}{Chong Teng}.} \bibinfo{year}{2023}\natexlab{}.
\newblock \showarticletitle{ECQED: emotion-cause quadruple extraction in dialogs}.
\newblock \bibinfo{journal}{\emph{arXiv preprint arXiv:2306.03969}} (\bibinfo{year}{2023}).
\newblock


\bibitem[Zhou et~al\mbox{.}(2022)]%
        {zhou2022case}
\bibfield{author}{\bibinfo{person}{Jinfeng Zhou}, \bibinfo{person}{Chujie Zheng}, \bibinfo{person}{Bo Wang}, \bibinfo{person}{Zheng Zhang}, {and} \bibinfo{person}{Minlie Huang}.} \bibinfo{year}{2022}\natexlab{}.
\newblock \showarticletitle{Case: Aligning coarse-to-fine cognition and affection for empathetic response generation}.
\newblock \bibinfo{journal}{\emph{arXiv preprint arXiv:2208.08845}} (\bibinfo{year}{2022}).
\newblock


\bibitem[Zhu et~al\mbox{.}(2023)]%
        {zhu2023minigpt}
\bibfield{author}{\bibinfo{person}{Deyao Zhu}, \bibinfo{person}{Jun Chen}, \bibinfo{person}{Xiaoqian Shen}, \bibinfo{person}{Xiang Li}, {and} \bibinfo{person}{Mohamed Elhoseiny}.} \bibinfo{year}{2023}\natexlab{}.
\newblock \showarticletitle{Minigpt-4: Enhancing vision-language understanding with advanced large language models}.
\newblock \bibinfo{journal}{\emph{arXiv preprint arXiv:2304.10592}} (\bibinfo{year}{2023}).
\newblock


\end{thebibliography}

\newpage

\appendix

\section{Ethic Considerations}

We can identify the following potential ethical considerations for our work:

\paratitle{Privacy and Data Protection.}
Empatheia relies on multimodal inputs, including text, voice, and video, which contain highly sensitive personal data. It is essential to ensure that all data collected and processed by the system adheres to strict privacy and data protection regulations, such as the GDPR or CCPA. The system must implement strong encryption techniques for storage and transmission, while also ensuring user data is anonymized where possible. Users should have full control over their data, including the ability to delete their inputs and outputs from the system. Regular audits of data handling and retention practices should be conducted to maintain compliance with privacy standards.

\paratitle{Bias and Fairness.}
Empatheia’s ability to generate empathetic responses relies on large language models and multimodal data, which can inherit biases from the training datasets. It is crucial to ensure that the system does not perpetuate harmful stereotypes or exhibit bias based on gender, race, age, or any other demographic characteristics. Diverse and inclusive training data, as well as regular testing for fairness, should be a priority. Additionally, measures should be taken to mitigate biases, such as using techniques like debiasing algorithms, and continuously refining the dataset to minimize any unintentional discrimination.

\paratitle{Emotional Manipulation and User Vulnerability.}
Since Empatheia is designed to interact empathetically with users, it may encounter individuals in emotionally vulnerable states. The system must avoid exploiting this vulnerability or manipulating emotions in harmful ways. Safeguards should be in place to ensure that the chatbot’s responses are supportive but do not give inappropriate advice or encourage dependency. Ethical guidelines should be established to prevent the misuse of the chatbot, and users should be made aware that it is a machine-generated response system and not a substitute for professional psychological help. Where appropriate, the system could be designed to refer users to human professionals in cases of serious emotional distress.

\paratitle{Autonomy and Transparency.}
The nature of Empatheia’s multimodal empathetic responses might blur the lines between human and machine interaction. It is essential to maintain transparency about the system’s limitations and make users fully aware that they are interacting with an AI. Users should also have the autonomy to make informed decisions about using the system and be provided with clear options to opt-out or disengage at any time. Regular disclosures about the system’s AI-driven nature, its data collection practices, and its purpose should be communicated transparently.

\paratitle{Potential for Misuse.}
As with any open-source system, there is a risk of Empatheia being misused in ways that could harm individuals or communities. Bad actors might leverage the system’s empathetic capabilities for malicious purposes, such as manipulating others through emotion-driven content or creating deepfakes for deceptive purposes. To mitigate this, the development of Empatheia should include security measures to prevent exploitation, such as limiting the use of avatars and ensuring that any generated content is watermarked or traceable. The open-source release should come with strict usage guidelines and community oversight to ensure responsible use of the system.

\paratitle{Long-Term Psychological Effects.}
The long-term effects of interacting with an empathetic AI system like Empatheia on human users should be carefully considered. While the system aims to foster deeper emotional connections, there is a risk that users may become overly reliant on AI for emotional support, potentially leading to social isolation or reduced human empathy. Further research should be conducted to assess the psychological impact of prolonged use of such systems, and regular evaluations should be made to ensure that the system enhances human emotional well-being rather than detracting from it.

\section{Future Work with AvaMERG}

In this paper, we present a comprehensive exploration of multimodal empathetic response generation.

We believe this work lays the foundation for future advancements in the field of multimodal sentiment analysis and empathetic interaction. 
From our practice, several promising directions for future research can be identified.

\paratitle{Exploring Higher Performance of MLLMs and Efficient Training Methods.} 
Future work can investigate the performance of various MLLMs in the generation of empathetic responses, particularly their advantages and limitations when processing multimodal inputs. 
Currently, we utilize state-of-the-art speech and avatar generators; however, their performance remains limited. Therefore, it is essential to enhance the quality of multimodal generation.
Also, more efficient training methods, such as transfer learning, few-shot learning, or self-supervised learning, can be explored to improve the training efficiency and performance of these models. Through systematic experimental comparisons, the aim is to identify best practices that enhance the quality and responsiveness of empathetic response generation.

\paratitle{Developing Multidimensional Evaluation Methods.} 
Currently, the evaluation of the multimodal generation component of MERG relies solely on human evaluations, which introduces significant uncertainty.
Future research should aim to establish multidimensional evaluation methods to comprehensively and automatically assess the effectiveness and quality of multimodal empathetic responses. 
This can be achieved by combining automated evaluations with human assessments. Specifically, deep learning-based evaluation models can be developed to automatically analyze the semantic consistency of generated responses, the accuracy of emotional conveyance, and the synergistic effects of multimodal inputs. Additionally, emotional analysis tools and semantic understanding techniques should be utilized to conduct detailed emotional depth analyses of the generated responses. Furthermore, the research should explore how to assess cross-modal correlations, such as evaluating the consistency between text, audio, and video, to further enhance the comprehensiveness and accuracy of the evaluation.

\paratitle{Enhancing the Model's Contextual Understanding.} 
Future research can focus on improving the model’s understanding of conversational context, particularly in retaining and utilizing historical information during long dialogues. Consideration could be given to incorporating more complex memory mechanisms or contextual attention mechanisms to enhance the model’s contextual awareness.

\paratitle{Exploring Cross-cultural Expressions of Empathy.} 
Future work can investigate how to effectively generate empathetic responses across different cultural contexts. The research could focus on analyzing the impact of cultural differences on emotional expression and communication styles, adjusting the model based on these findings to better accommodate users from diverse cultural backgrounds.

\paratitle{Improving Dataset Diversity and Quality.} 
Future work can focus on collecting and constructing larger-scale, more diverse multimodal datasets to encompass a wider range of emotional expressions and conversational scenarios. By enhancing the representativeness of the dataset, the model's generalization ability and robustness in diverse emotional interaction contexts can be further improved.

\section{More Details of Datasets}
\label{More Details of Datasets}

\subsection{Extended Details of Data Constructions}

\subsubsection{Dialogue Enriching}
\noindent\\\paratitle{Augmenting the Empathetic Dialogue (ED) Dataset.} 
We begin by augmenting the existing pure-text Empathetic Dialogue (ED) dataset to construct our AvaMERG dataset. The ED dataset consists of dialogues aimed at empathetic response generation (ERG) but lacks multimodal and identity-specific information essential for Multimodal Empathetic Response Generation (MERG). To address this, we first enrich each textual empathetic response \( t_i \in R_i \) with corresponding emotion chain, thereby constructing an emotional chain of thought (CoT).

Leveraging the advanced contextual understanding capabilities of OpenAI’s GPT-4, we annotate each utterance in the ED dataset with emotion chain. We define an emotion CoT: \textit{emotion} $\rightarrow$ \textit{emotion\_cause} $\rightarrow$ \textit{goal\_to\_response}. GPT-4 assigns an appropriate emotion chain to each utterance based on the dialogue context.
\vspace{-2mm}
\begin{tcolorbox}[breakable, colback=gray!10, colframe=black, title=An example of our prompt template:]
\vspace{-1.5mm}
As an expert in empathetic dialogue analysis, your task is to analyze the emotional dynamics and intentions behind a conversation between two participants: a 'speaker' and a 'listener'. 
The goal is to first consider the event scenario in which the conversation takes place, and then identify the 'Emotion Cause' for the speaker and the 'Goal to Response' for the listener's final reply.

\textbf{Task Overview:}
1. \textbf{Emotion Scenario}: Identify the specific event that occurs within the context of the conversation, which serves as the backdrop for the emotional dynamics and interactions between the speaker and listener.
2. \textbf{Emotion Cause}: Based on the conversation context, sentiment, and dialogue history, analyze and identify the underlying emotional cause or trigger for the speaker.
3. \textbf{Goal to Response}: Analyze the last response from the listener and identify the intended goal behind that response. The goal should relate to how the listener is attempting to address the speaker's emotional state.

\textbf{Input JSON Field Descriptions:}

- \textbf{dia\_id}: A unique identifier for the dialogue.

- \textbf{sentiment}: The emotional tone of the conversation.

- \textbf{context}: The background information of the conversation, describing the environment, setting, or situation in which the dialogue occurs.

- \textbf{dialogue}: The actual conversation or exchange of dialogue between speaker and listener.

\textbf{Expected Output:}
Please provide the following in **concise** JSON format:

- **Event Scenario**: A **short description** summarizing the main context or situation of the dialogue (e.g., 'The speaker is expressing fear after experiencing something unsettling').

- **Emotion Cause**: A **brief** explanation of the **specific event** or experience that triggers the speaker's emotion (e.g., 'Elevator game brings horror experience').

- **Goal to Response**: A **concise** goal that describes the **specific emotional state** the listener is attempting to address (e.g., 'Alleviating fear').

\textbf{Example Output:}
\begin{verbatim}
{ 
  "dia_id": "<dia_id>",
  "event_scenario": "The speaker is expressing 
  fear after experiencing something unsettling",
  "emotion_cause": "Elevator game brings horror 
          experience",
  "goal_to_response": "Alleviating fear"
}
\end{verbatim}

\end{tcolorbox}

To ensure the accuracy of the emotion annotations, we implement a validation step where human annotators review the GPT-4 assigned emotions. This process involves cross-referencing the emotion labels with the dialogue content to verify consistency and appropriateness. Any discrepancies are resolved through discussion among annotators, ensuring high-quality emotion annotations.

\paratitle{Enriching Identity Information.}
To enable MERG models to generate appropriate avatar profiles for both audio and video modalities, we further annotate each utterance with identity information for both participants in the dialogue. This includes:
\setdefaultleftmargin{0.5em}{1em}{}{}{}{}
\begin{itemize}
  \item \textbf{Age}: We define four age periods—\textit{child} (0-15 years), \textit{young} (16-34 years), \textit{middle-aged} (35-59 years), and \textit{elderly} (60+ years).
  \item \textbf{Gender}: Binary genders—\textit{male} and \textit{female}.
  \item \textbf{Timbre}: Three vocal timbres—\textit{low}, \textit{mid}, and \textit{high}.
\end{itemize}

GPT-4 is utilized to determine the above labels for each utterance, ensuring that the dialogue reflects realistic interactions between participants with diverse profiles. The identity annotations are critical for training models to generate contextually appropriate and personalized empathetic responses in multiple modalities.

\paratitle{Data Balancing and Expansion.}
Observing that the raw ED dataset is imbalanced (e.g., most dialogues involve young or middle-aged participants), we employ GPT-4 to generate additional dialogues that include underrepresented age groups and genders, as well as a balanced distribution of timbres. GPT-4 also detects and labels the dialogue topics, covering 10 primary common topics: [Social Issues and Moral Dilemmas, Achievements and Self-Realization, Support and Comfort, Emotions and Feelings, Disappointments and Expectations, Life Events, Interpersonal Relationships, Health and Well-being, Uncertainty About the Future, Personal Struggles and Challenges]. 

To enhance diversity, GPT-4 is instructed to generate dialogues that based on various races, cultural backgrounds, and socio-economic statuses, reflecting a realistic and inclusive range of human experiences. This process results in a more balanced and representative dataset.
\vspace{-2.5mm}
\begin{tcolorbox}[breakable, colback=gray!10, colframe=black, title=Enriching Identity for Dialogue Generation, width=\linewidth]
\vspace{-1.5mm}
You are an AI language model tasked with generating a dialogue between two participants, incorporating detailed identity information and topic annotations.

\textbf{Requirements:}
\setdefaultleftmargin{0.5em}{1em}{}{}{}{}
\begin{itemize}
  \item \textbf{Dialogue Structure:} The dialogue contains 3 turns, with alternating participants (max 6 utterances). Each turn includes the full dialogue history and the listener’s empathetic response.
  \item \textbf{Identity Information:} Profiles for both participants should include: \textbf{Age} (child: 0-15, young: 16-34, middle-aged: 35-59, elderly: 60+); \textbf{Gender} (male, female); \textbf{Timbre} (low, mid, high). Dialogue content should reflect these identities.
  \item \textbf{Emotion and Empathy Chain:} Each turn should include the speaker’s emotion (constant across turns) and evolving fields for \textbf{event scenario}, \textbf{emotion cause}, and \textbf{goal to response}.
  \item \textbf{Data Balancing:} Ensure representation of underrepresented age groups, genders, and timbres.
  \item \textbf{Topics:} The conversation should revolve around one of these topics: Social Issues and Moral Dilemmas; Achievements and Self-Realization; Support and Comfort; Emotions and Feelings; Disappointments and Expectations; Life Events; Interpersonal Relationships; Health and Well-being; Uncertainty About the Future; Personal Struggles and Challenges.
\end{itemize}

\textbf{JSON Format Example:}
\begin{verbatim}
{"conversation_id": "string",
  "speaker_profile": { "age": "string", "gender": 
  "string", "timbre": "string" },
  "listener_profile": { "age": "string", "gender"
  : "string", "timbre": "string" },
  "topic": "string",
  "turns": [{"turn_id": "string", "context":
  "string",  "dialogue_history": [
    { "index": int, "role": "string", "utterance"
    : "string" }], 
  "response": "string",
  "chain_of_empathy": {
  "speaker_emotion": "string",
  "event_scenario": "string",
  "emotion_cause": "string",
  "goal_to_response": "string"} 
  ]
} 
\end{verbatim}

\end{tcolorbox}

\vspace{-2mm}
\noindent\\ \paratitle{Human Annotation and Cross-Checking.}\vspace{-0.5em}
To ensure the quality and accuracy of the augmented dataset, we recruit human annotators for rigorous manual checking. Each dialogue undergoes a 3-person cross-checking process where annotators verify:

\begin{itemize}
  \item \textbf{Content Accuracy}: Whether the dialogue content is coherent, contextually appropriate, and free from biases or offensive language.
  \item \textbf{Meta-Profile Consistency}: Whether the assigned identity information and emotion labels are accurate and consistent with the dialogue content.
  \item \textbf{Topic Relevance}: Whether the dialogue topics are correctly identified and relevant.
\end{itemize}

Annotators receive detailed guidelines and training to ensure consistency in their evaluations. Discrepancies among annotators are discussed and resolved collectively. 
Only dialogues that receive unanimous approval from all three annotators are included in the AvaMERG dataset.

\subsubsection{Audio \& Video Recording.}

\noindent\\ \paratitle{Volunteer Recruitment.}\vspace{-0.5em}
We recruit a large and diverse group of English-speaking volunteers representing different ages, genders, vocal characteristics, and races (Asian, Caucasian, African, Latino, Indian). 
Recruitment is conducted through community outreach, social media, and collaboration with institutions to ensure diversity. 
All volunteers provide informed consent and are compensated for their participation. We standardized the filming environment requirements. 
Due to the need for diversity, 30\% of volunteers with certain characteristics could not be found offline, so they communicated with us online and submitted their recorded results. 
Similarly, they adhered to consistent environment requirements during their recording.

\noindent\\ \paratitle{Participant Pairing and Assignment.}\vspace{-0.5em}
Volunteers are paired and grouped according to the profiles determined in the annotated AvaMERG dialogues. 
Each pair corresponds to the identity profiles of the dialogue participants, ensuring that the multimodal data accurately reflects the textual annotations. 
Care is taken to match volunteers to profiles that they can authentically portray, enhancing the realism of the dataset.

\noindent\\ \paratitle{Recording Sessions.}\vspace{-0.5em}
During the recording sessions, volunteers perform the dialogues by carefully reading the utterance text. 
They are provided with context about the dialogue, including the emotional state and background of the characters. 
Instructions are given to exhibit the correct emotional performance, paying close attention to:

\begin{itemize}
  \item \textbf{Vocal Attributes:} Tone, pitch, and timbre corresponding to the annotated vocal timbre and emotion.
  \item \textbf{Facial Expressions:} Micro-facial expressions that align with the emotional content, captured using high-resolution cameras.

\end{itemize}

Professional recording equipment is used to ensure high-quality audio and video data. 
Sessions are supervised by directors who provide guidance to volunteers to achieve the desired performances.

\noindent\\ \paratitle{Post-processing.}\vspace{-0.5em}
Recorded data undergoes post-processing to enhance quality. 
This includes noise reduction in audio files, color correction in videos, and synchronization of audio and video streams. 
Metadata is added to files to link them with the corresponding textual annotations and identity profiles.
After processing, considering storage space and the input requirements of our model, we standardized the format to meet both our storage capacity and the model's needs, while ensuring the audio and audiovisual content remained unchanged.

\subsubsection{Manual Annotation Verification}

\noindent\\ \paratitle{Verification Process.}\vspace{-0.5em}
To ensure the highest quality of our dataset, we implement a thorough manual annotation verification process following the post-processing stage.

Each recorded dialogue, along with its corresponding annotations and multimodal content, undergoes a comprehensive review by at least two independent annotators. The verification process focuses on:
\begin{itemize}
  \item \textbf{Content Alignment:} Checking that the spoken words and visual expressions in the recordings accurately match the textual utterances and annotated emotion chains.
  \item \textbf{Profile Consistency:} Ensuring that the age, gender, timbre, and emotional expressions portrayed by the volunteers align with the assigned profiles.
  \item \textbf{Technical Quality}: Verifying that the audio and video recordings meet the required technical standards, including clarity, resolution, and synchronization.
\end{itemize}
\noindent\\ \paratitle{Quality Metrics Calculation.}\vspace{-0.5em}
We calculate the Cohen's Kappa Score to measure the agreement between annotators. Achieving a score of 0.78 indicates a high level of consistency and reliability in the annotation process. Any instances where annotators disagree or identify potential issues are reviewed collectively, and problematic data is either corrected or discarded.

\noindent\\ \paratitle{Finalization.}\vspace{-0.5em}
Only dialogues that pass the manual verification process with unanimous approval are included in the final AvaMERG dataset. This rigorous quality control ensures that the dataset is both reliable and suitable for training and evaluating MERG models.

\begin{table*}[!htbp]
\fontsize{8}{10.5}\selectfont
\setlength{\tabcolsep}{1.9 mm}
\centering
\caption{A snippet of an annotated data instance based on provided dialogue data.}
\vspace{-3mm}
\begin{tabular}{m{0.15\textwidth}m{0.8\textwidth}}
\hline
\textbf{KEY} & \textbf{VALUE} \\ \hline
Conversation-ID & 01797 \\ \hline
Speaker Profile & Age: young, Gender: male, Tone: mild, ID: 35 \\ \hline
Listener Profile & Age: young, Gender: female, Tone: emphatic, ID: 20 \\ \hline
Topic & Personal Struggles and Challenges \\ \hline
Dialogue & 
1. Speaker: When I left the bathroom in high school once I had toilet paper stuck to my shoe. \newline
2. Listener: I bet it was very embarrassing? \newline
3. Speaker: Yeah it sure was, you know how mean teenagers can be. It's like they have no empathy or think about what if it was them. \newline
4. Listener: It's alright, we've all been there many times. \\ \hline
Speaker Emotion & Embarrassed \\ \hline
Event Scenario & The speaker experienced embarrassment after accidentally leaving the bathroom with toilet paper stuck to their shoe. \\ \hline
Emotion Cause & Embarrassing public incident involving toilet paper. \\ \hline
Goal of Response & Validate the speaker's embarrassment, providing reassurance and comfort. \\ \hline
\end{tabular}
\label{dialogue_instance}
\end{table*}

\begin{table*}[!htbp]
\fontsize{8}{10.5}\selectfont
\setlength{\tabcolsep}{1.9 mm}
\centering
\caption{Detailed information of dialogue topic.}
\vspace{-3mm}
\begin{tabular}{m{0.2\textwidth}m{0.75\textwidth}}
\hline
\textbf{Topic} & \textbf{Keywords} \\ \hline
Social Issues and Moral Dilemmas & income inequality, climate change activism, ethical consumerism, human rights violations, systemic racism, gender discrimination, animal welfare concerns, mental health stigma, refugee crisis, digital privacy issues \\ \hline
Achievements and Self-Realization & bar exam, college graduation ceremony, soccer match, kitchen cooking test, financial success, project completion, professional certification, art exhibition, personal fitness milestone, community service award \\ \hline
Support and Comfort & dealing with burnout, feeling overwhelmed at work, losing a pet, struggling with loneliness, facing a tough decision, breakup with partner, family argument, lossing a job \\ \hline
Emotions and Feelings & obtaining driver's license, reckless driving behavior, forgotten debt, unexpected compliment, cherished memory, disappointing news, long-awaited vacation, stressful deadline \\ \hline
Disappointments and Expectations & missed opportunity, unfulfilled promise, low performance review, delayed project completion, unmet goals, failed partnership, incomplete application, rejected proposal, unsatisfactory results, broken commitment \\ \hline
Life Events & flight costs, unexpected surprise, enjoyable day trip, receiving unexpected flowers, mother's death, job interview, significant life transition, family gathering, grocery shopping \\ \hline
Interpersonal Relationships & marriage relationship, emotional distress, providing support, misbehaving with friends, overwhelming academic pressure, unexpected friendship formation, trust building, mutual Respect, forgiveness and reconciliation, social connection, shared experiences \\ \hline
Health and Well-being & mental resilience, physical fitness routine, healthy eating habits, stress management techniques, mindfulness practice, regular health check-ups, emotional balance, quality sleep, work-life harmony, social connection \\ \hline
Uncertainty About the Future & career instability, financial insecurity, unpredictable life events, fear of change,unfulfilled dreams \\ \hline
Personal Struggles and Challenges & financial stress, unexpected romantic encounter, feeling lonely, career uncertainty, self-doubt , mental health struggles, parental pressure \\ \hline
\end{tabular}
\label{detailed_topic_keywords}
\end{table*}

\subsection{Detailed Data Highlights}

Here, we extend the content of Dataset Construction from the main article to provide a more comprehensive introduction to all the highlights of our AvaMERG dataset.
AvaMERG boasts a diverse array of avatar profiles. 
This diversity ensures that models trained on AvaMERG can generalize across various demographic profiles, promoting inclusivity and reducing biases. AvaMERG contains a total of \textbf{33,048} dialogue samples and \textbf{152,021} dialogue utterances, providing ample usable data for the field. AvaMERG contains the following notable strengths:

\begin{figure}[!t]
\centering
\vspace{-2mm}
\includegraphics[width=0.90\columnwidth]{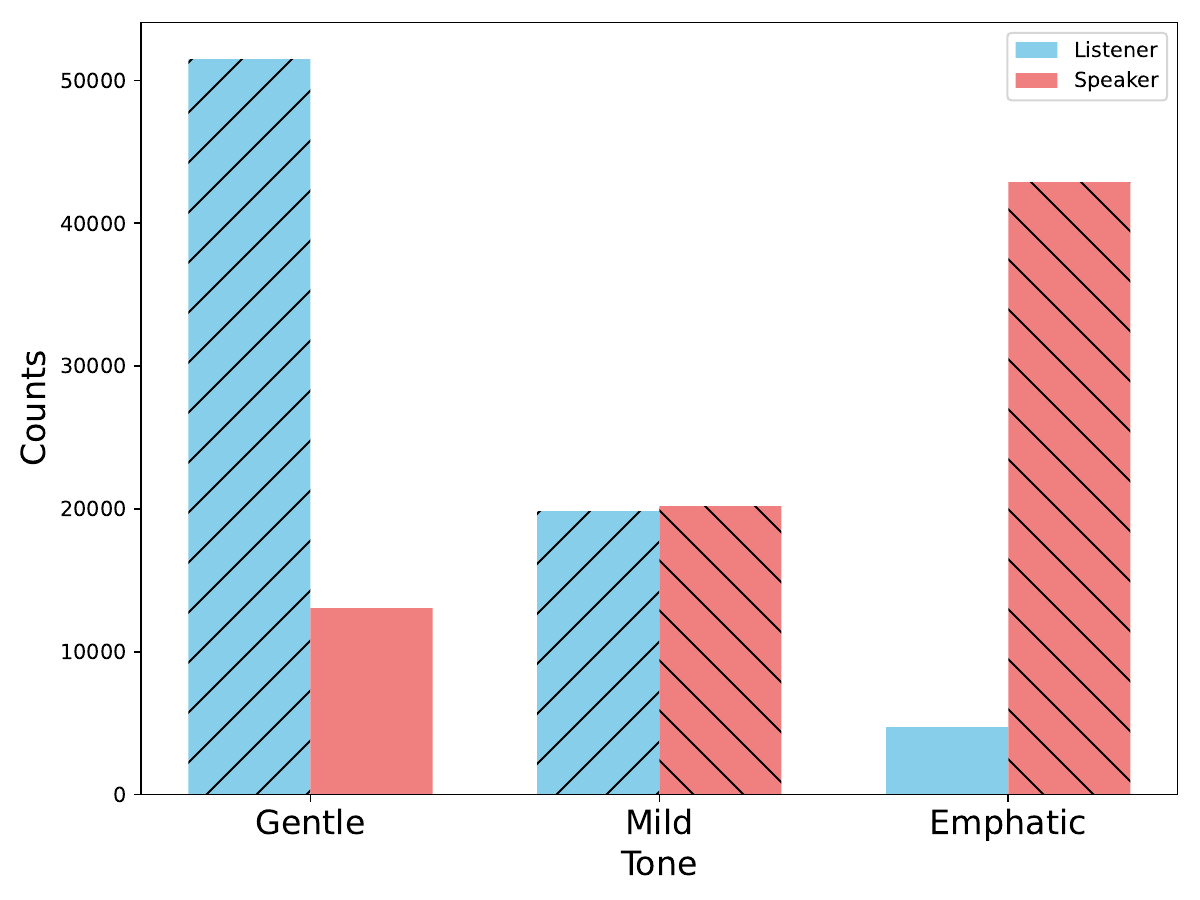}
\vspace{-3mm}
\caption{The distribution of tone.}
\label{distribution_Tone}
\vspace{-3mm}
\end{figure}
\noindent\\ \paratitle{A reasonable and well-balanced tone distribution.}\vspace{-0.5em}
In terms of Tone shown in Figure \ref{distribution_Tone}, Gentle tones were the most common (\textbf{42.43\%}, \textbf{64,504}), followed by Emphatic (\textbf{31.28\%}, \textbf{47,553}) and Mild (\textbf{26.29\%}, \textbf{39,964}) tones. Regarding Tone, listeners were mostly Gentle (\textbf{67.73\%}, \textbf{51,481}), followed by Mild (\textbf{26.05\%}, \textbf{19,801}) and Emphatic (\textbf{6.22\%}, \textbf{4,728}),  tones. Speakers were mostly Emphatic (\textbf{56.34\%}, \textbf{42,825}), followed by Mild (\textbf{26.53\%}, \textbf{20,163}) and Gentle (\textbf{17.13\%}, \textbf{13,023}) tones. AvaMERG provides a comprehensive distribution of tones between speakers and listeners, promoting balanced dialogue interactions, and its tone distribution is designed to align with the tones that different speaker identities would adopt in real-world scenarios.

\noindent\\ \paratitle{A diverse and balanced age-gender distribution for real-world applicability.}
AvaMERG sets itself apart with a well-balanced distribution across both age groups and gender, as illustrated in the figure \ref{fig_statistic}. The dataset includes avatars representing different life stages: Child, Young, Middle-aged, and Elderly, with a nearly equal gender representation within each category. This ensures that the models trained on AvaMERG can generalize effectively to a wide variety of demographic profiles, promoting inclusivity and reducing bias in real-world dialogue systems. For children, there is an almost equal split between male (\textbf{12,563}) and female (\textbf{12,577}) utterances. Among the young adult group, the dataset includes \textbf{37,010} male utterances and \textbf{37,488} female utterances. In the middle-aged category, there are \textbf{15,208} male utterances and \textbf{13,818} female utterances. Lastly, the elderly group shows an equal number of utterances for both genders, with \textbf{11,763} male and \textbf{11,763} female utterances. This diverse age and gender representation ensures that dialogue systems trained on AvaMERG can perform robustly across various demographic profiles, enhancing fairness and inclusivity in real-world applications.

\noindent\\ \paratitle{Detailed and rich emotional design for real-world applicability.}
We adopt 32 fine-grained textual emotions and 7 coarse-grained multimodal emotions, and map them appropriately, as shown in Figure \ref{emotion_map}. As shown in the emotional distribution in Figure \ref{fig_statistic}, AvaMERG covers a wide range of emotional expressions. Among the 33,048 dialogues, the most prevalent emotion is sadness, accounting for \textbf{56.7\%} of the samples, which reflects the empathetic nature of many conversations, especially in scenarios requiring emotional support or assistance. This is followed by happiness (\textbf{20.3\%}), which captures the dialogues involving positive reinforcement or joyful interactions.
Other emotions such as anger (\textbf{7.9\%}), contempt (\textbf{7.9\%}), and surprise (\textbf{6.3\%}) are also represented in reasonable proportions, ensuring that the dataset includes not only empathetic responses but also situations where the user expresses negative or unexpected emotions. Fear and disgust appear less frequently, with \textbf{5.1\%} and \textbf{1.8\%} respectively, but still provide valuable instances for training models that can handle a full spectrum of emotional states. 
This rich emotional diversity ensures that models trained on AvaMERG are capable of understanding and generating appropriate responses to a broad range of emotional expressions, improving the model's ability to handle real-world interactions where emotions play a critical role.

\noindent\\ \paratitle{A comprehensive topic design.}
The AvaMERG dataset showcases a meticulously designed mapping between various real-world topics and the emotional responses they elicit, providing an invaluable resource for training empathetic response models. As shown in the heatmap Figure \ref{topics_emotions_map}, topics such as "Achievements and Self-Realization", "Disappointments and Expectations", "Health and Well-being", and "Personal Struggles and Challenges" elicit a wide range of emotional responses, including anger, happiness, and sadness.
For instance, "Health and Well-being" is associated with a significant number of utterances reflecting fear (\textbf{1,158}), sadness (\textbf{2,214}), and disgust (\textbf{810}), which mirrors real-world conversations where health concerns often evoke complex emotions. Similarly, "Achievements and Self-Realization" is more frequently linked to positive emotions, with \textbf{3,214} utterances expressing happiness, demonstrating the natural alignment between positive life events and joyful emotions.
In contrast, "Social Issues and Moral Dilemmas" and "Uncertainty About the Future" are characterized by more negative emotions, such as fear and contempt, reflecting the inherent challenges and anxieties that arise when discussing social complexities or future uncertainties.
By covering a wide spectrum of topics and emotions, the dataset enables the development of empathetic models capable of understanding and responding to the nuanced emotional underpinnings of different conversations.

\begin{figure}[!t]
\centering
\vspace{-2mm}
\includegraphics[width=0.90\columnwidth]{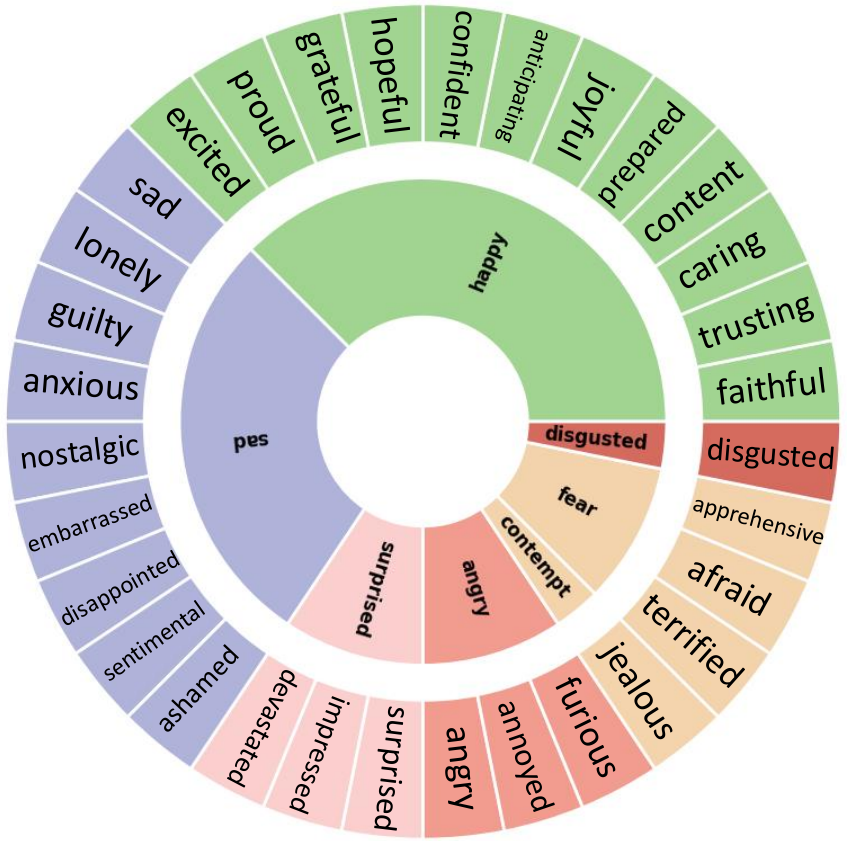}
\vspace{-3mm}
\caption{The mapping of fine-grained textual emotions to coarse-grained multimodal emotions.}
\label{emotion_map}
\vspace{-3mm}
\end{figure}

\begin{figure}[!t]
\centering
\vspace{-2mm}
\includegraphics[width=1.0\columnwidth]{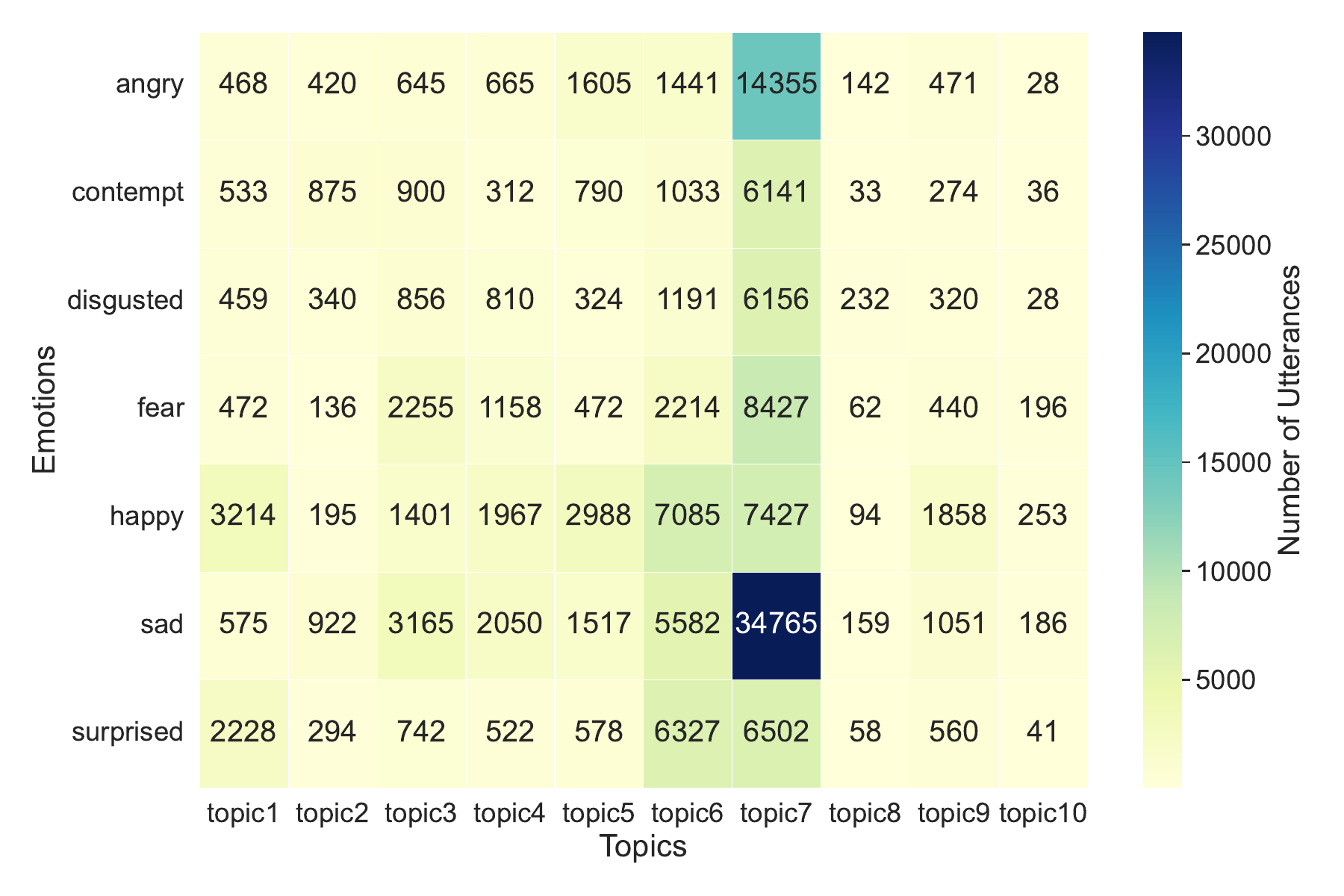}
\vspace{-3mm}
\caption{Emotion-topic heatmap for dialogue utterances. Topics 1 to 10 represent achievements and self-realization, disappointments and expectations, emotions and feelings, health and well-being, interpersonal relationships, life events, personal struggles and challenges, social issues and moral dilemmas, support and comfort, uncertainty about the future, respectively.}
\label{topics_emotions_map}
\vspace{-3mm}
\end{figure}

\section{More Details of Methods}
In this part, we provide an extension to our Empatheia system, including the input construction format, the specifics of the generator, and the training details.

\subsection{Empatheia Text Input}
Empatheia is designed with four complementary training stages, each of which employs a distinct input structure. For the CoE learning and Content Consistency Learning (CCL) stages, the input format is as follows:
\vspace{-2mm}
\begin{tcolorbox}[breakable, colback=gray!10, colframe=black, title=An Input Example for the CoE and CCL]
\vspace{-2mm}
\textbf{Input:}\\
Provide an empathetic response based on the given dialogue context below. 
Don't rush to give the response, thinking step by step.\\
Dialogue Context:\\
  \{\\
     Speaker:I paid all my bills today, I feel great!\\
     Assistant:Every little accomplishment counts! What bills did you have to pay?\\
     Listener:Rent and electricity. We've been struggling financially so it's such a relief to pay bills.\\
  \}\\
\textbf{Target:}\\
Firstly, the event scenario of this conversation is: Paying off overdue rent and electricity bills amid financial struggles.\\
Secondly, the emotion of the speaker is: content\\
Thirdly, the emotion cause is: Relief from the burden of financial stress after successfully paying bills.\\
Fourthly, the goal to response is:Providing support and validation for the speaker's sense of accomplishment.\\
Finally, the response is: Sorry to hear that. Well, at least this will be a weight off your shoulders.\\
\vspace{-4mm}
\end{tcolorbox}

\vspace{-2mm}
For different modalities of dialogue input, we standardize them into feature vectors, which can then be fed into the LLM for multimodal information comprehension. 
For the stages of style consistency learning, our input consists solely of multimodal speech and video. 
This setup encourages the model to learn how to generate multimodal responses with consistent emotion and style.
Specifically, the input format to the LLM is structured as follows:

\vspace{-2mm}
\begin{tcolorbox}[breakable, colback=gray!10, colframe=black, title=An Input Example for Style Learning]
\vspace{-2mm}
\textbf{Input:}\\
Provide an multimodal empathetic response based on the given dialogue context below. \\
Dialogue Context:\\
  \{\\
     Speaker:\textcolor{red}{<Aud>} \textcolor{blue}{<Vid>}.\\
     Listener:\textcolor{red}{<Aud>} \textcolor{blue}{<Vid>}.\\
     Speaker:\textcolor{red}{<Aud>} \textcolor{blue}{<Vid>}.\\
  \}\\
\textbf{Target:}\\
\textcolor{red}{<AUD1><AUD2><AUD3><AUD4><AUD5><AUD6>
<AUD7><AUD8><AUD9><AUD10><AUD11><AUD12>
<AUD13><AUD14><AUD15><AUD16>}\\
\textcolor{blue}{<VID1><VID2><VID3><VID4><VID5><VID6><VID7>}
\textcolor{blue}{<VID8><VID9><VID10><VID11><VID12><VID13>}
\textcolor{blue}{<VID13><VID14><VID15><VID16>}
\end{tcolorbox}

Here, <Aud> and <Vid> are special placeholders, and before being input into the LLM, the token embeddings at these positions will be replaced with the corresponding audio and video features. 
<AUDi> and <VIDi> are used as multimodal generation signals.

Combining the first two phases, the final overall training input consists of three modalities. We define the input format for the LLM in this phase as follows:

\vspace{-2mm}
\begin{tcolorbox}[breakable, colback=gray!10, colframe=black, title=An Input Example for Overall Training]
\vspace{-2mm}
\textbf{Input:}\\
Provide a multimodal empathetic response based on the given dialogue context below. 
Don't rush to give the response, thinking step by step.\\
Dialogue Context:\\
  \{\\
     Speaker:I paid all my bills today, I feel great!\textcolor{red}{<Aud>} \textcolor{blue}{<Vid>}.\\
     Listener:Every little accomplishment counts! What bills
did you have to pay?\textcolor{red}{<Aud>}\textcolor{blue}{<Vid>}.\\
     Speaker:Rent and electricity. We’ve been struggling finan-
cially so it’s such a relief to pay bills.\textcolor{red}{<Aud>} \textcolor{blue}{<Vid>}.\\
  \}\\
\textbf{Target:}\\
Firstly, the event scenario of this conversation is: Paying off overdue rent and electricity bills amid financial struggles.\\
Secondly, the emotion of the speaker is: content \\
Thirdly, the emotion cause is: Relief from the burden of
financial stress after successfully paying bills.\\
Fourthly, the goal to response is:Providing support and
validation for the speaker’s sense of accomplishment.\\
Finally, the response is: Sorry to hear that. Well, at least this will be a weight off your shoulders.\\
\textcolor{red}{<AUD1>...AUD16>}
\textcolor{blue}{<VID1>...<VID16>}
\end{tcolorbox}

\subsection{Technical Details of Two Generators}

Next, we will introduce the details of the speech generator and video generator we used, as well as how they receive the features passed from the CS and SD modules.

\subsubsection{Speech Generator.}

\noindent\\\vspace{1mm}For our system’s speech generation, we employ the state-of-the-art TTS model, StyleTTS2. 
Leveraging advanced diffusion models and adversarial training, StyleTTS2 is able to produce speech with emotions that are more authentic and natural than earlier TTS models, which is critical for our application. 
The input to StyleTTS2 includes the target empathic response text, denoted as $t$, and an optional reference mel-spectrogram, denoted as $x$. StyleTTS requires the input of the text script to be converted into speech, along with a reference audio containing emotional cues to serve as the style template. The modules responsible for reconstructing $x$ in StyleTTS2 are listed as follows.
\vspace{2mm}
\noindent\\ \paratitle{Text Encoder. }
The acoustic text encoder $E_{aco}$ encodes the input phonemes into hidden representations:
\begin{equation}
    \bm{h}_\text{text} = E_{aco}(\bm{t})
\end{equation}

\noindent\\ \paratitle{Style Encoder. } 
The style encoder $E_{ref}$ encodes the input reference mel-spectrogram $x$ into a styled vector:
\begin{equation}
    \bm{ref_s} = E_{ref}(\bm{x}) 
\end{equation}
where $ref_s$ encapsulates style information such as timbre, emotion, and other stylistic characteristics present in the reference audio.

\noindent\\ \paratitle{Text Aligner. }
The text aligner $A$ utilizes a dot product to extract aligned phoneme representations:
\begin{equation}
    \bm{h}_{algn} = \bm{h}_{text}\cdot \bm{a}_{pred} 
\end{equation} 
from the input speech $x$ and phonemes $t$. 
Here, $a_pred$ denotes the duration prediction, which is computed by
\begin{equation}
    \bm{h}_{bert} =BERT(\bm{t}), 
\end{equation} 
\begin{equation}
\bm{a}_{pred} =E_{dur}(h_{bert})
\end{equation}
where BERT is pre-trained on extensive corpora of Wikipedia articles as a prosodic text encoder. $D_{dur}$ is the duration predictor.

\noindent\\ \paratitle{Duration Predictor. }
The duration predictor $S$ predicts the duration of the reconstructed phonemes by:
\begin{equation}
    \bm{d}_{pred} = S(\bm{h}_{text}, \bm{s}) \,
\end{equation}

\noindent\\ \paratitle{Prosody Predictor. }
The prosody predictor $P$ predicts the pitch and energy of the reconstructed phonemes by:
\begin{equation}
    \hat{p}_{\bm{x}},\hat{n}_{\bm{x}} = P(\bm{h}_{text},\bm{s})
\end{equation}
Finally, the reconstructed speech is obtained by a speech decoder $G$:
\begin{equation}
    \hat{x} = G(\bm{h}_{algn},ref_s,\hat{p}_{\bm{x}},\hat{n}_{\bm{x}})
\end{equation}

\subsubsection{Talking Head Generator.}

\noindent\\\vspace{1mm}DreamTalk is a sophisticated framework for generating expressive talking heads, utilizing diffusion models to deliver high-quality performance while minimizing reliance on expensive style references.
The framework is comprised of a denoising network, a style-sensitive lip expert, and a style predictor. 
The denoising network employs diffusion models to create audio-driven facial movements that reflect the speaking style indicated by a reference video. 
The style-sensitive lip expert guarantees accurate lip synchronization and dynamic facial expressions, while the style predictor derives personalized speaking styles directly from the audio input.

Here, we will focus on introducing the denoising network $E_{\theta}$, which learns to denoise the noisy motion $m$ to obtain the predicted motion $m^*(0)$ under the conditions of audio window $A_w$ and reference video $R$:
\begin{equation}
     m^*(0) = E_{\theta}(A_w, R, m, t)
\end{equation}
where $t$ represents the time step, and  $E_{\theta}$ consists of two encoders: the audio encoder $E_{aud}$ and the style encoder $E_{sty}$.
We denote the style code obtained from the style encoder as $s$.
During inference, based on the style code $s$, DreamTalk employs the DDPM sampling algorithm to generate predicted facial motions.The generated facial motions are subsequently rendered into videos by the PIRenderer.

\subsection{Training Details}
.
Empatheia comprises four complementary training stages, each playing a crucial role in the overall process.
In the following sections, we will provide a more detailed overview of the specific training details for each stage.

\noindent\\ \paratitle{\(\blacktriangleright\) Training step1: CoE Learning Stage}

\(\triangledown\) Training Data: As shown in D.1, the CoE training data consists of dialogue context and targets. 
The dialogue context includes both the dialogue history and the user’s current query input, while the target represents the CoE reasoning process and the empathetic response.

\(\triangledown\)Training Objective: In this phase, we aim for the LLM to learn how to engage in step-by-step reasoning based on CoE to generate high-quality empathetic responses, effectively completing the ERG task. 
This stage lays the groundwork for subsequent content consistency learning.

\(\triangledown\)Training Method: During this phase, we employ LoRA and Negative Log-Likelihood (NLL) loss to fine-tune Vicuna, training it to engage in empathetic reasoning by calculating the loss associated with the target component.

\noindent\\ \paratitle{\(\blacktriangleright\)Training step2: Content Consistency Learning} 

\(\triangledown\) Training Data: The training data for this phase not only includes the dialogue context from the first phase but also incorporates the pre-extracted audio and video content representations corresponding to each response text. 
Specifically, for speech, we input the response text into the two text encoders, $E_{aco}$ and BERT, in StyleTTS2. 
The resulting embeddings, $h_{text}$ and $h_{bert}$, are concatenated to produce the gold content representation for speech. 
For video, we utilize the audio encoder $E_{aud}$ in DreamTalk to obtain the gold content representation for video. 

\(\triangledown\)Training Objective: The training objective for this phase is to align the speech and video representations output by the content synthesizer with the gold content representation. 
This alignment enables the CS to learn to produce outputs that are both accurate and exhibit consistent content signal features. 

\(\triangledown\)Training Method: In this phase, we freeze the parameters of the LLM and only fine-tune the parameters of the CS module. We calculate the L2 loss between the predicted content representation and the ground truth content representation.

\noindent\\ \paratitle{\(\blacktriangleright\)Training step3: Style Aligning and Consistency Learning} 

\(\triangledown\) Training Data: The input data format for this phase is illustrated in D.1, which uses multimodal special token placeholders in a dialogue format. 
The target training data consists of the audio and video style representations. The gold audio style representation is pre-extracted through the $E_{ref}$ encoder of StyleTTS2, which processes the gold speech, while the gold video style representation is pre-extracted using DreamTalk’s $E_{sty}$ encoder, which processes the gold video.
To ensure style consistency, in this stage, we also apply supervised constraints to the SD module using emotion and profile labels.

\(\triangledown\)Training Objective:The goal of this phase is to align the speech and video style representations output by the SD module with the gold style representations, allowing the SD module to learn to produce accurate and consistent style signals.

\(\triangledown\)Training Method: During training, we use ImageBind and a mapping layer to extract audio and video features from the multimodal dialogue, which are then used to replace the special token placeholders. As in the previous phase, we freeze the parameters of the LLM and train the SD module using L2 loss and classification loss.

\noindent\\ \paratitle{\(\blacktriangleright\)Training step4: Overall Training} 

\(\triangledown\) Training Data: As illustrated in D.1, the training data format for this stage encompasses all three modalities, integrating the training data from previous steps. 
The target data also includes all pre-extracted multimodal content and style representations.

\(\triangledown\)Training Objective: This stage constitutes a comprehensive fine-tuning process aimed at equipping the model with the full capability of multimodal empathetic responses. 
It aims to provide not only sufficiently empathetic textual responses but also synchronized, content- and style-accurate multimodal responses.

\(\triangledown\)Training Method: In this stage, we employ LoRA for fine-tuning the LLM while simultaneously updating the CS and SD modules. 
The overall loss is calculated as the weighted sum of the losses from the previous three stages.

\section{Extended Experiment Settings}
\label{Extended Experiment Settings}
In this section, we provide more detailed experimental settings, encompassing hyperparameters, the construction of the pipeline, and the specifics of the evaluation.

\subsection{Hyper-Parameter Settings}
The basic hyperparameter settings for our training process are shown in \autoref{tab:hyper-1}.
\vspace{-3mm}
\begin{table}[htbp]
\fontsize{8}{8}\selectfont
\setlength{\tabcolsep}{4mm}
\centering
\caption{Some Basic Hyper-Parameter Settings
}
\vspace{-3mm}
\begin{tabular}{lc}
\toprule
 \bf Basic Hyper-Parameters & \bf Value\\
\midrule
$\alpha$ & 0.2 \\
$\beta$ & 0.3 \\
lora\_r & 16 \\
lora\_alpha & 32 \\
Num of Generate Tokens & 16\\
Num of layers in Transformer Block & 4 \\
\bottomrule
\end{tabular}
\vspace{-3mm}
\label{tab:hyper-1}
\end{table}

In order to reduce the usage of GPU memory and accelerate the training speed, we utilized Deepspeed to train our model, with some of the parameter settings presented in the \autoref{tab:hyper-2}.
\vspace{-3mm}
\begin{table}[htbp]
\fontsize{8}{8}\selectfont
\setlength{\tabcolsep}{4mm}
\centering
\caption{Some Deepspeed Hyper-Parameter Settings
}
\vspace{-3mm}
\begin{tabular}{lc}
\toprule
 \bf Deepspeed Hyper-Parameters & \bf Value\\
\midrule
fp16 & True\\
lr & 5e-5\\
weight decay & 0.001\\
train batch size & 32 \\
train micro batch size per gpu & 4 \\
gradient accumulation steps & 8 \\
zero optimization & stage2 \\
\bottomrule
\end{tabular}
\vspace{-5mm}
\label{tab:hyper-2}
\end{table}

\subsection{Pipeline Baseline Implementation Details}
We utilize the same LLM backbone as Empatheia to construct our pipeline baseline, also adopting StyleTTS2 and DreamTalk as multimodal generators. 
Unlike Empatheia which employs implicit instructional features embedding for end-to-end signal passing, the pipeline solely relies on explicit meta-response texts to propagate generated instructional signals.
Also, the pipeline system will not be equipped with the CoE strategy.
For the training of our pipeline, we exclusively utilize text data and additionally introduced profile attributes. 
Consequently, after training, we can guide the subsequent generation of speech and talking face videos based on the response text, emotion, and profile information outputted by the model.

\subsection{Evaluation Details}

Here, we detail how we conduct the evaluation for the various tasks, including the textual ERG task, speech generation task, and talking head generation task on the AvaMERG dataset.

\subsubsection{Text ERG Evaluation Metrics}

\noindent\\\vspace{1mm}For the text ERG task, we employ three evaluation metrics: Emotion Accuracy (Acc), and Distinct metrics (Dist-1 and Dist-2). These metrics are designed to evaluate both the emotional correctness of the generated responses and their lexical diversity.

\paratitle{Emotion Accuracy (Acc).}
Acc measures the percentage of correctly predicted emotions in the generated responses. The correct emotion is defined as the exact match with the ground truth emotion label. Acc is computed as follows:
\begin{equation}
Acc = \frac{\# \text{correct emotions}}{\# \text{total emotions}}
\end{equation}
where `correct emotions' represents the total number of responses where the predicted emotion matches the gold label.

\paratitle{Distinct-1 (Dis-1).}
This evaluates the diversity of unigrams (single words) in the generated responses. It is defined as the ratio of unique unigrams to the total number of unigrams in the generated text:
\begin{equation}
Dis\text{-}1 = \frac{\# \text{unique unigrams}}{\# \text{total unigrams}}
\end{equation}

A higher Distinct-1 score indicates more lexical diversity and reduces the likelihood of repetitive responses.

\paratitle{Distinct-2 (Dis-2).}
This is analogous to Distinct-1, but it measures the diversity of bigrams (two consecutive words). It is computed as follows:
\begin{equation}
Dis\text{-}2 = \frac{\# \text{unique bigrams}}{\# \text{total bigrams}}
\end{equation}

A higher Distinct-2 score indicates that the model is generating more contextually diverse phrases.

\paratitle{Human Evaluation of Textual ERG.} For human evaluation, we randomly selected 200 dialogues from the test dataset. Taking into account both the cost of human labor and the reliability of the results, we chose competitive models from the last year as representative baselines. Given the dialogue context and the responses generated by these models, we engaged three annotators to score the responses using a majority voting system. They rated each response on a scale from 1 to 5 (1: not at all, 3: adequate, 5: excellent) based on four key criteria: Empathy, Coherence, Informativity, and Fluency. Specifically, these criteria are:

\begin{itemize}
    \item[1)] \textbf{Empathy} (Emp): whether the response demonstrates an understanding of the user’s emotions and experiences, and responds appropriately.
    
    \item[2)] \textbf{Coherence} (Coh): whether the response is logically consistent and contextually relevant.
    
    \item[3)] \textbf{Informativity} (Inf): whether the response provides useful and meaningful information.
    
    \item[4)] \textbf{Fluency} (Flu): whether the response is grammatically well-formed and easy to read.
    
\end{itemize}

\subsubsection{Speech Generation Evaluation Metrics}

\noindent\\\vspace{1mm}For the speech generation component of MERG, we use subjective and objective metrics to assess the quality and emotional expressiveness of the generated speech.

\paratitle{Mean Opinion Score (MOS).}
MOS is a subjective evaluation metric where human evaluators rate the naturalness of the generated speech on a scale of 1 to 5, with 5 indicating highly natural speech. MOS is computed as:
\begin{equation}
MOS = \frac{\sum \text{ratings}}{\# \text{evaluators}}
\end{equation}

\paratitle{Similarity MOS (SMOS).}
SMOS measures how similar the generated speech is to a reference speech sample in terms of emotional tone and expressiveness. Like MOS, it is rated on a 5-point scale by human evaluators.
\begin{equation}
SMOS = \frac{\sum \text{similarity ratings}}{\# \text{evaluators}}
\end{equation}

\subsubsection{Talking Head Avatar Generation Evaluation Metrics}

\noindent\\\vspace{1mm}For evaluating the quality of the generated talking head avatars, we employ a combination of perceptual and geometric metrics to measure the visual fidelity and synchronization of the avatar’s lip movements with the speech.

\paratitle{Cumulative Probability of Blur Detection (CPBD).}
CPBD quantifies the perceptual sharpness of the generated video frames by estimating the probability that an observer would detect blurring based on edge width analysis. Higher CPBD values indicate sharper images and fewer perceived blurs. The CPBD score is calculated by first analyzing the cumulative distribution of edge widths in an image. For each edge in the image, the probability that the edge is perceived as blurred is computed. The overall CPBD score is the average of these probabilities over all edges. 
The formula can be expressed as:
\begin{equation}
CPBD = \frac{1}{N} \sum_{i=1}^{N} P(e_i)
\end{equation}

where \( P(e_i) \) is the probability of blur detection for edge \( e_i \), and \( N \) is the total number of detected edges in the image. This metric gives a perceptual estimate of how likely it is for human viewers to notice blur across the video frames.

\paratitle{Structural Similarity Index Measure (SSIM).}
SSIM assesses the visual similarity between the generated avatar video frames and the ground truth video frames. It measures the perceived quality in terms of luminance, contrast, and structure:
\begin{equation}
SSIM = \frac{(2 \mu_x \mu_y + C_1)(2 \sigma_{xy} + C_2)}{(\mu_x^2 + \mu_y^2 + C_1)(\sigma_x^2 + \sigma_y^2 + C_2)}
\end{equation}

where \( \mu_x \) and \( \mu_y \) are the means of the generated and reference images, \( \sigma_x^2 \) and \( \sigma_y^2 \) are the variances, and \( \sigma_{xy} \) is the covariance of the two images. \( C_1 \) and \( C_2 \) are constants to stabilize the division.

\paratitle{SyncNet Confidence Score (Sync$_{cf}$).}
Sync$_{cf}$ is used to evaluate the synchronization between the generated speech and the avatar’s lip movements. It measures the alignment between the visual lip movements and the audio, with higher scores indicating better synchronization:
\begin{equation}
Sync_{cf} = \frac{\text{aligned speech and lip movements}}{\text{total frames}}
\end{equation}

\subsubsection{Human Evaluation of MERG}

\noindent\\\vspace{1mm}To better evaluate the performance of models on the MERG task, we have newly defined six human evaluation metrics, specifically:

\textbullet\ \textbf{Speech Content Accuracy (SCA):} Assesses whether the content in the generated audio is complete and consistent with the response text.
\begin{equation}
SCA = Consistency(C_{gen}^s, C_{gold}^s)
\end{equation}
where $C_{gene}^s$ represent the generation speech's content, and the $C_{gold}^s$ represent the gold speech's content.

\textbullet\ \textbf{Video Content Accuracy (VCA):} Evaluates whether the face in the generated video accurately and fluently reads out the response text.
\begin{equation}
VCA = Consistency(C_{gen}^v, C_{gold}^v)
\end{equation}
where $C_{gene}^s$ represent the generation video's content, and the $C_{gold}^s$ represent the gold video's content.

\textbullet\ \textbf{Speech Style Accuracy (SSA):} Determines the accuracy of the emotion conveyed in the generated speech and whether the voice matches the intended character profile.
\begin{equation}
SSA = Similarity(S_{gen}^s,S_{gold}^s)
\end{equation}
where $S_{gene}^s$ represent the generation speech's style, and the $S_{gold}^s$ represent the gold speech's style.

\textbullet\ \textbf{Video Style Accuracy (VSA):} Assesses the accuracy of the emotion expressed by the generated avatar and whether the avatar's appearance and behavior match its intended profile.
\begin{equation}
VSA = Similarity(S_{gen}^v,S_{gold}^v)
\end{equation}
where $S_{gene}^v$ represent the generation video's style, and the $S_{gold}^v$ represent the gold video's style.

\textbullet\ \textbf{Multimodal Content Consistency (MCC):} This is a comprehensive comparison that evaluates the consistency of content across the three modalities (speech, video, and text).
\begin{equation}
MCC = Consistency(C_{gen}^s, C_{gen}^v)
\end{equation}

\textbullet\ \textbf{Multimodal Style Consistency (MSC):} Provides an overall evaluation of the consistency of style across the three modalities.
\begin{equation}
MSC = Similarity(S_{gen}^s,S_{gold}^v)
\end{equation}

During the testing process, we deliberately engaged three experienced evaluators to ensure the comprehensiveness and accuracy of the evaluation.
The task of these three evaluators was to conduct meticulous assessments of 200 test cases based on the aforementioned six aspects, grading them from 1 to 5 according to the model's performance (where 1 represents extremely poor performance and 5 represents excellence).
Prior to the testing, we organized a detailed training session to clarify the scoring criteria, grading rules, and handling methods for potential special cases, thereby further ensuring the reliability of the scores.

\section{More Experiments and Analysis}
In this part, we present additional experimental results and analyses to further demonstrate the performance of Empatheia.

\subsection{Impact of Hyperparameter Settings}
Here we study the impact of various hyperparameters on model performance, including the number of transformer blocks, the number of special signal tokens used for multimodal generation, and the loss weights alpha and beta.

\begin{figure}[!t]
\centering
\includegraphics[width=0.99\columnwidth]{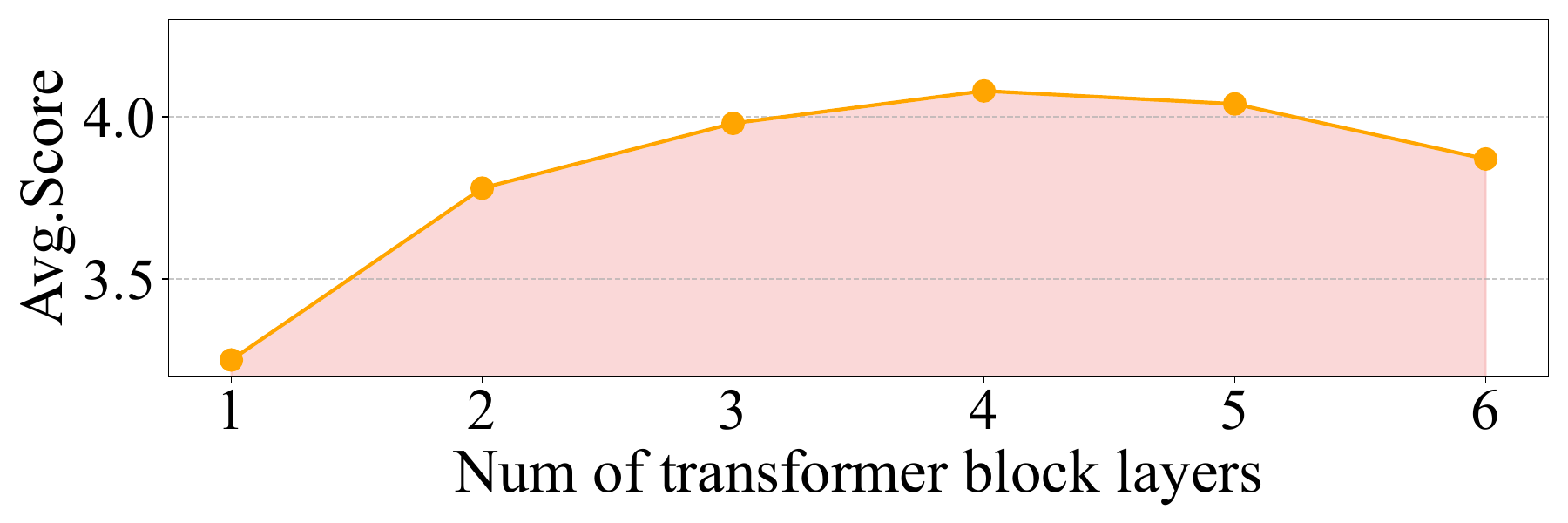}
\vspace{-3mm}
\caption{Impact of the transformer layer number.}
\label{fig:hypara1}
\vspace{-3mm}
\end{figure}

\begin{figure}[!t]
\centering
\includegraphics[width=0.99\columnwidth]{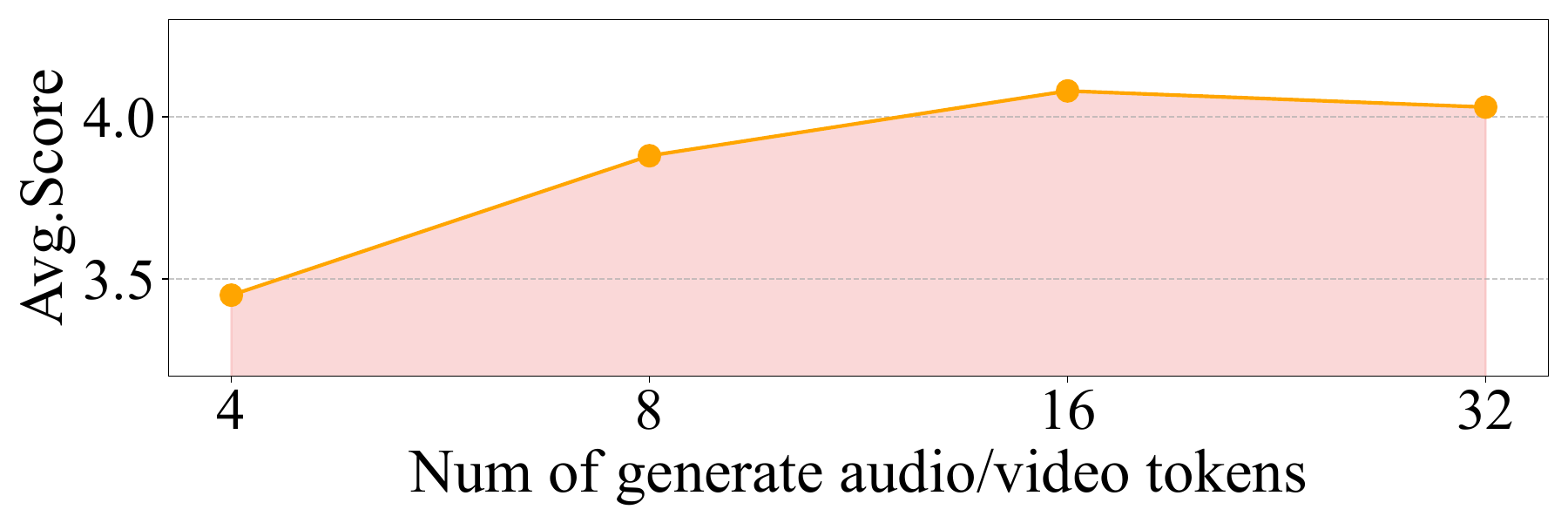}
\vspace{-3mm}
\caption{Impact of special token numbers.}
\label{fig:hypara2}
\vspace{-3mm}
\end{figure}

\autoref{fig:hypara1} illustrates the model’s performance when varying the number of transformer blocks in CS and SD modules, specifically evaluating the model’s output on multiple metrics. 
The “Avg. Score” reflects the average performance across six carefully crafted manual evaluation metrics, which together capture various aspects of the model’s ability to generate multimodal empathetic responses. 
The results show that the transformer block reaches its peak performance at the 4 layers. Deeper layers can, to some extent, enhance the model's learning ability by capturing more complex interactions between patterns. 
However, further increasing the number of transformer layers did not lead to significant improvement. 
Therefore, Empatheia sets the number layers of transformer blocks to 4.
\autoref{fig:hypara2} illustrates the impact of different numbers of audio-visual special tokens on model performance. As shown, the model's performance peaks when the number of tokens reaches 16.

\begin{figure}[!t]
\centering
\includegraphics[width=0.70\columnwidth]{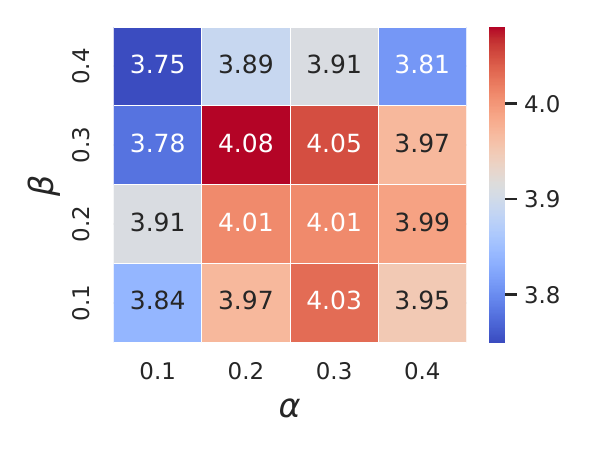}
\vspace{-5mm}
\caption{Impact of loss weights.}
\label{fig:hypara3}
\vspace{-3mm}
\end{figure}

We scale the content and style learning losses using \(\alpha\) and \(\beta\) to bring them closer to the magnitude of the Vicuna's inherent loss, aiming to achieve balanced training. 
The experiments on loss hyperparameters are visualized in a heatmap in \autoref{fig:hypara3}, the model performs better when \(\alpha\) and \(\beta\) are relatively balanced, indicating that a more balanced loss contributes to the model's convergence to an optimal solution.

\subsection{Impact of Training Data Amounts}
\begin{figure}[!t]
\centering
\hspace{-4mm}
\subfloat{
		\includegraphics[width=0.5\columnwidth]{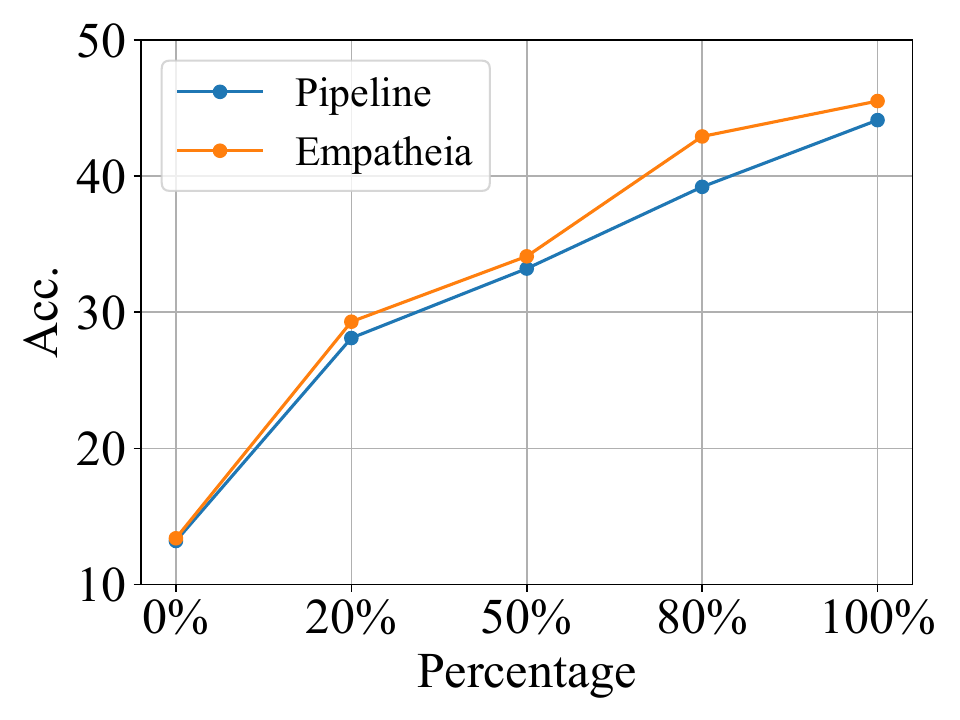}}
\subfloat{
		\includegraphics[width=0.5\columnwidth]{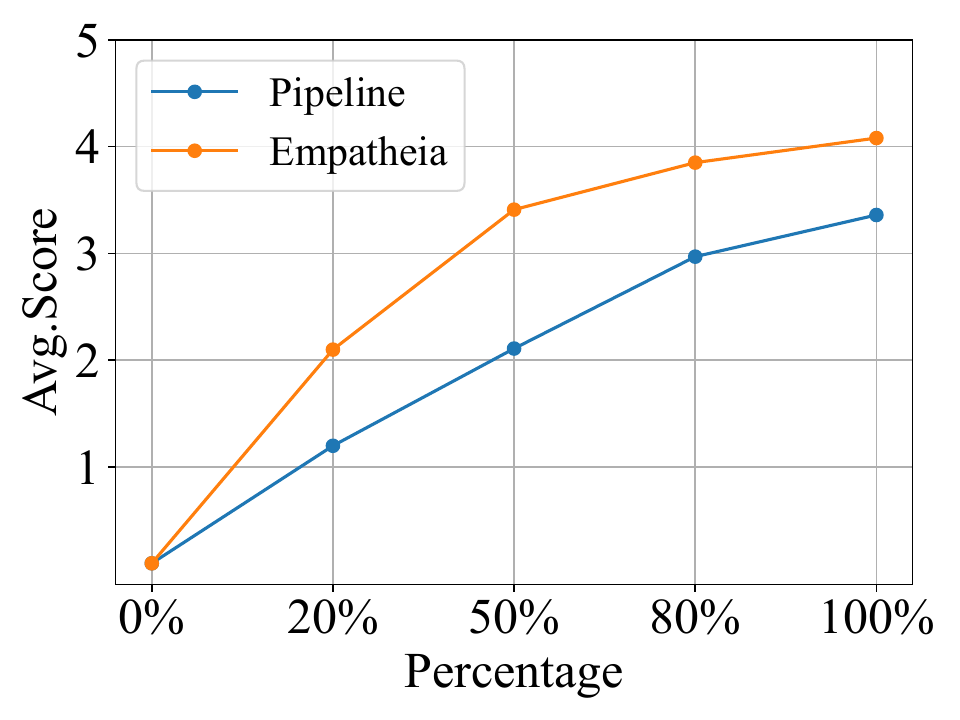}}
\vspace{-2mm}
\caption{Performance comparison under different quantities of training data.}
\label{fig_data_amounts}
\end{figure}

\autoref{fig_data_amounts} presents the performance of Empatheia and the pipeline under varying proportions of training data.
We incrementally increased the proportion of the training dataset from 0\%, 20\%, 50\%, to 100\%. 
It is observable that both the Pipeline and Empatheia exhibit enhanced performance as the volume of data increases.
Notably, due to a lack of understanding of multimodality and a synchronous module for multimodal generation, the average score growth rate of the Pipeline in terms of accuracy and consistency in multimodal generation lags significantly behind Empatheia. 
This once again demonstrates the robust potential of Empatheia.

\subsection{Impact of COE steps}
\begin{figure}[!t]
\centering
\vspace{-2mm}
\includegraphics[width=0.96\columnwidth]{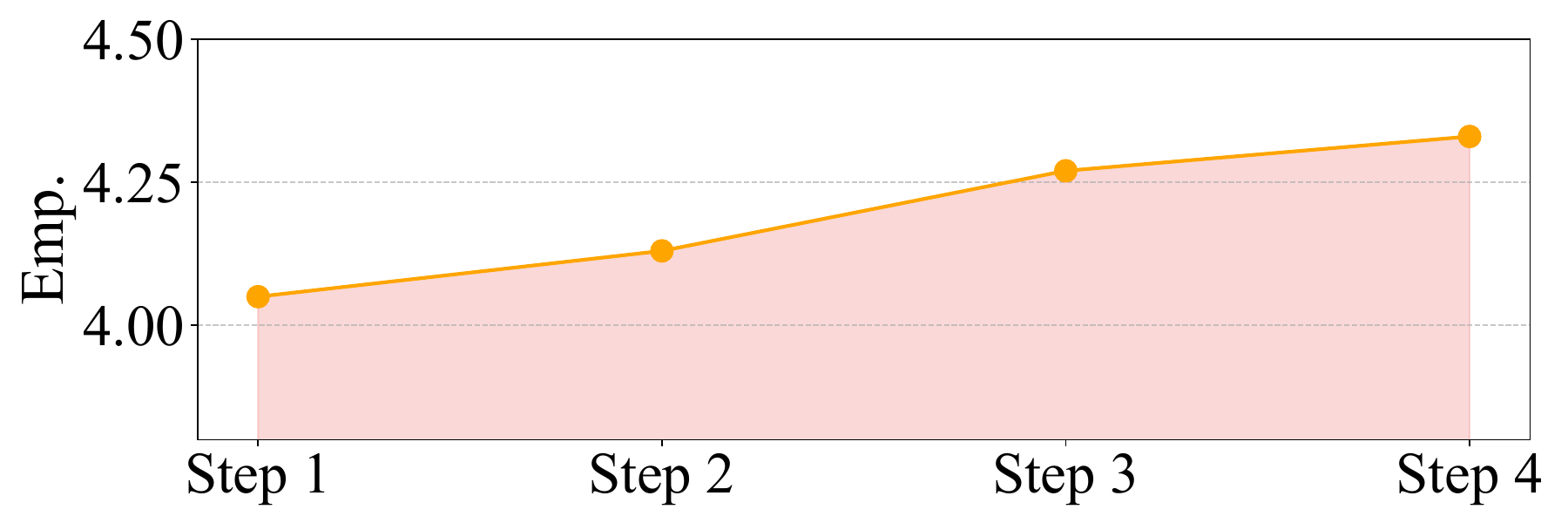}
\vspace{-3mm}
\caption{In-Depth Analysis of the Effectiveness of CoE.}
\label{CoE_step_analysis}
\vspace{-3mm}
\end{figure}

CoE works by breaking down the overall hard problem into semantically coherent and easier sub-problems.
To gain a deeper understanding of how each step of the CoE reasoning process contributes to the ERG, we conduct ablation experiments on CoE, by progressively using more leveled steps of CoE prompts. 
The results, as illustrated in \autoref{CoE_step_analysis}, show that as Empatheia advances through the steps of CoE, the model's empathetic performance improves significantly, where the capacity becomes stronger with each level of complexity, leading to more accurate and contextually appropriate responses.
This indicates that each individual step in CoE plays a vital role, enhancing the model's ability to comprehensively analyze and understand user's emotion and intention. 
By systematically breaking down the CoE into distinct phases, we highlight that the model’s empathetic performance has been progressively enhanced.

\subsection{More Case Study}
\label{More Case Study}

We further provide several examples to demonstrate the performance differences between our model and pipeline.

\begin{figure}[t]
\centering
\includegraphics[width=1\columnwidth]{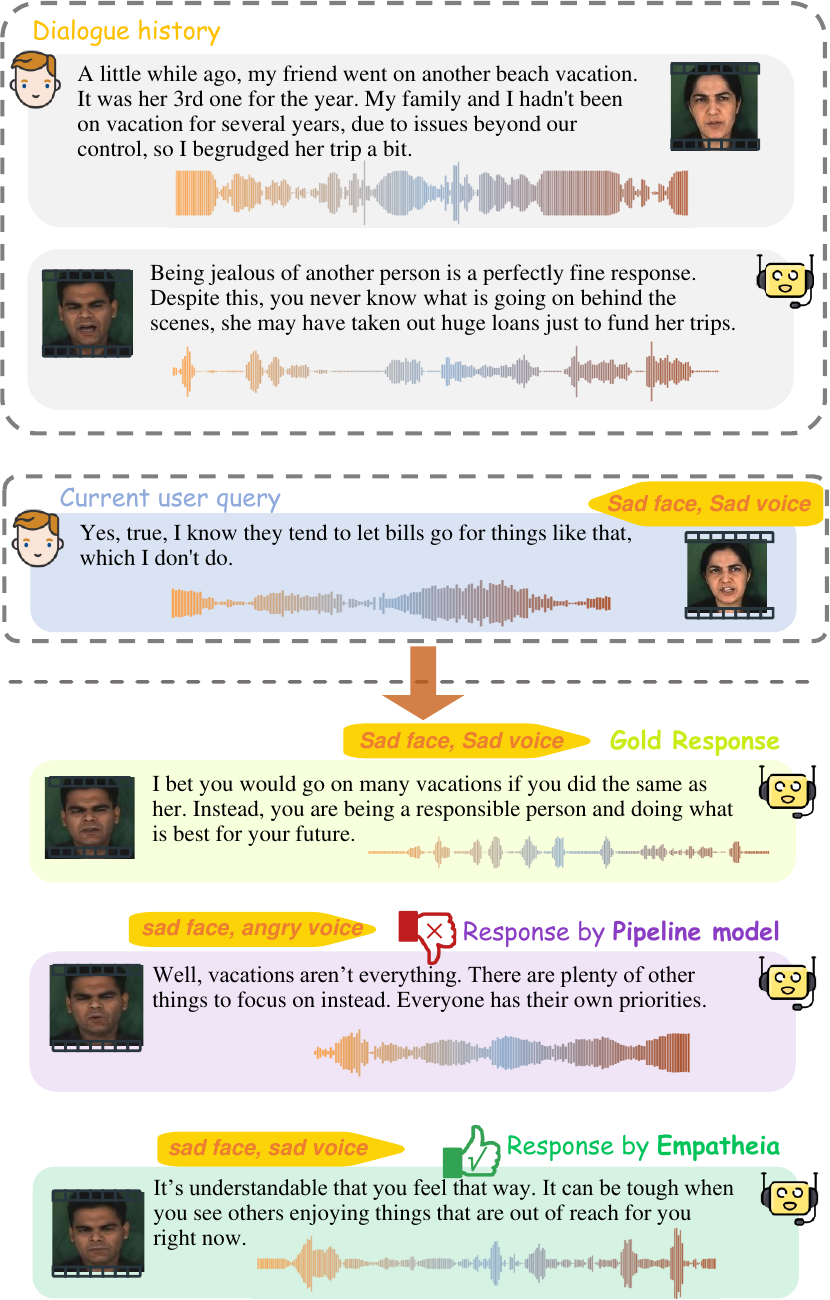}
\caption{Qualitative results A of MERG.}
\label{fig:appendix-1}
\vspace{\fill}
\end{figure}

\begin{figure}[t]
\centering
\includegraphics[width=1\columnwidth]{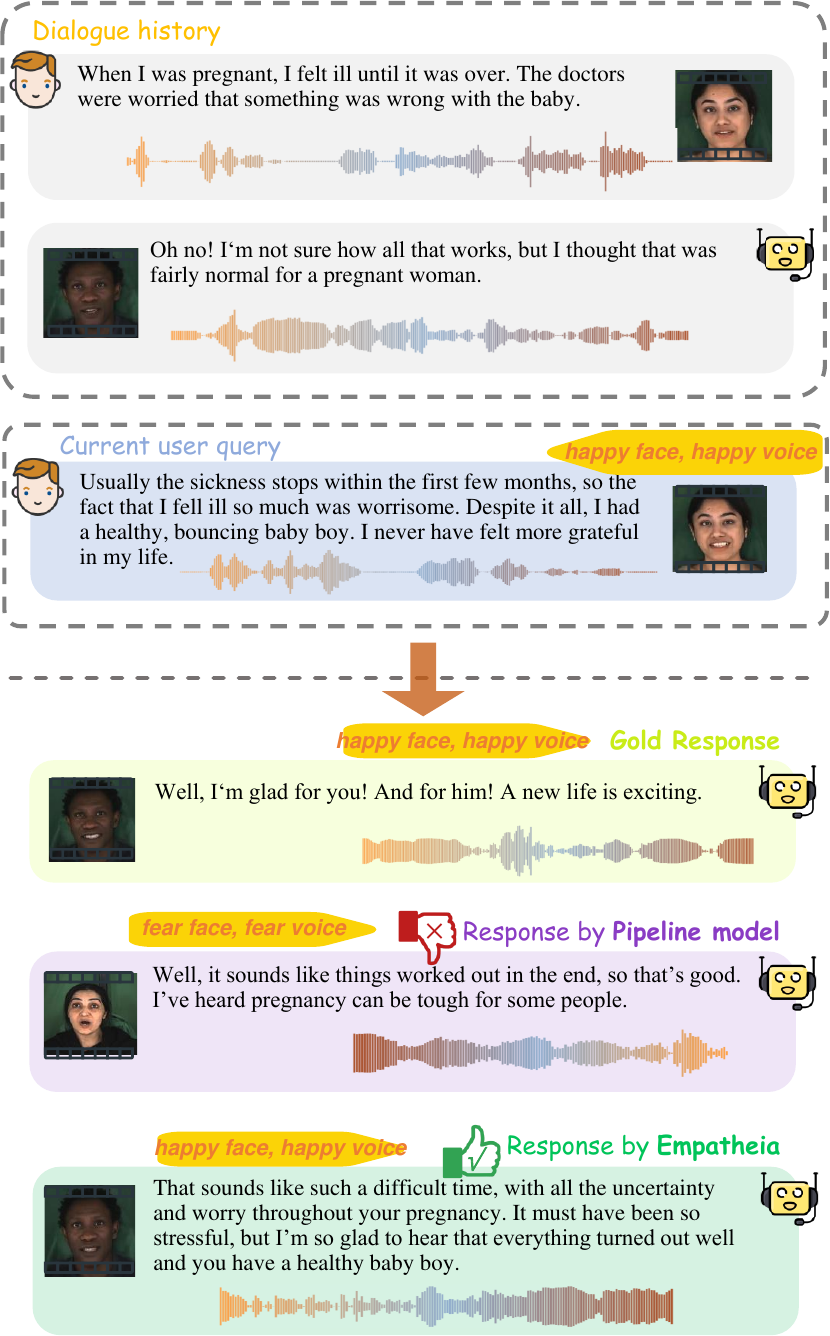}
\caption{Qualitative results B of MERG.}
\label{fig:appendix-2}
\end{figure}

\begin{figure}[!t]
\centering
\includegraphics[width=1\columnwidth]{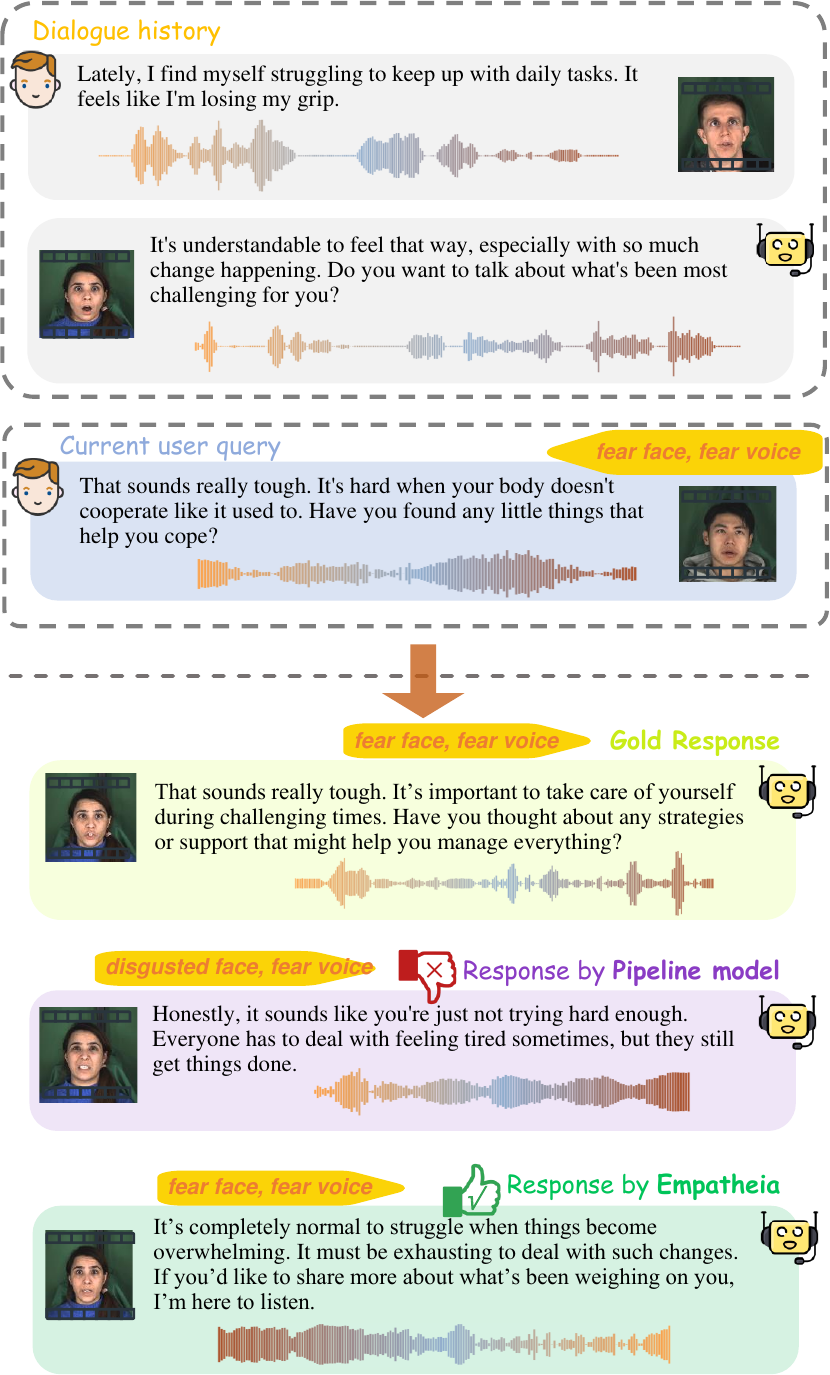}
\caption{Qualitative results C of MERG.}
\label{fig:appendix-3}
\end{figure}

\begin{figure}[!t]
\centering
\includegraphics[width=1\columnwidth]{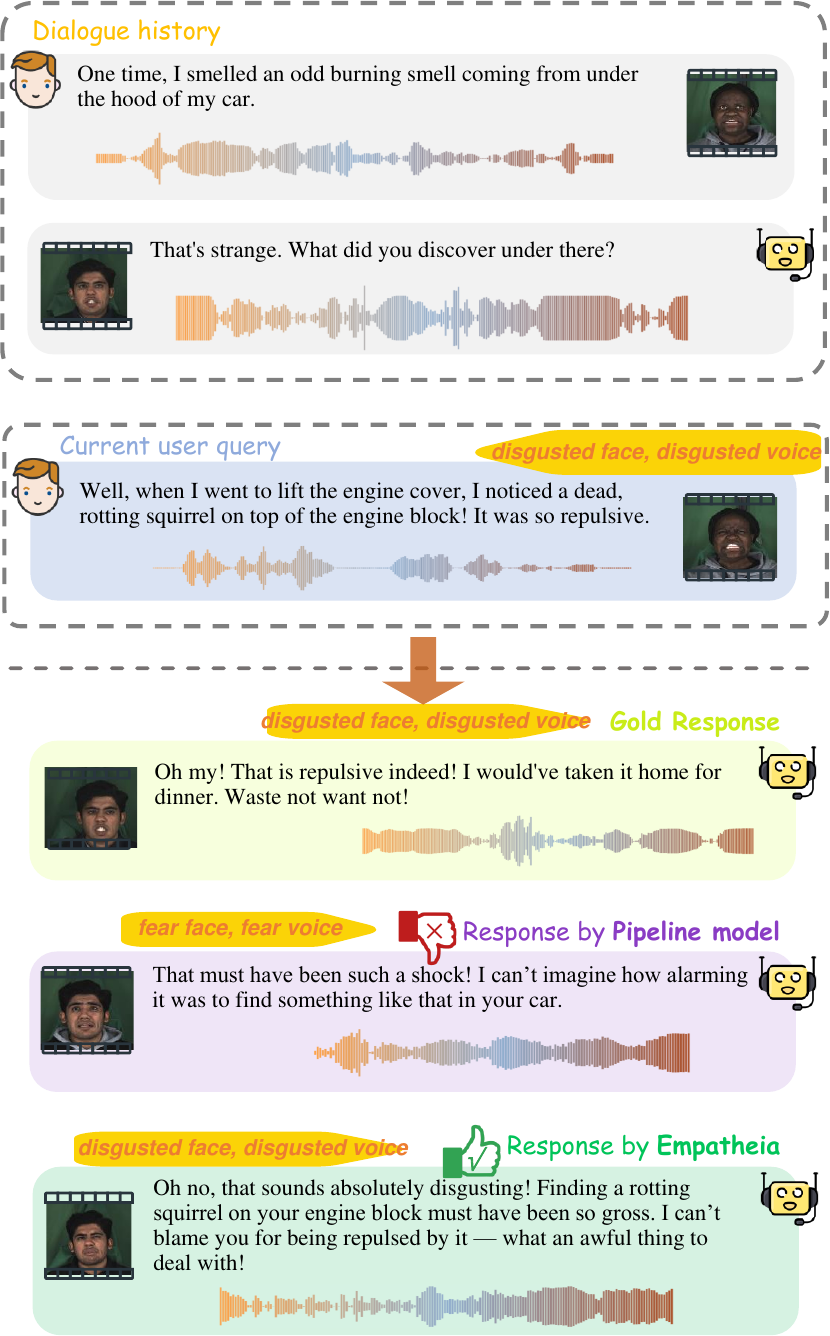}
\caption{Qualitative results D of MERG.}
\label{fig:appendix-4}
\end{figure}

\paratitle{MERG Performance Comparisons.}
As illustrated in \autoref{fig:appendix-1}, \autoref{fig:appendix-2}, \autoref{fig:appendix-3} and \autoref{fig:appendix-4}, our model exhibits more consistent and empathetic multimodal responses, indicating Empatheia's deeper understanding of multimodal contexts. This is attributed to our multi-stage training approach, as well as the advanced content synchronizer and style deconstructor employed.

\begin{figure}[!t]
\centering
\includegraphics[width=1\columnwidth]{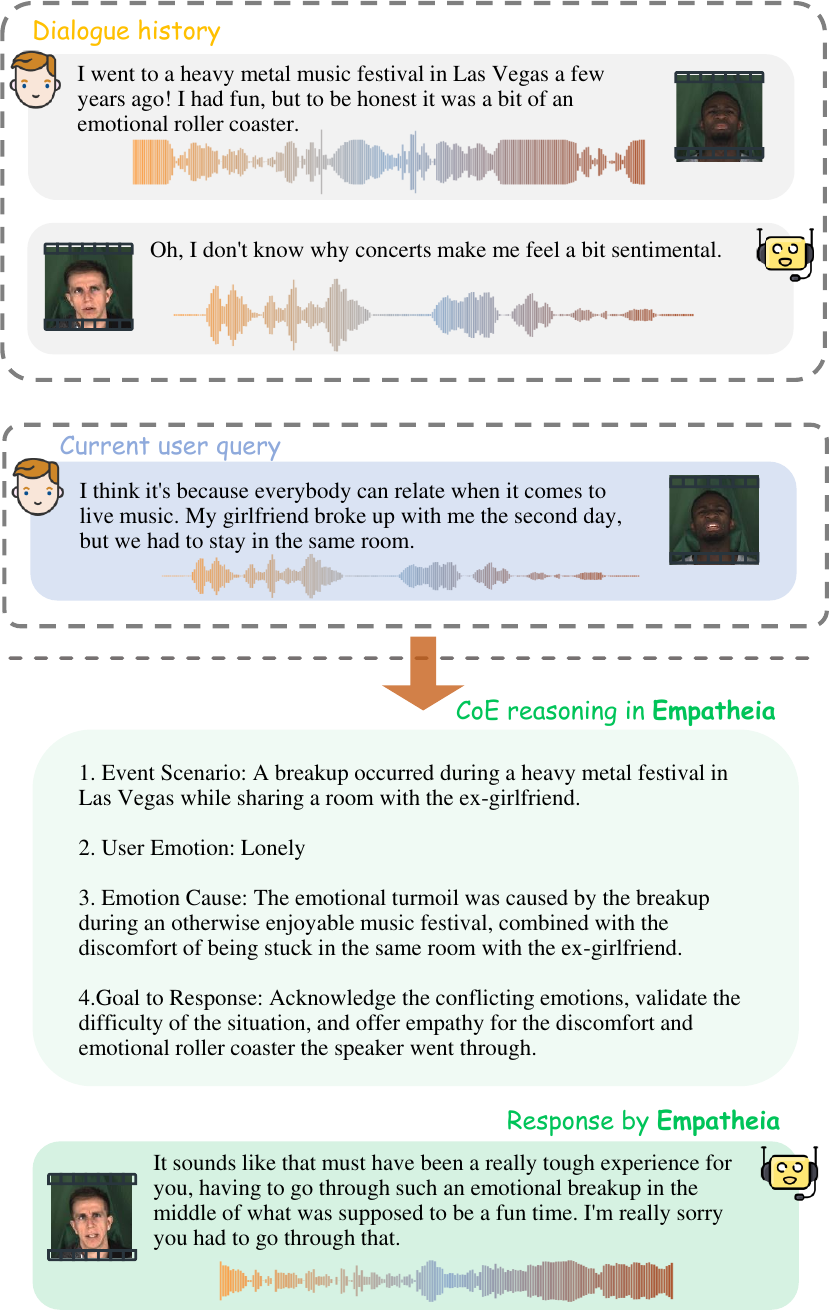}
\caption{Qualitative results A of CoE mechanism.}
\label{fig:coe-1}
\end{figure}

\begin{figure}[!t]
\centering
\includegraphics[width=1\columnwidth]{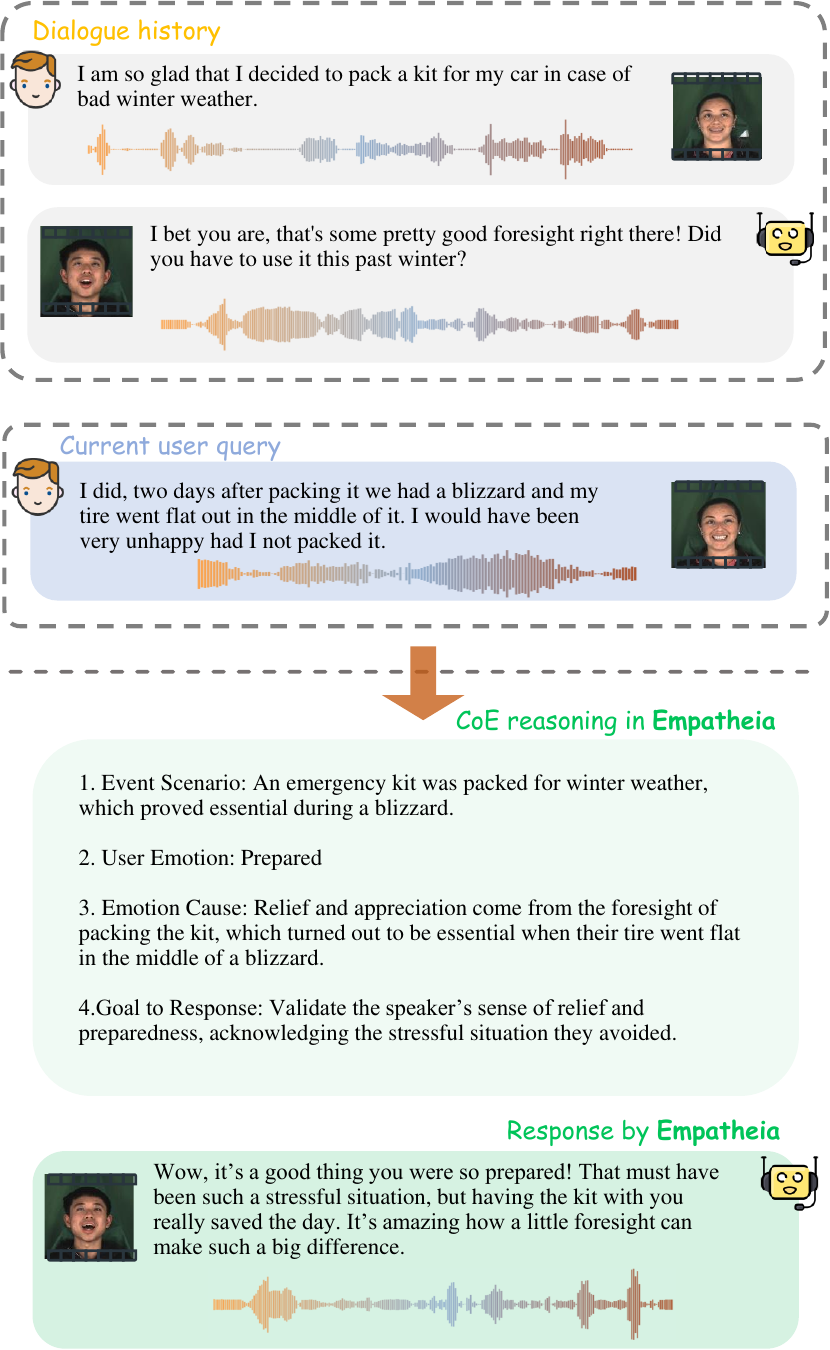}
\caption{Qualitative results B of CoE mechanism.}
\label{fig:coe-2}
\end{figure}

\begin{figure}[!t]
\centering
\includegraphics[width=1\columnwidth]{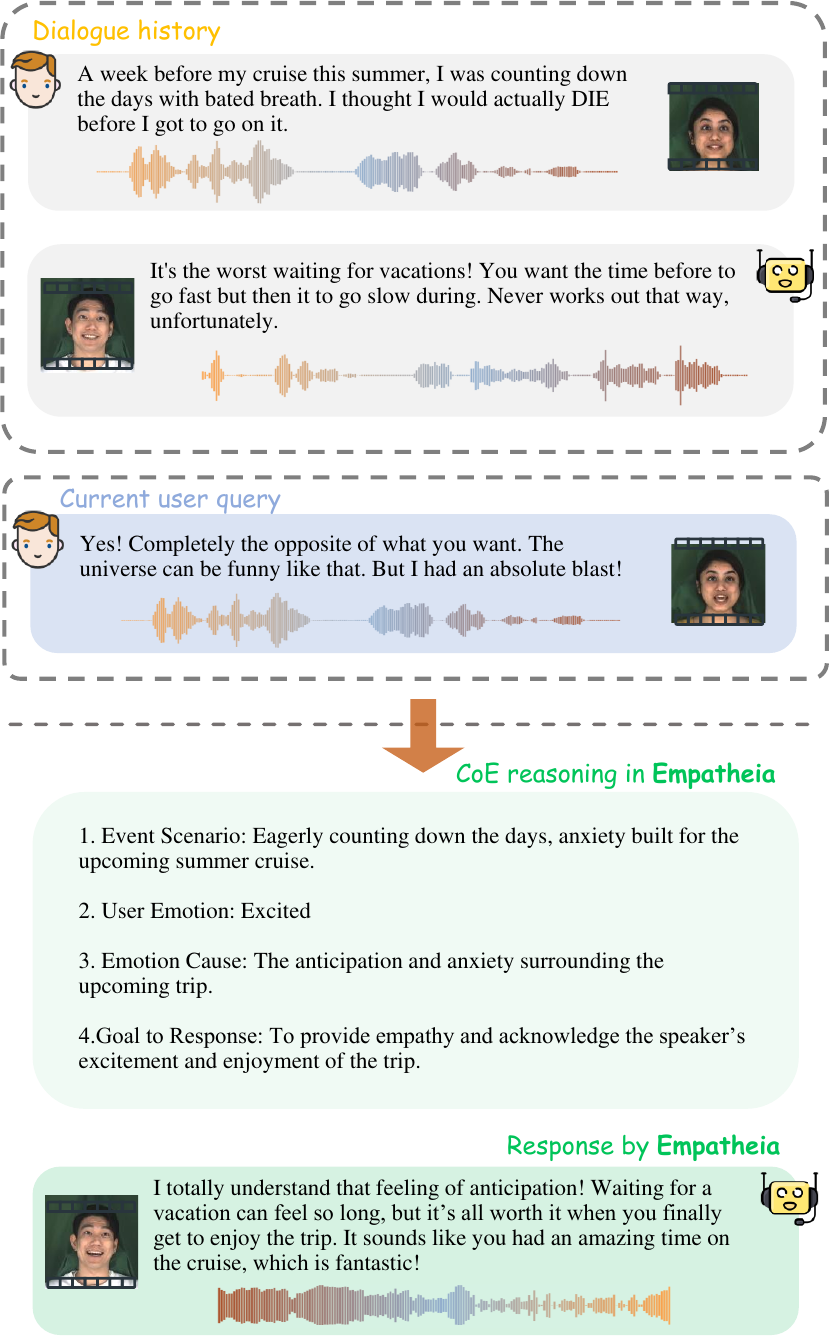}
\caption{Qualitative results C of CoE mechanism.}
\label{fig:coe-3}
\end{figure}

\begin{figure}[!t]
\centering
\includegraphics[width=1\columnwidth]{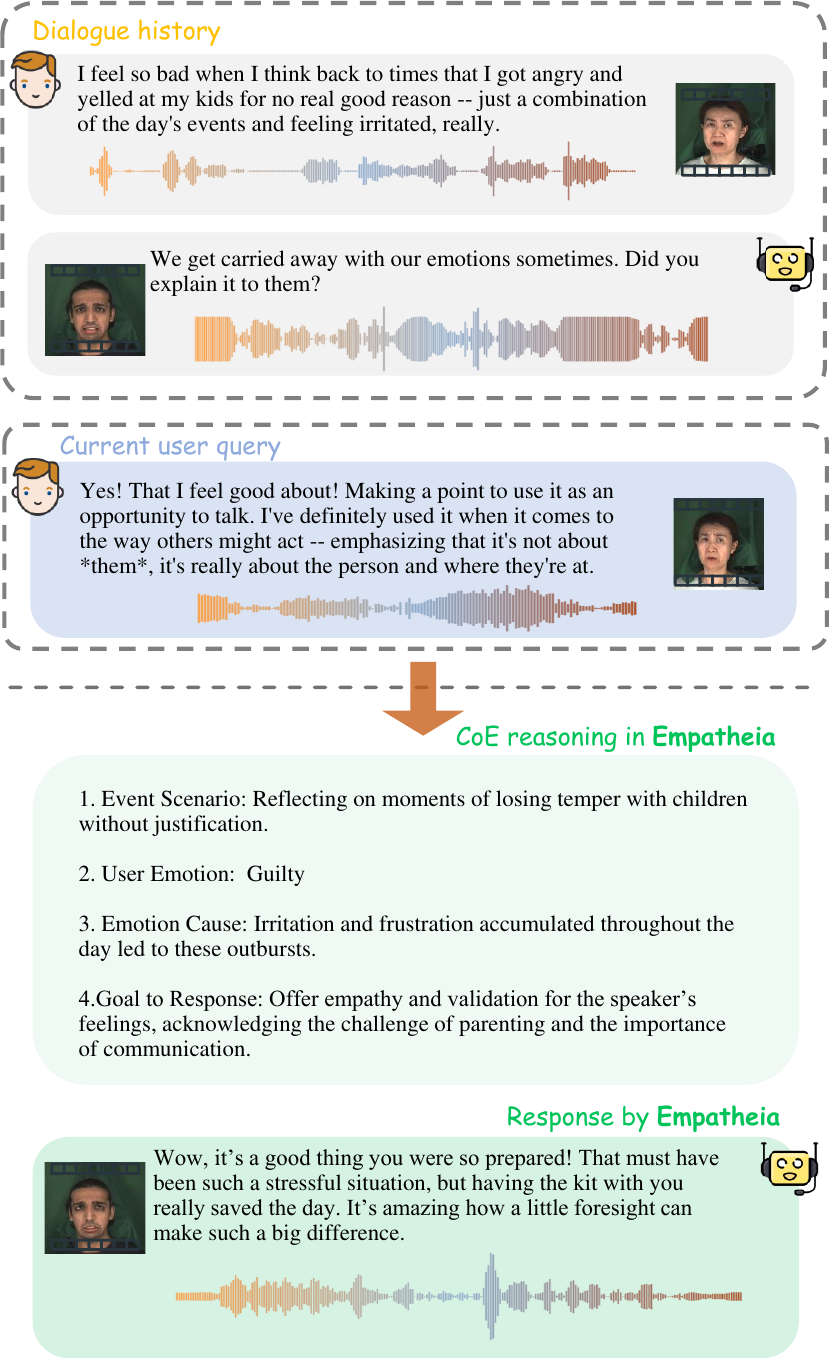}
\caption{Qualitative results D of CoE mechanism.}
\label{fig:coe-4}
\end{figure}

\paratitle{CoE Qualitative Results.}
We also present several examples to visualize the reasoning process of CoE. As shown in \autoref{fig:coe-1}, \autoref{fig:coe-2}, \autoref{fig:coe-3} and \autoref{fig:coe-4}, CoE can significantly enhance the model's empathy capabilities and the accuracy of perceiving users' emotions, the underlying intention, and finally the correct response.

\newpage

\color{white}{xxxxxx}

\newpage

\color{white}{xxxxxx}

\newpage

\color{white}{xxxxxx}

\newpage

\color{white}{xxxxxx}

\newpage

\color{white}{xxxxxx}

\newpage

\color{white}{xxxxxx}

\newpage

\color{white}{xxxxxx}

\newpage

\color{white}{xxxxxx}

\newpage

\end{document}